\documentclass[]{caps}
\usepackage{psfig,latexsym,longtable,lscape}
\def\ed{\end{document}}
\def\hide#1{}

\hide{
\def\rem#1{{\bf(#1)}}
\def\editnote#1{{\it\bf(editnote: #1)}}
\def\mylabel#1{\label{#1}\ {\tt #1}}
\def\myref#1{\ref{#1}\ {\tt(#1)}}
}
\def\rem#1{}
\def\editnote#1{}
\def\mylabel#1{\label{#1}}
\def\myref#1{\ref{#1}}
\def\b{$\bullet$}
\def\c{$\circ$}
\def\ms#1#2{\multispan#1#2}
\def\bd{\begin{document}}
\def\ed{\end{document}}
\def\cl#1{\centerline#1}
\def\about{$\sim$}
\def\p{\ifmmode\pm\else$\pm$\fi}
\def\approxgt{\ifmmode \rlap{$>$}{}_{{}_{{}_{\textstyle\sim}}} \else%
$\rlap{$>$}{}_{{}_{{}_{\textstyle\sim}}}$\fi}
\def\approxlt{\ifmmode \rlap{$<$}{}_{{}_{{}_{\textstyle\sim}}} \else%
$\rlap{$<$}{}_{{}_{{}_{\textstyle\sim}}}$\fi}
\def\msun{\ifmmode \rm{M}_\odot \else M$_\odot$\fi}     
\def\mdot{\ifmmode \dot M \else $\dot M$\fi}    
\def\Lx{$L_x$}
\def\LEdd{$L_{Edd}$}
\def\Lu{L$_u$}
\def\Ll{L$_\ell$}
\def\Llow{$L_{\ell ow}$}
\def\Lb{L$_b$}
\def\Lbt{L$_{b2}$}
\def\Lh{L$_h$}
\def\LHF{L$_{HF}$}
\def\LLF{L$_{LF}$}
\def\LLFt{L$_{LF/2}$}
\def\LhHz{L$_{_{hHz}}$}
\def\usec{$\mu$sec}
\def\E#1;{$10^{#1}$}
\def\Sz{$S_z$}
\def\Sa{$S_a$}
\def\rxrv{rapid X-ray variability}
\def\mc{\multicolumn{2}{c}}
\def\rb{\raisebox{1.5ex}[0pt]}
\def\da{$\downarrow$}
\def\rr{\raggedright}
\def\mco{\multicolumn}
\def\n#1 {$^{#1}$}
\def\NCA{\em Nuovo Cimento}
\def\NIM{\em Nucl. Instrum. Methods}
\def\NIMA{{\em Nucl. Instrum. Methods} A}
\def\NPB{{\em Nucl. Phys.} B}
\def\PLB{{\em Phys. Lett.}  B}
\def\PRL{\em Phys. Rev. Lett.}
\def\PRD{{\em Phys. Rev.} D}
\def\ZPC{{\em Z. Phys.} C}
\renewcommand{\topfraction}{1.}
\renewcommand{\textfraction}{0.}
\begin{document}
\pagenumbering{roman}
\tableofcontents
\cleardoublepage
\pagenumbering{arabic}
\setcounter{chapter}{1}
\setcounter{table}{0}
\author[M. van der Klis]{M. VAN DER KLIS \\Astronomical Institute
``Anton Pannekoek'', University of Amsterdam, \\ Kruislaan 403, 1098
SJ Amsterdam}
%
\chapter{Rapid X-Ray Variability}
%
%
\section{Introduction}\mylabel{s:intro}
One of the principal motivations for studying X-ray binaries is the
unique window that accretion onto neutron stars and black holes
provides on the physics of strong gravity and dense matter.  Our best
theory of gravity, general relativity, while tested, and confirmed,
with exquisite precision in weak fields ($GM/R\ll c^2$; e.g., Taylor
et al.\ 1992) has not yet been tested by direct observation of the
motion of particles  
in the strong gravitational field near compact objects,
where the gravitational binding energy is of order the rest mass.  Among the
extreme predictions relativity makes for these regions are the
existence of event horizons, i.e., black holes (\S\myref{s:nsvsbh}),
the existence of an inner radius within which no stable orbits exist,
strong dragging of inertial frames, and general-relativistic
precession at rates similar to the orbital motion itself,
\about10$^{16}$ times as fast as that of Mercury.

In a neutron star the density exceeds that in an atomic nucleus. Which
elementary particles occur there, and what their collective properties
are, is not known well enough to predict the equation of state (EOS),
or compressibility, of the matter there, and hence the mass-radius
($M$-$R$) relation of neutron stars is uncertain.  Consequently, by
measuring this relation, the EOS of supra-nuclear density matter is
constrained. As orbital motion around a neutron star constrains both
$M$ and $R$ (\S\myref{s:grorbits}), measurements of such motion bear
on the fundamental properties of matter. Likewise, such motion near
black holes constrains the size and spin of black holes of given mass.

For addressing these issues of strong gravity and dense matter, we
need to study motion under the influence of gravity within a few
Schwarzschild radii\footnote[1]{The radius of a zero angular-momentum
black hole, $R_{Schw} = 2r_g = 2GM/c^2\approx3$\,km $M/M_\odot$.} of
compact objects and map out the strongly curved spacetime there.  As
the characteristic velocities near the compact object are of order
$(GM/R)^{1/2}\sim0.5c$, the dynamical time scale $(r^3/GM)^{1/2}$ for
the motion through this region is short; $\sim$0.1\,ms at
\about15\,km, and \about2\,ms at 10$^2$\,km from a 1.4\,\msun\ neutron
star, and \about1\,ms at 3$R_{Schw}$ (\about10$^2$\,km) from a
10\,\msun\ black hole.  These millisecond dynamical time scales, the
shortest associated with any astrophysical object, form one of the
most basic expressions of the compactness of compact objects.

The accretion flow is expected to be turbulent and may show magnetic
structures.  Its emission will vary in time due to the motions of
inhomogeneities through, and with, the flow.  This variability can be
used to probe the accretion-flow dynamics.  For a 10-km object, 90\%
of the gravitational energy is released in the inner \about10$^2$\,km,
hence the bulk of the emission likely comes from within the
strong-field region from where we expect the millisecond variability.
Temperatures here are $\approxgt$\E7;\,K, so most of this emission is
in X-rays.

The transfer of matter towards the compact object usually occurs by
way of an accretion disk in which the matter moves in near-Keplerian
orbits (Ch.\,13.2).  However, the geometry of the innermost part of
the flow is uncertain.  In many models the Keplerian disk extends down
to well into the strong-field region. It is terminated at an inner
radius $r_{in}$ of a few $R_{Schw}$ by for example relativistic
effects, radiation drag, a weak magnetic field (or the neutron-star
surface).  Advective, or in the case of strongly magnetic neutron
stars, magnetically dominated flows feature larger
(\about10$^2$\,$R_{Schw}$) inner disk radii.  Within $r_{in}$ the flow
is no longer Keplerian, and may or may not be disk-like.  Both inside
and outside $r_{in}$ matter may leave the disk plane and either flow
in more radially, or be expelled.  Together these flows constitute
what is called the ``accretion flow'' in this chapter.

The radiation from the Keplerian disk and from a neutron-star surface
are expected to be basically thermal, and observations indeed show
such thermal X-rays. In addition there is ubiquitous evidence for
non-thermal spectral components which may originate in one of the
non-disk flows and/or in an energetic 'corona' of uncertain geometry
(\S\myref{s:statesbh}) associated with the disk.

Observations of X-ray binaries show considerable variability on a wide
range of time scales in all wavelengths, and down to less than a
millisecond in X-rays. The study of this variability is called
'timing'. In this chapter we focus on aperiodic phenomena (QPOs and
noise, \S\myref{s:timing}) that are potential probes of the
strong-gravity dominated flow dynamics, i.e., millisecond aperiodic
phenomena in weakly magnetic compact objects, with particular
attention to the potential for measuring fundamental properties of
spacetime and matter. As observationally there are correlated spectral
and timing phenomena covering a range of time scales, we look at
longer time-scale phenomena and relations with X-ray spectral
properties as well.

That millisecond variability will naturally occur in the process of
accretion of matter onto a stellar-mass compact object is an insight
that dates back to at least Shvartsman (1971). Sunyaev (1973) noted
that clumps orbiting in an accretion disk closely around a black hole
could cause quasi-periodic variability on time scales of about a
millisecond.  Twenty-five years after these early predictions,
millisecond variability was finally discovered, with NASA's
breakthrough Rossi X-Ray Timing Explorer (RXTE; Bradt et al.\ 1993,
see \S\myref{s:nskhz}, \S3.4\hide{burstosc}, \S\myref{s:bhhf},
\S1\editnote{mspuls}).  To this day, RXTE is providing a veritable
flood of timing information that is still only partially digested.
Hence, contrary to the situation ten years ago it is no longer
possible to be exhaustive when reviewing X-ray binary timing (see,
e.g., Lewin et al.\ 1988, Stella 1988, Hasinger 1988, Miyamoto 1994,
van der Klis 1986, 1989a, 1995a,b, 2000 for preceding timing
reviews). In this chapter \S\S\myref{s:timing}--\myref{s:freqcorr}
provide a broad overview of the phenomenology.  We look for common
traits in the phenomena of different classes of objects indicating
common physics.  In \S\S\myref{s:generic} and
\myref{s:othermodels} we examine the ideas that have been put forward
to explain the phenomena by, respectively, orbital motions and
accretion-flow instabilities.  Sections \myref{s:ns} and
\myref{s:bh} give more detail on individual objects and phenomena,
with \S\S\myref{s:nskhz} and \myref{s:bhhf} dealing with the most
rapid (millisecond) phenomena and the remaining sections summarizing
the work on slower variability.
\section{Timing}\mylabel{s:timing}
The rapid variations diagnosing the inner accretion flow are
stochastic, and most effectively dealt with using statistical
(random-process) techniques.  Fourier analysis is the dominant tool,
and the one we focus on here.  Some other techniques are mentioned as
well.
\paragraph{Fourier analysis.} The Fourier power spectrum of the X-ray 
flux time series provides an estimate of the variance as a function of
Fourier frequency $\nu$ in terms of the {\it power density}
$P_\nu(\nu)$ (van der Klis 1989b for details).  The usual range of
$\nu$ is mHz to kHz; slower variations are usually studied in the
time-domain (but see Reig et al.\ 2002, 2003a), as on longer time scales
source-state changes and data gaps cause trouble for Fourier
techniques. Variations faster than those in the kHz range have not
(yet) been detected.

A number of {\it variability components} or {\it power-spectral
components} together make up the power spectrum (see, e.g.,
Fig.\,\myref{f:psa}). An aperiodic component by definition covers
several, usually many, frequency resolution elements. Broad structures
are called {\it noise} and narrow features {\it quasi-periodic
oscillations} (QPOs); 'broad-band noise' and 'QPO peaks' are common
terms. Least-squares fitting techniques are used to measure these
components.  When a series of power spectra is calculated from
consecutive chunks of data, usually the components change: they move
through the power spectrum (change frequency from one power spectrum
to the next), vary in width and strength, etc., but they remain
identifiable as the changes are gradual. This is the empirical basis
for the concept of a power-spectral component. The shortest time scale
on which changes can typically be followed is seconds to minutes.

The signal in the time series is not completely specified by the power
spectrum (and the signals are usually too weak to recover the Fourier
phases).  The same QPO peak could be due to, e.g., a damped harmonic
oscillator, randomly occurring short wave trains, a
frequency-modulated oscillation, an autoregressive signal, white noise
observed through a narrow passband filter or even a closely spaced set
of periodic signals.  {\it Time lags} (delays) between signals
simultaneously detected in different energy bands can be measured
using the cross-correlation function (CCF; Brinkman et al.\ 1974,
Weisskopf et al.\ 1975), but if the lags at different time scales
differ, the {\it cross-spectrum} (van der Klis et al.\ 1987c, Miyamoto
et al.\ 1988, Vaughan et al.\ 1994, Nowak et al.\ 1999a) performs
better.  This is the Fourier transform of the CCF and in a sense its
frequency-domain equivalent.  It measures a {\it phase lag} (time lag
multiplied by frequency) at each frequency.  The term {\it hard lag}
means that higher energy photons lag lower energy ones, and vv. for
{\it soft lag}. {\it Cross-coherence}\footnote[2]{Often just called
``coherence'', a term that is also used for the sharpness of a QPO
peak, below.} is a measure for the correlation between the signals
(Vaughan \& Nowak 1997).

{\it Power-law noise} is noise that (in the frequency range
considered) follows a power law $P_\nu\propto\nu^{-\alpha}$. The
power-law index (also 'slope') $\alpha$ is typically between 0 and 2;
for $|\alpha|>2$ Fourier analysis suffers from power leakage, so
measurements of noise steeper than that are suspect (e.g., Bracewell
1986, Deeter 1984). '1/f noise' has $\alpha=1$, and {\it white noise}
is constant ($\alpha=0$). {\it Red noise} is a term variously used for
either $\alpha=2$ power-law noise or any kind of noise whose $P_\nu$
decreases with $\nu$.

{\it Band-limited noise} (BLN) is defined here as noise that steepens
towards higher frequency (i.e., its local power-law slope $-d\log
P_\nu/d\log\nu$ increases with $\nu$) either abruptly (showing a
``break'' at {\it break frequency} $\nu_{break}$) or gradually. BLN
whose power density below a certain frequency is approximately
constant (white) is called {\it flat-topped noise}. The term {\it
peaked noise} is used for noise whose $P_\nu$ has a local maximum at
$\nu>0$).  Various modified power laws (broken, cut-off) as well as
broad Lorentzians are used to describe BLN. The precise value of the
characteristic frequency associated with the steepening (e.g.,
$\nu_{break}$) differs by factors of order unity depending on the
description chosen (e.g., Belloni et al.\ 2002a).

A {\it quasi-periodic oscillation} (QPO) is a finite-width peak in the
power spectrum. It can usually be described with a Lorentzian
$P_\nu\propto\lambda/[(\nu-\nu_0)^2+(\lambda/2)^2]$ with {\it centroid
frequency} $\nu_0$ and full width at half maximum (FWHM) $\lambda$.
This is the power spectrum of an exponentially damped sinusoid
$x(t)\propto e^{-t/\tau}\cos(2\pi\nu_0t)$, but the underlying signal
may well be different from this.  $\lambda$ is related to the {\it
coherence time} $\tau = 1/\pi\lambda$ of the signal, and is often
reported in terms of the {\it quality factor} $Q\equiv\nu_0/\lambda$,
a measure for the {\it coherence} of the QPO.  Conventionally, signals
with $Q>2$ are called QPOs and those with $Q<2$ peaked noise.  A {\it
sharp} QPO peak is one with high Q.

The strength (variance) of a signal is proportional to the integrated
power $P=\int{P_\nu d\nu}$ of its contribution to the power spectrum, 
and is usually reported in terms of its {\it fractional
root-mean-squared (rms) amplitude} $r\propto P^{1/2}$, which is a
measure for signal amplitude as a fraction of the total source
flux. It is often expressed in percent, as in ``2\% (rms)''.

The signal-to-noise of a weak QPO or noise component is $n_\sigma =
{1\over 2} I_x r^2 (T/\lambda)^{1/2}$ (van der Klis 1989b, see van der
Klis 1998 for more details), where $I_x$ is the count rate and $T$ the
observing time (assumed $\gg1/\lambda$). As $n_\sigma$ is proportional
to signal amplitude {\it squared}, if a clear power-spectral feature
``suddenly disappears'' it may have only decreased in amplitude by a
factor of two (and gone from, say, 6 to 1.5$\sigma$).

Red and flat-topped BLN may have only one characteristic frequency
(e.g., $\nu_{break}$), but peaked noise and a QPO have two ($\nu_0$ and
$\lambda$).  This leads to difficulties in describing the
phenomenology when, over time, power-spectral components change in $Q$
between noise and QPO.  For this reason, defining the characteristic
frequency as $\nu_{max}$, the
frequency at which power-density times frequency ($\nu P_\nu$,
equivalent to $\nu S_\nu$ in spectroscopy) reaches its maximum, has
gained some popularity (Belloni et al.\ 2002a).  This method is equally
applicable to centroids of narrow peaks and breaks in broad-band
noise, and smoothly deals with intermediate cases (see
\S\myref{s:decoherence} for an interpretation).  For a Lorentzian,
$\nu_{max}=\sqrt{\nu_0^2+\Delta^2}$, where $\Delta\equiv\lambda/2$, so
a narrow QPO peak has $\nu_{max}\approx\nu_0$ and a BLN component
described by a zero-centered Lorentzian $\nu_{max}=\Delta$.

Power spectra are in practice presented in a variety of ways, each more
suitable to emphasize particular aspects (displaying either $P_\nu$ or $\nu
P_\nu$, linearly or logarithmically, subtracting the white Poisson
background noise or not; e.g., Figs.\,\myref{f:hf} and
\myref{f:psa}).
\paragraph{Time-series modeling.\mylabel{s:timeseriesmodels}} 
Time-series modeling goes beyond Fourier analysis in an attempt to
obtain more detailed information about what is going on in the time
domain. One approach is to invent, based on physical hunches,
synthetic time series that reproduce observed statistical properties
of the variability.  Shot noise and chaos (below) are common
hunches. Another approach is to refine statistical analysis beyond
Fourier analysis. Here, systematic approaches involving higher order
statistics, such as skewness and bi-spectra (e.g., Priedhorsky et
al.\ 1979, Elsner et al.\ 1988, Maccarone \& Coppi 2002b) as well as
more heuristic ones (e.g., shot alignment, below) have been
tried. 'Variation functions' measuring variance as a function of time
scale $\tau$ (e.g., Ogawara et al.\ 1977, Maejima et al.\ 1984, Li \&
Muraki 2002) do not contain additional information as is sometimes
claimed; they basically provide power integrated over frequencies
$<1/\tau$.  The calculation of Fourier power spectra on short time
scales is a useful technique that has produced interesting results
(e.g., Norris et al.\ 1990, Yu et al.\ 2001, Yu \& van der Klis 2002, Uttley \& McHardy
2001). \editnote{if time: NS en theory papers nog doorkijken op rare
methodes -- wavelets?}

Shot noise is a time-series model of randomly occurring identical
discrete finite events called shots (Terrell 1972, Weisskopf et al.\
1975, Sutherland et al.\ 1978, Priedhorsky et al.\ 1979, Lochner et
al.\ 1991\editnote{if time: cir x-1 refs, sco x-1 refs}).  The power
spectrum is that of an individual shot.  Mathematically, the power
spectrum of a random process can always be modeled in this way (Doi
1978), so for modeling power spectra the method has little predictive
power, but it can be physically motivated in various settings
(\S\myref{s:shots}).  Heuristic 'shot alignment' techniques (Negoro et
al.\ 1994, 1995, 2001, Feng et al.\ 1999), like any 'fishing in the
noise' technique must be applied with care: they can be misleading if
the underlying assumption (i.e., shot noise) is incorrect.  Various
modifications of pure shot noise have been explored, e.g., involving
distributions of different shot profiles.  As pointed out by Vikhlinin
et al.\ (1994), if the shot occurrence times are correlated (e.g., a
shot is less likely to occur within a certain interval of time after
the previous shot, certainly plausible in, e.g., magnetic flare
scenarios; \S\myref{s:shots}), then peaked noise or a QPO is produced
whose $Q$ increases as the correlation between the shot times
increases.  Oscillating shots, short wave trains with positive
integrated flux, produce a QPO (due to the oscillation) and BLN (due
to the shot envelope) in the power spectrum (Lamb et al.\ 1985, Alpar
1986, Shibazaki \& Lamb 1987, Elsner et al.\ 1987, Shibazaki et al.\
1987, 1988, Elsner et al.\ 1988), and fit well within models involving
short-lived orbiting clumps (\S\myref{s:rpm}).

Given our limited a priori knowledge of the physical processes
producing the rapid variability, mostly mathematically-motivated
time-series models such as autoregressive, linear phase-state and
chaos models (e.g., Lochner et al.\ 1989, Unno et al.\ 1990, Scargle
et al.\ 1993, Pottschmidt et al.\ 1998, Timmer et al.\ 2000) usually
do not suffiently constrain the physics to conclude much from them
(but see \S\myref{s:shots}).
\section{Spectroscopy}\mylabel{s:spectra}
Recent work with Chandra and XMM-Newton suggests that, as previously
suspected (e.g., Barr et al.\ 1985, White et al.\ 1985),
relativistically broadened Fe lines near 6.5~keV similar to those
inferred in AGNs (e.g., Fabian et al.\ 2000) occur in some
neutron-star and black-hole binaries as well (e.g., Miller et
al.\ 2002, Parmar et al.\ 2002; \S4.2.3).  The gravitational and Doppler
distortions of these lines diagnose the dynamics of the same
strong-field region as millisecond timing (e.g., Reynolds \& Nowak
2003).  Combining timing with such spectroscopic diagnostics can
enormously improve the grip we have on what is going on in the inner
disk, but such work is still in a very early stage, and here we
concentrate on the better-explored link between timing and {\it
broad-band spectroscopy}.

Variations in broad-band ($\Delta E/E>0.1$, usually
continuum-dominated) X-ray spectral shape usually just become
detectable on the same seconds to minutes time scales on which
power-spectral changes are detectable (\S\myref{s:timing}).  Two
different techniques of broad-band spectroscopy are used to diagnose
these changes: multi-band photometry and spectral fitting.  The
spectral band used varies somewhat but is usually in the 1--60 keV
range and nearly always covers at least the 3--8 keV band.  

{\ \ \ \ \it Photometric method\ \ \ \ } One approach to quantifying
broad-band X-ray spectral shape uses {\it X-ray colors}. An X-ray
color is a 'hardness' ratio between the photon counts in two broad
bands; it is a rough measure for spectral slope. By calculating two
X-ray colors (a {\it hard color} in a higher energy band and a {\it
soft color} in a lower band) as a function of time, a record is
obtained of the broad-band X-ray spectral variations that is
well-matched to the power-spectral variations.  Plotted vs.\ one
another in a {\it color-color diagram} (CD), one can observe the
source to move through the diagram and, nearly always, create a
pattern.  It is then possible to study the relation between timing and
location in the CD. A {\it hardness-intensity diagram} (HID) or
color-intensity diagram is a similar diagram with a color vs.\
'intensity'; {\it X-ray intensity} in this context is nothing but a
count rate in some broad X-ray spectral band. Figs.\,\myref{f:cds} and
\myref{f:transients} provide examples of CD and HID patterns.  Whether
CD or HID presents the 'cleanest' pattern depends on source, and on
the quality of the data.  There are distinct advantages to working
with the logarithm of colors and intensities, like magnitudes in the
optical, but this is not general practice.

For distinguishing between source states (\S\myref{s:states}) the
photometric method combined with timing performs well, and the method
provides excellent sensitivity to subtle spectral variations. However,
this is at the expense of detector dependence. All X-ray detectors are
different and change over time, and contrary to optical photometry, in
X-rays there is only one bright standard star (Crab).  For the low
spectral resolution detectors typically used for timing (proportional
counters) it is not possible to completely correct X-ray colors for
the detector response: In the absence of {\it apriori} knowledge of
the intrinsic spectral shape all correction methods intended to derive
'intrinsic' colors (scaling by Crab colors, unfolding through the
detector response matrix either directly or by fitting an arbitrary
model) are mathematically imperfect (Kahn \& Blisset 1980, Kuulkers et
al.\ 1994, Kuulkers 1995, Done \& Gierli\'nski 2003).

{\ \ \ \ \it Spectral fitting method\ \ \ \ } There are considerable
interpretative advantages in describing the X-ray spectral variations
instead in terms of physical models fitted to the observed spectra.
The drawback is that this involves a description of spectral shape in
terms of more numbers (the spectral parameters) than just two X-ray
colors, and that the correct models are unknown.  This, plus the
tendency towards spurious results related to statistical fit-parameter
correlations masking true source variations, usually means that to
obtain sensible results, fitting requires longer integration times
than the time scale on which we see the power spectra, and the X-ray
colors, change, and that measured parameter-value changes obtained are
hard to interpret.  Nevertheless, for sufficiently long integration
times, and particularly in the case of the black-hole candidates,
where the X-ray spectral variations are clearer than in neutron stars,
this approach has met with some success.  Because both methods have
limitations, in practice the interpretation of X-ray spectral
variations tends to occur in a back-and-forth between CD/HIDs and
spectral fitting.  Techniques such as plotting the predicted colors of
model spectra in CDs together with the data, fitting spectra obtained
by averaging data collected at different epochs but in the same part
of the CD, and approximate intrinsic colors (e.g., Belloni et al.\
2000, Done \& Gierli\'nski 2003), are all being employed to make the
link between the two methods.

{\null\ \ \ \ \it Presentation and parametrization of CD/HIDs\ \ \ \ }
There is unfortunately no uniformity in either the presentation or the
choice of X-ray spectral bands for CD/HIDs. In particular,
%
%
there is a tendency to present the black-hole diagrams transposed as
compared to the neutron-star ones.  As illustrated by Jonker et
al.\ (2002a) these differences can obscure some of the similarities
between neutron-star and black-hole X-ray spectral behavior.

For sources which trace out one-dimensional {\it tracks} in the
CD/HIDs along which a source moves smoothly, position in the track, in
a single datum, summarizes its spectral state.  Curve length along the
track (Hertz et al.\ 1992, Kuulkers et al.\ 1994) is conventionally
indicated with a symbol $S$ (\Sz, \Sa; see \S\myref{s:statesns}).  $S$
is defined relative to the track: if the track drifts, the colors of a
point with given $S_x$ drift along with it.  The patterns observed in
a CD are often also recognizable in the corresponding HID obtained by
replacing soft color by intensity.  Presumably this is the case
because intensity is dominated by the more numerous photons in the
lower bands, whose dominant thermal component(s) strongly correlate to
temperature and hence, color. 
\section{Source types}
\mylabel{s:types}
Different sources exhibit similar patterns of timing and spectral
properties, allowing to group them into a number of {\it source
types}.  Both spectroscopy and timing, preferably in several source
states (\S\myref{s:states}), are necessary to reliably identify source
type. The primary distinction is not between neutron stars and black
holes (\S\myref{s:nsvsbh}) but between high and low magnetic-field
strengths.  Black holes and low magnetic-field ($\approxlt$\E10;\,G)
neutron stars potentially have gravity-dominated flows down to the
strong-field region (although radiative stresses can also be
important, \S\myref{s:radii}).  It is these objects, mostly found in
low-mass X-ray binaries (LMXBs, \S1\editnote{LMXBs}), that we shall be
mostly concerned with.  Strongly magnetic ($\approxgt$\E12;\,G)
neutron stars are briefly discussed in \S\myref{s:hmfns}.
\subsection{Neutron stars {\it vs.}\ black holes}\mylabel{s:nsvsbh}
A neutron star is only a few times larger than its Schwarzschild
radius, so accretion onto black holes and neutron stars is expected to
show similarities.  Indeed, based on just the inner-flow diagnostics
(timing and spectrum), the distinction between neutron stars and black
holes is notoriously difficult to make, with some examples of
black-hole candidates turned neutron stars (e.g., Cir\,X-1, Jones et
al.\ 1974, Tennant et al.\ 1986; V0332+53, Tanaka et al.\ 1983, Stella
et al.\ 1985; GS\,1826--238, Tanaka 1989, Ubertini et al.\ 1999;
4U0142+61, White \& Marshall 1984, Israel et al.\ 1994), and in
practice it is not easy to prove a compact object is a black hole.
There are two levels of proof: (i) showing that, assuming general
relativity, the compact object must be a black hole, and (ii) showing
that such an object indeed has the properties general relativity
predicts for a black hole.  For (i) it is sufficient to prove that a
mass is concentrated within its Schwarzschild radius; currently, this
mostly relies on dynamical mass estimates combined with theoretical
arguments about the maximum mass of a neutron star (e.g., Srinivasan
2002, \S1.3.8) and not on direct, empirical measurements of radius.  For (ii)
it is necessary to observe the interaction of the compact object with
its surroundings, empirically map out its exterior spacetime, and
demonstrate properties such as extreme frame dragging, the existence
of an innermost stable orbit, and an event horizon.

In recent work, for well-studied sources, clear differences are
discerned between the patterns of correlated spectral and timing
behaviour (\S\myref{s:states}) of neutron stars and black holes which
agree with distinctions based on X-ray bursts, pulsations, and
dynamical mass estimates.  In particular the most rapid variability
(see \S\myref{s:hf}), presumably that produced closest to the compact
object and most affected by its properties, is clearly quite different
between neutron stars and black-hole candidates.  Evidently we are
learning to distinguish between black holes and neutron stars based on
the properties of the flow in the strong-field gravity region, so
there is progress towards the goal of testing general relativity.  On
the other hand, remarkable spectral and timing similarities exist
between certified neutron stars and black-hole candidates,
particularly in low luminosity states (\S\myref{s:lfc}).  To
demonstrate the existence of black holes in the sense of general
relativity based on an understanding of the accretion phenomena near
them is a goal that has not yet been reached. The study of \rxrv\ of
low-magnetic field compact objects in X-ray binaries is one of the
programs contributing towards this end.  Both the actual measurement
of compact-object radius, which is part of a level (i) proof, as well
as level (ii) proofs are addressed by the research described in this
chapter: timing to diagnose motion very near compact objects.  I shall
use the term {\it black hole} both for objects whose black-hole
candidacy is based on a measured mass as well as for those whose
patterns of timing and spectral behavior put them into the same
phenomenological category as these objects.
\subsection{Low magnetic-field object types}
\mylabel{s:lmfnstypes}
The low magnetic-field neutron-star systems are subdivided into Z
sources, atoll sources (Hasinger \& van der Klis 1989) and what I
shall call the 'weak LMXBs'.  These three main sub-types are closely
related (see also \S\myref{s:statesns}).  {\it Z sources} are the most
luminous (Fig.\,\myref{f:typesLx}), and accrete at an appreciable
fraction of the Eddington critical rate (perhaps 0.5--1~\LEdd).  
{\it Atoll sources}, many of which are X-ray burst sources, cover a much wider range in
luminosities, from perhaps 0.001\,\LEdd\ (much lower in transients,
but below this level timing becomes difficult) all the way up to the
range of the Z sources (e.g., Ford et al.\ 2000; \Lx\ overlaps may well
occur; distances are uncertain).  Ordinary atoll sources are usually
in the 0.01--0.2~\LEdd\ range, while the 'GX' atoll sources in the
galactic bulge (see \S\myref{s:statesns}) usually hover at the upper
end (perhaps 0.2--0.5\,\LEdd), and the {\it weak LMXBs} (see
\S\myref{s:statesns}) at the lower end ($<$0.01\,\LEdd) of that range.  
Weak LMXBs comprise the overlapping%
\begin{figure*}[htbp]
$$\psfig{figure=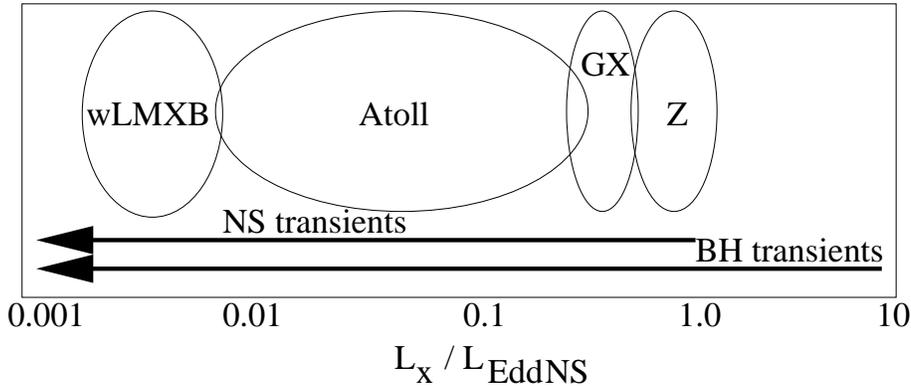,height=2.in} $$
\caption{Luminosities attained by Z sources, GX
atoll sources, ordinary atoll sources and weak LMXBs, respectively, as
well as by neutron-star  and black-hole transients. The extent of the
\Lx\ overlaps between these source types is undecided in detail, but
those shown here are likely. 
\mylabel{f:typesLx}}
\end{figure*}
\footnote[3]{Since the discovery in 1998 of millisecond pulsations 
in the thermonuclear burster and low magnetic field neutron star
SAX\,J1808.4--3658 (\S1\editnote{check}, \S\myref{s:nskhz}) the old adage
about the mutual exclusion between pulsations and type I X-ray bursts
``pulsars don't burst and bursters don't pulse'' (e.g., Lewin et
al.\ 1993), which was based on the dichotomy in neutron-star $B$ fields
(\S\myref{s:types}), no longer holds, because low magnetic-field
neutron stars now can be pulsars, too.
\editnote{ch 1 should also give the nrs of BH, lmf and hmf ns in lm
and hm systems}.}
groups of faint burst sources, millisecond pulsars and low-luminosity
transients (\S6\editnote{check}); many of them appear to be just atoll
sources stuck at low \Lx. Some faint LMXBs have luminosities that are
not well known due to uncertain distances or emission anisotropies
(e.g., in dippers, \S1\editnote{dippers}), but many of these are
probably weak as well.

While nearly all LMXBs whose X-ray emission is persistent contain a
neutron star, nearly all with a black hole (and many with a neutron
star as well) are transients, showing intermittent activity intervals
called outbursts usually lasting weeks to months and separated by long
quiescent intervals.  The black holes have no well-defined
subdivisions; although one might expect differences in spin (and
hence, frame dragging, \S\myref{s:grorbits}), accretion mode (wind or
Roche-lobe overflow, \S1\editnote{check}) or between transient and
persistent systems to show up, none of these lead to obvious
differences in the observable properties attributable to the inner
flow.  Some black-hole transients remain in the low hard state
(\S\myref{s:states}) during their entire outburst (e.g., Nowak 1995,
Brocksopp 2004), but this can vary from one outburst to another (e.g.,
Belloni et al.\ 2002b).

%
Some accreting objects that are {\it not} the topic of this chapter
(or even in some cases this book) notably the ultra-luminous X-ray sources observed
in some external galaxies (ULX), active galactic nuclei (AGN), both of
which are thought to contain black holes, and cataclysmic variables
(CV), which contain white dwarfs, can have accretion geometries that
are similar to those in X-ray binaries (XRB), and it is of interest to
compare their variability properties.  The ULX are discussed in 
\S9.8.4, \S12.3, 12.4.1 and \S13.8 (see also Strohmayer \& Mushotzky 2003,
Cropper et al.\ 2004) and the CV in
\S10.  In AGN, while QPO detections are still difficult (see Benlloch
et al.\ 2001 and references therein, and see also Halpern et al.\
2003) the best measurements now clearly show band-limited noise with
characteristic frequencies consistent with the idea that variability
time scales scale with mass (e.g., Edelson \& Nandra 1999, Czerny et al.\ 2001,
Uttley et al.\ 2002, Markowitz et al.\ 2003, McHardy et al.\ 2004).
The comparison of XRB with AGN is of particular interest as the
physics is likely similar but different observational regimes apply.
We typically receive more X-ray photons per dynamical time scale from
AGN than from XRB, making it easier, in principle, to study such fast
variability in the time domain.  X-ray binaries of course have much
higher photon fluxes so that background is less of an issue and it is
easy to cover very large numbers of dynamical time scales (a hundred
million per day) and reliably determine the {\it parameters} of the
stochastic process characterizing the variability rather than
observing just one particular {\it realization} of it.  The
possibility to compare neutron-star and black-hole systems is unique
to X-ray binaries as well.
\section{Source states}
\mylabel{s:states}
Time variability and spectral properties are found to be correlated,
presumably because of their common origin in physical processes in the
inner, X-ray emitting part of the accretion flow.  {\it Source states}
are qualitatively different, recurring patterns of spectral and timing
characteristics.  They are thought to arise from qualitatively
different, somewhat persistent, inner flow configurations.  Both
timing and spectroscopy are usually required to determine source
state.  Luminosity, and the way in which power spectrum and X-ray
spectrum vary on time scales of minutes and longer (called source {\it
behaviour}) can be used as additional state indicators, but note that
luminosity does {\it not} determine source state (below).

The qualitative changes in phenomenology used to define state include
the appearance of a spectral or variability component, a sudden step
in luminosity, or a clear bend in a CD/HID track, events which often
coincide, indicating a qualitative change in the flow.  As
observations improve, 'sudden' transitions become resolved, so
eventually somewhat arbitrary boundaries need to be set to make state
definitions precise.  Depending on which criteria are chosen, authors
may differ on what is the 'correct' subdivision of the phenomenology
into states. Nevertheless, the concept is central in describing the
behaviour of X-ray binaries, and in this chapter, the \rxrv\ is
discussed as a function of state.  How X-ray spectral fit parameters
depend on source state and what in detail this might imply is beyond
the scope of this chapter (see e.g., di Salvo \& Stella 2002, Done \&
Gierlinsky 2003, Gilfanov et al.\ 2003 for neutron stars and \S4 for
black holes).
%
%
\noindent\begin{table*}[htb]
\caption{Source states of low magnetic-field compact objects.\mylabel{t:states}}
\hskip1cm {\bf HIGH} frequencies \hskip2cm disk-dominated; thermal; {\bf SOFT} spectra
\begin{minipage}{6.2cm}
$$
{Q,\ \ L_{x,short}
\atop{\displaystyle {\rm generally}
\atop{\displaystyle {\rm increase\strut}
\atop{\displaystyle {\rm upward}}}}}
\left\uparrow
\quad 
{\bf\begin{tabular}{ccccc}
\hline 
\hline
 Z      & Atoll   & weak LMXB & Black hole  \\
\hline                              
FB      & UB      &            &            \\
---     & ---     &            &            \\
NB      & LB      &            & HS         \\
        &{\it 50} &            &            \\
{\it 20}& LLB     &            &            \\
HB      &{\it 10} &            & {\it 10}   \\
        & IS      &            &            \\
{\it 2} &{\it 2}  &            & IMS/VHS    \\
        & EIS     & {\it 1}    &            \\
        &         & EIS        & {\it 0.5}  \\
        &{\it 0.2}&            & LS         \\
        &         &{\it 0.1}   &            \\
        &         &            & {\it 0.01} \\
\hline 
\hline
\end{tabular}}
\quad
\right\downarrow
{rms,\ L_{radio}
\atop{\displaystyle {\rm generally}
\atop{\displaystyle {\rm increase\strut}
\atop{\displaystyle {\rm downward}}}}} 
$$
\end{minipage}
\medskip
\hfill\break\null\hskip1cm {\bf LOW} frequencies \hskip1cm
corona-dominated; non-thermal; {\bf HARD} spectra 
\begin{minipage}{\hsize}\scriptsize 
\null\bigskip\null $Q$: variability coherences; $L_{x,short}$: X-ray 
luminosity with long term variations filtered out, i.e., $L_x/\langle
L_x\rangle$ where $\langle L_x\rangle$ is \Lx\ averaged over a day,
cf. \S2.9.1.1; $rms$: variability amplitudes and $L_{radio}$: radio
flux. FB: flaring branch; NB: normal branch; HB: horizontal branch;
UB: upper banana; LB: lower banana; LLB: lower left banana; IS: island
state; EIS: extreme island state; HS: high state; IMS: intermediate
state; VHS: very high state; LS low state.  Numbers indicate typical
BLN frequencies $\nu_b$ (in Hz) delineating the states
(\S\myref{s:lfc}); '---' indicates the BLN is usually
undetected. Reversals in $\nu_b$ occur beyond the highest frequencies
in Z and atoll sources (\S\myref{s:lfc}).
\end{minipage}
\end{table*}

Table\,\myref{t:states} lists the source states distinguished in this
chapter ordered from spectrally hard, generally characterized by low
variability frequencies and luminosities at the bottom, towards soft,
with higher frequencies and often higher luminosities at the top.  Typical
band-limited noise (BLN, \S\myref{s:timing}) frequencies $\nu_b$
corresponding to the state transitions are given (see
\S\S\myref{s:lfc} and \myref{s:freqcorr}).  These frequencies are
indicative, and in practice depend somewhat on source, and on the
precise definition of each state.  The most striking variability
(strong sharp kHz QPOs (\S\myref{s:hf}), black-hole high-frequency
QPOs (\S\myref{s:hf}), sharp low-frequency QPOs (\S\myref{s:lfc})
tends to occur in the intermediate states (HB, LLB, IS, IMS, VHS, see
Table\,\myref{t:states}) both in neutron stars and in black holes.
Note that Z sources so far appear to lack true low-frequency states.
That a rough ordering of states such as in Table\,\myref{t:states} is
possible suggests, of course, that there are physical similarities
between the states in the different source types (e.g., van der Klis
1994a,b), but a considerable amount of uncertainty still surrounds
this issue, as well as the exact nature of the states themselves.
\begin{figure*}[htbp]
$$\psfig{figure=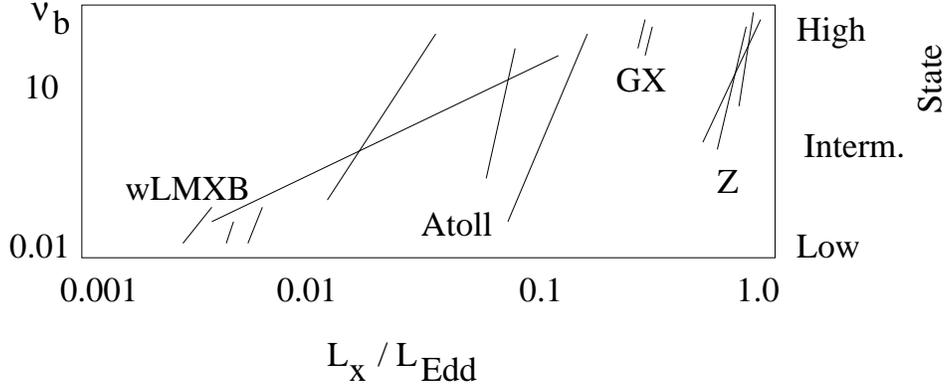,height=2.in} $$
\caption{States and luminosities attained by Z sources, GX atoll
sources, ordinary atoll sources and weak LMXBs, respectively. Similar
states occur at very different luminosities; not all states are seen
in all object types. The issue of \Lx\ overlaps between these source
types is undecided. \mylabel{f:statesLx}}
\end{figure*}

In neutron stars, X-ray luminosity \Lx\ tends to correlate with state
only within a source, not across sources, and, in a given source, much
better on short (hours to days), than on longer time scales
(\S\S\myref{s:statesns}, \myref{s:parlines} and, e.g., van der Klis
2000 and references therein); in black holes, the predictive value
\Lx\ has with respect to state is even more tenuous
(\S\myref{s:statesbh} and, e.g., Miyamoto et al.\ 1995, Nowak et
al.\ 2002, Maccarone \& Coppi 2003).  Note, however, that a really low
\Lx, $<$0.01\LEdd\ or so, will reliably produce a hard state in most 
sources.  In general, X-ray spectral shape (position in the colour
diagram) predicts timing characteristics much better than \Lx\ (e.g.,
van der Klis et al.\ 1990, Hasinger et al.\ 1990, Kuulkers et al.\
1994, 1996, van der Klis 1994a,b, 1995a, Ford et al.\ 1997b, Kaaret et
al.\ 1998, M\'endez \& van der Klis 1999; Homan et al.\ 2001, Wijnands
\& Miller 2002, Rossi et al.\ 2004, Belloni 2004).  The range of
correlations in neutron stars between state as defined in
Table\,\myref{t:states} and \Lx\ is schematically illustrated in
Fig.\,\myref{f:statesLx}; a full systematic population study has so
far only been performed for states with kHz QPOs (Ford et al.\ 2000).
Clearly, the overall correlation of state to \Lx\ is not good.  What
causes neutron stars to exhibit similar states at very different \Lx\
levels, while short-term state changes {\it correlate} to changes in
\Lx, and why some sources cover a wider range of states than others,
is uncertain.  Note that not only the differences between Z and atoll
sources need explanation, but also those {\it among} atoll sources.
Differences in magnetic field (e.g., Miller et al.\ 1998a) have been
considered.  Clearly, this would allow for intermediate cases, but
these are rare (\S\myref{s:nsother}).

If, as often assumed, accretion rate \mdot, varying due to processes
outside the strong-field region (e.g., instabilities further out in
the disk), and increasing from bottom to top in
Table\,\myref{t:states} would underly the state differences, then
\mdot\ would have to govern both timing and spectral properties, but
{\it not} the long-term changes in measured X-ray flux nor the \Lx\
differences between sources (e.g., van der Klis 1995a).  This could
arise if mass outflows or (in black holes) radiatively inefficient
(advective) inflows (\S4\editnote{check}) destroy the expected
\mdot-\Lx\ correlation by providing sinks of mass and (kinetic) energy
(e.g., Ford et al.\ 2000), or if the flux we measure is not
representative for the true luminosity because there are large and
variable anisotropies or bolometric corrections in the emission (e.g.,
van der Klis 1995a).  However, in neutron stars differences in X-ray
burst properties as a function of observed \Lx\ (\S3\editnote{check}),
suggest that the \mdot-\Lx\ correlation is at least fair, and hence
that the \mdot-state correlation is not so good.  Possibly, the \mdot\
that sets source state is not total \mdot, but only one component of
it, e.g., that through the X-ray emitting part of the disk, $\dot
M_d$, while there is also a radial inflow $\dot M_r$ (e.g., Fortner et
al.\ 1989, Kuulkers \& van der Klis 1995, Kaaret et al.\ 1998, van der
Klis 2000, 2001, Smith et al.\ 2002).  A more radical solution is that
\Lx\ does track \mdot\ but that source state is governed by a physical
parameter not correlating well to {\it any} \mdot\ (perhaps, inner disk
radius $r_{in}$, van der Klis 2000).  The question then becomes what,
if not a varying accretion rate, {\it does} cause the changes.  A clue
is the hysteresis observed in the state transitions of various sources
(\S\S\myref{s:statesbh}, \myref{s:statesns}), which suggests that the
history of a source's behaviour affects its current state.  S-curve
disk-flow solutions (\S13\editnote{check}) have this property, but a
smoothed response of $\dot M_r$ to variations in $\dot M_d$ where
$r_{in}$ is set by $\dot M_d/\dot M_r$ may cause hysteresis as well
(see \S\myref{s:parlines}).  So, while it seems out of the question
that the states {\it as well} as the differences between sources are
all just caused by differences in (instantaneous) \mdot, differences
in \mdot\ {\it and} in some time average over \mdot\ might still in
principle be sufficient.  However, the influence of other parameters
seems likely.  More than one mechanism may be at work.  In this
chapter, when referring to source states I use the terms {\it high} and
{\it soft} vs.\ {\it low} and {\it hard} in the general sense implied
by Table\,\myref{t:states}, with the understanding that the relation
to \Lx\ is complex.
\subsection{Black-hole states}\mylabel{s:statesbh}
In black holes the 1--20\,keV X-ray spectrum can be decomposed into a
{\it hard}, non-thermal, power-law component with photon index
typically 1.5--2 and a {\it soft} (also: 'ultrasoft'), thermal,
black-body like component with $kT<1$\,keV (\S1.3.4, \S4\editnote{check}).  The
latter is usually attributed to thermal emission from the accretion
disk, the former to a {\it corona} containing energetic electrons.
There is no agreement about the nature or energetics of this corona
(see e.g., Maccarone \& Coppi 2003): it could be located within the
inner disk edge or cover part of the disk, it could be
quasi-spherical, a thin layer on, or magnetic loops anchored in, the
disk, or be the base of the radio jets.  It could be quasi-static or
part of the accretion flow (azimuthally and/or radially).  Its
electrons' energy could be thermal or due to bulk motion, and the
radiation mechanism could be either Compton or synchrotron.

In the classic black-hole {\it low state} (LS) in the 2--20\,keV band,
the hard component dominates (hence it is also called 'low hard
state'), and strong (up to typically \about50\% rms) flat-topped BLN
(\S\myref{s:timing}) is present which has a low characteristic
frequency (down to typically $\nu_b$\about0.01\,Hz), in the {\it high
state} (HS) the soft component dominates (hence 'high soft state') and
weak ($\approxlt$3\%) power-law noise occurs, and in the {\it very
high state} (VHS) \Lx\ is high in the range of both spectral
components, and strong 3--12\,Hz QPOs and BLN that is weaker and
higher-frequency (up to \about10\,Hz) than in the LS as well as
power-law noise stronger than in the HS occur, with rapid transitions
between these two noise types (Miyamoto et al.\ 1991).  Both LS and HS
have turned out to occur over a wide and largely overlapping \Lx\
range in the spectrally hard and soft parts of the HID, respectively
(Figs.\,\myref{f:2d}, \myref{f:transients}), and an {\it intermediate
state} (IMS) with timing properties similar to the VHS and at
intermediate spectral hardness, but likewise covering a wide \Lx\
range (e.g., Homan et al.\ 2001; see Fig.\,\myref{f:2d}) was
identified.  To maintain continuity with existing literature (cf.,
Table\,\myref{t:bh}), I continue to refer to these states as LS for
the hard low-frequency states and HS for the soft states with weak
power-law noise, respectively.  I take the IMS to include the VHS as
the highest-\Lx\ intermediate state in a source (usually attained
early in a transient outburst).  Somewhat fortuitously, 'low', 'high'
and 'intermediate' can also be taken to refer to BLN frequency
(although in a full-blown HS the BLN is not detected).  Note that in \S4
McClintock and Remillard take another approach: they call HS 'thermal
dominant state', and depending on the results of X-ray spectral
continuum fits classify some instances of the IMS together with the LS
as 'low hard state' and others as 'steep power law state'.  Timing in
the low-\Lx\ quiescent or 'off' state, down to levels where this can
be checked, is consistent with the LS \hide{Kong claims a lot but
provides no rxrv data}but with $\nu_b$ down to perhaps 0.0001\,Hz
(\S\myref{s:bhlfc}).  QPOs in both the 1--30\,Hz and 100--450\,Hz
ranges are most prominent in the VHS/IMS, but occasional
\about0.01--20\,Hz QPOs are also seen in LS and HS.
\begin{figure*}[htbp]
\begin{center}
\begin{tabular}{c}
\psfig{figure=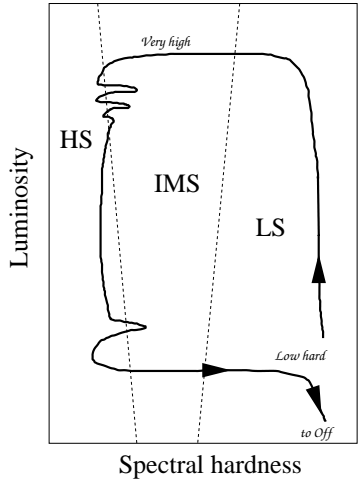,height=2.5in}
\end{tabular}
\hspace{1cm}
\begin{minipage}{5cm}
\caption{Black-hole states plane. Main states are LS, IMS and HS
(low, intermediate and high state) as indicated; locations of classic
low/hard, very high and off states are shown as well. Thick curve
shows path of typical black-hole transient outburst, inspired by
observations of XTE\,J1550--564 (Homan et al.\ 2001) and GX\,339--4
(Belloni 2004).  Contours of $\nu_b$ might be a way to delineate the
three states in this plane; as these lines are not accurately known,
the dashed lines shown are only indicative. See Fig.\,\myref{f:psa}
for typical power spectra observed in these states.  \mylabel{f:2d}}
\end{minipage}
\end{center}
\end{figure*}

In a 'standard' black-hole transient outburst (Figs.\,\myref{f:2d},
\myref{f:transients}) sources tend to follow a harder trajectory
from low to high luminosity than vice versa, and make the main
hard-soft (LS$\rightarrow$HS) transition at much higher luminosity
than the soft-hard (HS$\rightarrow$LS) one, behaviour that is often
called {\it hysteresis} (Miyamoto et al.\ 1995, Nowak 1995, van der
Klis 2001, Nowak et al.\ 2002, Smith et al.\ 2002, Maccarone \& Coppi
2003), expressing the idea of a driving force and a memory effect in
the response to that force (\S\myref{s:states}); whether this is truly
what is going on requires further investigation.  Transitions to and
from the IMS can be rapid, suggesting an obvious delineation between
states, but slower transitions occur as well, and in these cases the
exact transition point becomes a matter of definition
(\S\myref{s:bhlfc}).  It might be possible to base such definitions on
values of $\nu_b$ (cf., Table\,\myref{t:states}), perhaps defining
$\nu_b$ contours in CD/HIDs as in Fig.\,\myref{f:2d}), but it should
be noted that the issue is currently in flux and that in particular
the best way to deal with the IMS/VHS is surrounded by some
controversy (see also \S4).  Further subdivisions probably exist in
the IMS (e.g., Miyamoto et al.\ 1994), but different boundaries are
drawn depending on whether spectral fit parameters (as in \S4) or
rapid variability (Belloni et al.\ 2004, Klein-Wolt et al.\ 2004b) are
given precedence in defining the sub-states.

Various spectral characteristics can serve to define the black hole
states plane.  Flux, intensity or soft color can track the
'luminosity' parameter, which physically might be dominated by the
accretion rate through the inner disk $\dot M_d$.  The 'state'
parameter, which could be a measure for the strength or size of the
corona can be defined as, e.g., hard colour or a spectral-component
luminosity ratio ('thermal fraction'; $L_{soft}/L_{total}$, 'power-law
ratio'; $L_{hard}/L_{total}$, etc.; e.g., Miyamoto et al.\ 1994, Nowak
1995, Rutledge et al.\ 1999, Remillard et al.\ 2002c).

On physical grounds one might expect spectral and timing properties to
be affected not only by state, but also by the large ($>$ factor 10)
differences in \Lx\ within each state, but certainly in HS (e.g.,
Homan et al.\ 2001) and LS (e.g., Belloni 2004) the \Lx\ effect
appears to be modest.  Historically, after LS and HS (Tananbaum et
al.\ 1972), the VHS was identified first (Miyamoto et al.\ 1991); the
intermediate state was initially noticed as a HS$\rightarrow$LS
transitional state with 10-Hz BLN similar to that in the VHS in the
decay of the transient GS\,1124--68 (Belloni et al.\ 1997), and as a
10-Hz BLN state at much lower \Lx\ than the VHS in GX\,339--4
(M\'endez \& van der Klis 1997).  Intermediate states with 10-Hz BLN
and/or strong power-law noise were then also seen in Cyg\,X-1 (Belloni
et al.\ 1996) and a number of other sources (e.g., Kuulkers et al.\
1997b), and in XTE\,J1550--564 during excursions at various \Lx\
levels from the HS to a somewhat harder state characterized by 10-Hz
BLN, LF QPOs and occasional rather strong power-law noise
(\S\myref{s:bhpl}, Homan et al.\ 2001).  This led to the
two-dimensional paradigm of Fig.\,\myref{f:2d}.  Weak QPOs $>$100\,Hz
occur in the VHS (\S\myref{s:hf}) but were also found at least once in
the IMS at lower \Lx\ (Homan et al.\ 2001).

That the variability frequencies decrease (\S\myref{s:components})
when the spectrum gets harder suggests a relation with inner disk
radius $r_{in}$ (stronger corona for larger $r_{in}$) and some
spectroscopic work points into the same direction
(\S4\editnote{check}), but this entire picture is firmly within the
realm of the working hypotheses.  Nevertheless, most modeling of
black-hole states approximately conforms to this generic framework.
The strength of the corona has a relation to that of the radio jets,
and as both rely on energetic electrons for their emission, it is
natural to suspect a physical relation between these structures
(\S9\editnote{check}).  Even at a purely empirical level, there is
still considerable uncertainty associated with the motion of the
sources through Fig.\,\myref{f:2d}, e.g., why do many sources
approximately take the depicted 'canonical' path while some deviate
from this, are all areas in this plane accessible or are some
forbidden, what kind of state transitions are possible at each
luminosity level?  Yet the diagram and its segmentation into three
areas representing three main source states provides a useful template
embodying the broad correlation between variability frequencies and
spectrum occurring largely irrespectively of \Lx\ level, against which
black-hole behaviour can be matched.
\subsection{Low magnetic-field neutron-star states}\mylabel{s:statesns}
In low magnetic-field neutron stars spectral decomposition is much
less obvious than in black holes, perhaps because of the presence of
two thermal emission sites (disk and star) and cooling of the hot
electrons in the corona by the stellar flux.  The spectra become
neither as hard nor as soft as in black holes, but weak hard
components are sometimes seen and similarly explained by
Comptonization (e.g., Barret \& Vedrenne 1994 , di Salvo \& Stella
2002).  Contrary to the case in black holes, CD/HIDs show rather
reproducible tracks that embody a one-dimensional states sequence
(except at the lowest luminosity levels, see \S\myref{s:statesa}).
\begin{figure*}[htbp]
$$\psfig{figure=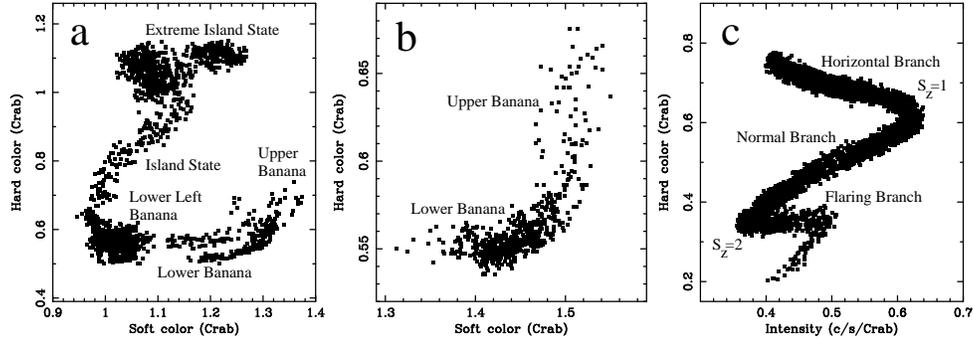,width=\linewidth}$$
\caption{Spectral branches of neutron stars.  {\it (a)} CD of the atoll 
source 4U\,1608$-$52, {\it (b)} CD of the 'GX' atoll source
GX\,9+1, and {\it (c)} HID of the Z source GX\,340+0.  RXTE/PCA
data; soft colour 3.5--6/2--3.5\,keV, hard colour 9.7--16/6--9.7\,keV,
intensity: 2--16\,keV, all normalized to Crab.  Conventional branch
names and \Sz\ values are indicated; for \Sa\ the choice of values
differs between authors (though not in sense, see text). Compare van
Straaten et al.\ (2003), Reerink et al.\ (2004), Jonker et
al.\ (2000a). \mylabel{f:cds}}
\end{figure*}
\subsubsection{Z sources}
Z sources on time scales of hours to a day or so trace out roughly Z
shaped tracks (Fig.\,\myref{f:cds}{\it c}) in CD/HIDs consisting of
three branches connected end-to-end and called {\it horizontal
branch}, {\it normal branch} and {\it flaring branch} (HB, NB, FB).
Curve length \Sz\ (\S\myref{s:spectra}), defined to increase from HB
via NB to FB (Fig.\,\myref{f:cds}{\it c}) performs a random walk; it
varies stochastically but shows no jumps.  kHz QPOs and a 15--60\,Hz
QPO called HBO occur on the HB and upper NB, an \about6\,Hz QPO called
NBO on the lower NB, and mostly power-law noise $<$1\,Hz on the FB;
see \S\myref{s:components}.  With increasing \Sz\ the band-limited
noise (LFN, see \S\myref{s:lfc}) becomes weaker and flux and
frequencies generally increase, although at high \Sz\ reversals occur
(\S\myref{s:lfc}).  Z tracks differ somewhat between sources (e.g., Hasinger
\& van der Klis 1989, Kuulkers 1995, Muno et al.\ 2002), and also show
slow drifts ('secular motion') that do not much affect the variability
and its strong correlation with \Sz\ (e.g., Hasinger et al.\ 1990,
Kuulkers et al.\ 1994, Jonker et al.\ 2002a); some occasionally show
shape changes which {\it do} affect the variability (Kuulkers et al.\
1996).  There is no evidence for hysteresis in either the Z tracks or
the rapid variability.
\subsubsection{Atoll sources and weak LMXBs}\mylabel{s:statesa}
At high \Lx\ atoll sources trace out a well-defined, curved {\it
banana branch} in the CD/HIDs (Fig.\,\myref{f:cds}{\it a,b}; Hasinger
\& van der Klis 1989, Reerink et al.\ 2004) along which, like in Z
sources, sources move back and forth with no hysteresis on time scales
of hours to a day or so, and which sometimes shows secular motion not
affecting the variability (e.g., van der Klis et al.\ 1990, di Salvo
et al.\ 2003, Schnerr et al.\ 2003).  The banana branch is further
subdivided into the {\it upper banana} (UB) where the $<$1\,Hz power
law noise (VLFN, \S\myref{s:nsotherrxrv}) dominates, the {\it lower
banana} (LB) where dominant several 10-Hz BLN occurs
(\S\S\myref{s:lfc}, \myref{s:nslfc}), and the {\it lower left banana}
(LLB) where twin kHz QPOs (\S\S\myref{s:hf}, \myref{s:nskhz}) are
observed.

The spectrally harder parts of the CD/HID patterns are traced out at
lower \Lx.  CD motion is often much slower here (days to weeks), and
observational windowing can cause isolated patches to form, which is
why this state is called {\it island state} (IS; Hasinger \& van der
Klis 1989); it is characterized by dominant BLN (\S\myref{s:lfc}),
becoming stronger and lower-frequency as flux decreases and the
$>$6\,keV spectrum gets harder.  The hardest, lowest luminosity island
states (the {\it extreme island state}; EIS; e.g., Prins \& van der
Klis 1997, Reig et al.\ 2000a; van Straaten et al.\ 2003 and
references therein) are similar to the black-hole LS, with strong,
low-frequency flat-topped noise (\S\myref{s:lfc}) and a hard power-law
X-ray spectrum.  Curve length \Sa\ (\S\myref{s:spectra}) increases
from EIS to UB, generally anti-clockwise (Fig.\,\myref{f:cds}{\it a}).
With increasing \Sa\ the BLN tends to become weaker; frequencies
generally increase up to the LB, from where in bright sources $\nu_b$
decreases again (\S\myref{s:lfc}).  Most atoll sources show a banana
branch as well as island states but the four GX atoll sources
(\S\S\myref{s:lmfnstypes}, \myref{s:states}; Reerink et al.\ 2004)
are (nearly) always in the banana branch (LB and UB,
Fig.\,\myref{f:cds}{\it b}), and the weak LMXBs (nearly) always in the
EIS (e.g., Barret et al.\ 2000, Belloni et al.\ 2002a).
\begin{figure*}[htbp]
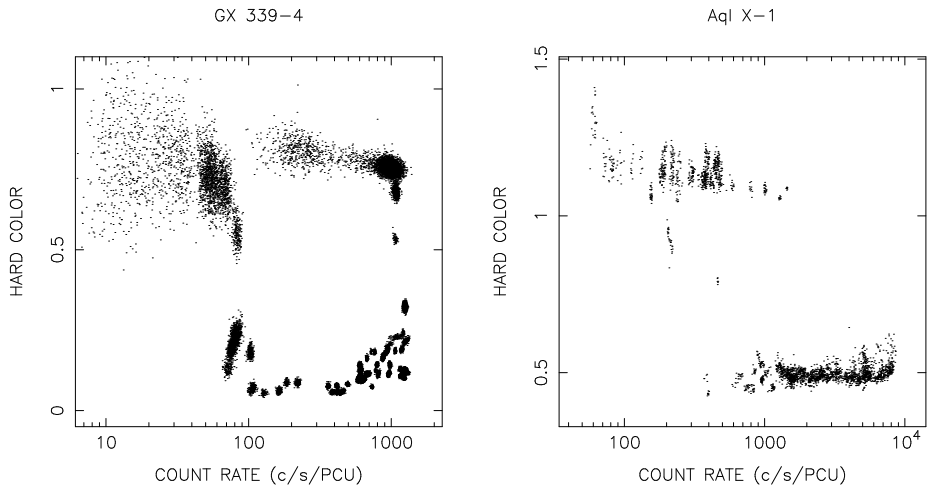

$$\psfig{figure=339_lines3_symbols2_characters14.ps,height=2.5in}
\qquad
\psfig{figure=aql.ps,height=2.5in}  $$
\caption{Hardness-intensity diagrams of the black hole GX 339--4 and the
neutron-star LMXB transient Aql X-1.  Note the orientation of these
diagrams, which is similar to that of Fig.\,\myref{f:cds} and
transposed compared to Fig.\,\myref{f:2d}. Compare Belloni (2004), Reig
et al.\ (2004). \mylabel{f:transients}}
\end{figure*}

In the island states two-dimensional motion (i.e., not following one
well-defined track) is often observed, perhaps because secular motion
and motion along the branch happen on similar time scales.  In
transient atoll sources this takes the form of hysteresis
(\S\myref{s:statesbh}) during relatively rapid (\about1\,day)
transitional IS episodes between EIS and banana state
(Fig.\,\myref{f:transients}): similarly to black-hole transients, hard
to soft transitions occur at higher luminosity than the reverse, but
neutron stars lack the hard, high-\Lx\ states (e.g., Barret et al.\
1996, Olive et al.\ 2003, Maccarone \& Coppi 2003).  Persistent
sources can transit to and from the EIS tracing out a one-dimensional
IS branch (Fig.\,\myref{f:cds}{\it a}), suggesting hysteresis is
related to the abruptness or speed of the transition.  The full
picture concerning the EIS, of which only glimpses were seen before
(e.g., Mitsuda et al.\ 1989, Langmeier et al.\ 1989, Yoshida et al.\
1993), is only now becoming clear.  In some cases several parallel
horizontal EIS branches form, where $>$6\,keV spectral hardness, not
luminosity or the $<$6\,keV spectrum, determines the source state (van
Straaten et al.\ 2003 \editnote{add his figure?}).  Similarities
between this CD behaviour in the EIS and the Z-source HB were pointed
out by a number of authors (Muno et al.\ 2002, Gierli\'nski \& Done
2002b, see also Langmeier et al.\ 1989), but do not extend to the way
in which the branches connect, nor to the timing behaviour (Barret \&
Olive 2002, van Straaten et al.\ 2003, Olive et al.\ 2003, Reig et
al.\ 2004, \S\myref{s:lfc}), where the HB is more like the IS/LLB than
like the EIS (Table\,\myref{t:states}).

Similarities of atoll CD/HID behaviour with that of Z sources are
mostly confined to the banana branch. Similar one-dimensional motion
on similar (hrs-day) time scales and, in the persistent sources,
covering similar \Lx\ ranges takes place there through similarly
slowly drifting tracks (which, however, are shaped differently and
also intrinsically wider in atoll sources). In island states the CD
behaviour is quite different from that of Z sources: it shows
hysteresis and other forms of two-dimensional motion, and often takes
place on longer (days to weeks) time scales over much larger \Lx\
ranges (factors of 5--10, up to 10$^3$ for transients).  This
behaviour is more similar to that of black holes than that in other
neutron-star states, possibly because in low states the inner-disk
radius is larger, further away from the compact object.
\section{Variability components}\mylabel{s:components}
Table\,\myref{t:matrix} summarizes the \rxrv\ components seen in low
magnetic-field neutron stars and black holes that we will discuss, grouped
into high-frequency phenomena ($\approxgt$100\,Hz), the low-frequency
complex (a group of correlated 10$^{-2}$--10$^2$\,Hz phenomena),
power-law components (usually dominating mainly the low frequencies),
and other phenomena.  Typical characteristic frequency ranges are also
given, as well as names used for phenomena in particular source types.
Here we introduce the high-frequency and low-frequency-complex
components, emphasizing relations across source types.  Further
details, and discussion of the power-law and other components are
provided in \S\S\myref{s:ns}--\myref{s:bh}.
\begin{table*}[h]
\begin{center}
\centering
\caption{Variability components \mylabel{t:matrix}}
\begin{tabular}{lcccccc}
\hline 
\hline 
          &             &\multispan{3}{\hfil Low magnetic-field neutron stars\hfil}&& Typical      \\
          &             &\multispan{3}{\hrulefill}    & Black                       & frequency    \\
          &             & Z        & Atoll  & Weak    & holes                       & range$^a$        \\
          & Name        & sources  & sources& LMXBs   &                             & (Hz)         \\
\hline                  
\hline                  
High-     & kHz QPOs    & \b       &  \b    & ---     &                             & 300--1200    \\
frequency & HF QPO      &          &        &         & \b                          & 100--500     \\
          & hHz QPO     & \c       &  \b    & ---     &                             & 100--200     \\
\hline                                                                                               
Low-      & BLN         & \b\,$^1$ & \b$^b$ & \b      & \b\,$^2$                    & 0.01--50     \\
frequency & LF hump/QPO & \b\,$^3$ &  \b    & \b      & \b                          & 0.1--50      \\
complex   & 3d Lor.     & ---      &  \b\,$^4$& \b\,$^4$ & \b                       & 1--100       \\
          & 4th Lor.    & ---      &  \b\,$^5$& \b\,$^5$ & \c                       & 10--500$^c$  \\
\hline
Power law & VLFN        & \b$^d$   & \b$^d$ & ---     & \b\,$^{6,e}$                & ---$^f$         \\
\hline                                                                                               
Other     & N/FBO       & \b       & \c     & ---     & ---                         & 4--20        \\
          & 1 Hz QPOs   & ---      &  ---   & \b$^g$  & ---                         & 0.5--2       \\
          & mHz QPOs    & ---      & \b     & ---     & \b                          & 0.001--0.01  \\
\hline                                                                               
\hline                                                                               
\end{tabular}
\end{center}
\begin{minipage}{\hsize}
\b: observed; \c: some doubt (uncertain, ambiguous, atypical, rare 
and not clearly seen, etc.); ---: variability of similar type as in
the other source types not reported. HF: high frequency; LF:
low-frequency; BLN: band limited noise; Lor.: Lorentzian; VLFN:
very-low frequency noise; N/FBO: normal/flaring branch oscillation.
Alternative names: $^1$\,LFN (low frequency noise), $^2$\,LS (low
state) noise, $^3$\,HBO (horizontal branch oscillation), $^4$\,\Llow,
$^5$\,\Lu, $^6$\,HS (high state) noise.  Notes: $^a$\,$\nu_{max}$,
cf.,
\S\myref{s:timing}, $^b$\,Sub-components, see \S\myref{s:lfc},
$^c$\,See note to Table\,\myref{t:lfc}, $^d$\,Sub-components, see
\S\myref{s:nsotherrxrv}, $^e$\,Stronger power law noise in IMS, see
\S\myref{s:bhpl}, $^f$\,Often only detected $<$ a few Hz, $^g$\,e.g., dipper
QPOs; 1~Hz flaring in SAX\,J1808.4--3658 \S\myref{s:nsotherrxrv}.
\end{minipage}
\end{table*}
\subsection{High-frequency phenomena}\mylabel{s:hf}
\paragraph{Neutron-star kilohertz QPOs}
\begin{figure*}[htbp]
$$
\psfig{figure=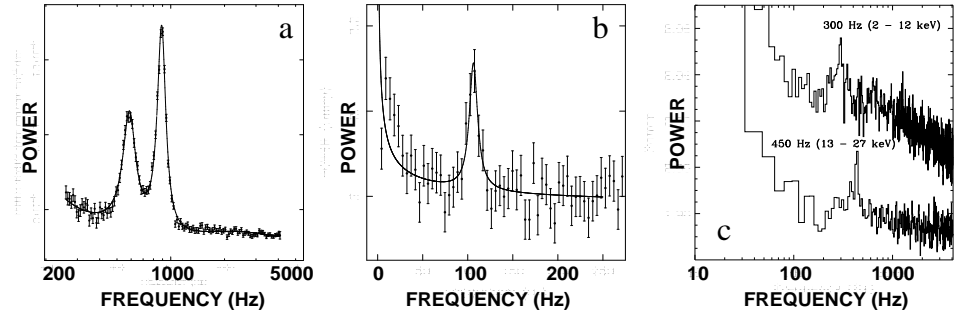,width=\linewidth}
$$
\caption{{\it (a)} Twin kHz QPOs in Sco\,X-1 (van der Klis et al.\ 1997), 
{\it (b)} hectohertz QPO in 4U\,0614+09, {\it (c)} HF
QPOs in GRO\,J1655--40 (Strohmayer 2001a).
\mylabel{f:hf}}
\end{figure*}
The fastest variability components in X-ray binaries are the {\em
kilohertz quasi-periodic oscillations} (kHz QPOs; van der Klis et al.\
1996, Strohmayer et al.\ 1996; van der Klis 1998 for a historical
account) seen in nearly all Z sources (in HB and upper NB) and atoll
sources (in IS and LLB) as well as a few weak LMXBs, including msec
pulsars (Wijnands et al.\ 2003).  Two QPO peaks (the {\it twin peaks})
occur in the power spectrum (Fig.\,\myref{f:hf}{\it a}) and move up
and down in frequency together in the 200--1200\,Hz range in
correlation with source state (cf., Fig.\,\myref{f:termscheme}). The
higher-frequency of these two peaks is called the {\it upper kHz QPO},
frequency $\nu_u$, the other the {\it lower kHz QPO} at $\nu_\ell$;
towards the edges of their observed frequency range peaks also occur
alone.  The several 100-Hz {\it peak separation}
$\Delta\nu\equiv\nu_u-\nu_\ell$ is typically within 20\% of the
neutron-star spin frequency, or half that, depending on source, and
usually decreases by a few tens of Hz when both peaks move up by
hundreds of Hz.  Most models involve orbital motion in the disk at one
of the kHz QPO frequencies (\S\myref{s:generic}).  Weak sidebands in
the kHz domain have been reported in some kHz QPO sources
(\S\myref{s:nskhz}).
\paragraph{Black-hole high-frequency QPOs}
The fastest black-hole phenomenon are the {\it high-frequency} (HF)
QPOs (Remillard et al.\ 1999c) seen in the IMS (usually
VHS).  Frequencies $\nu_{HF}$ range from 100 to 450\,Hz (their
relation to QPOs in the 27--67\,Hz range, Morgan et al.\ 1997, is unclear), 
and are reported to usually occur at fixed
values different in each source, perhaps inversely proportional
to black-hole mass.  In a few cases harmonically related (2:3;
\S\myref{s:bhhf}) frequencies have been seen (Fig.\,\myref{f:hf}{\it c}).  The phenomenon is weak
and transient so that observations are difficult, and discrepant
frequencies occur as well.  The constant frequencies might indicate a
link with neutron-star hectohertz QPOs (below), but it is not excluded
that black-hole HF and neutron-star kHz QPOs can be reconciled within
a single explanation (\S\myref{s:bhhf}).
\paragraph{Neutron-star hectohertz QPO}
The {\it hectohertz} (hHz) QPO (Ford \& van der Klis 1998) is a
peaked noise phenomenon (sometimes coherent enough to be called a QPO,
\S\myref{s:timing}) with a frequency $\nu_{_{hHz}}$ in the 100--200\,Hz
range that is seen in atoll sources in most states
(Fig.\,\myref{f:hf}{\it b}). It stands out from
all other neutron-star components by its approximately constant
frequency (Fig.\,\myref{f:termscheme}) which is quite similar across
sources, perhaps because $\nu_{_{hHz}}$ derives from compact-object
properties and the neutron stars in these systems are all similar
(\S\myref{s:generic}). In addition to a possible link with black-hole
HF QPOs, a link with the features between 10 and 100\,Hz reported by
Nowak (2000) in Cyg\,X-1 and GX\,339--4 has been suggested (van
Straaten et al.\ 2002) which may allow a $1/M$ scaling of frequency .
\bigskip

This concludes the summary of the $>$100\,Hz phenomena.  Neutron stars
have much more broad-band power in the kHz range than black holes
(Sunyaev \& Revnivtsev 2000, Klein-Wolt et al.\ 2004b), but there is no
indication that this is due to other components than those already
mentioned here.  The 'high-frequency noise' reported from neutron
stars in earlier work (e.g., Dieters \& van der Klis 2000) may variously be
due to the (low $Q$) upper kHz QPO at low frequency, hHz QPO, an HBO
harmonic (below) and/or instrumental effects (cf., Berger \& van der
Klis 1994).  Sporadic strong millisecond time-scale bursts were
reported from Cyg X-1 (Rothschild et al.\ 1974, 1977, Meekins et
al.\ 1984, Gierli\'nski \& Zdziarski 2003) but detecting such
phenomena in a strongly variable source is fraught with statistical
difficulties (e.g., Press \& Schechter 1974, Weisskopf \& Sutherland
1978, Giles 1981); instrumental problems caused some of the reported
detections (Chaput et al.\ 2000).
\subsection{The low-frequency complex}\mylabel{s:lfc}
In the 0.01--100 Hz range a set of usually two to five band-limited
noise, peaked-noise and QPO components is observed whose frequencies
all correlate.  This {\it low-frequency complex} is often dominated by
strong (up to 60\% rms), {\it flat-topped BLN} with a break at
frequency $\nu_b$ in the \about0.01--50\,Hz range and peaked noise
(the {\it LF hump}) at frequency $\nu_h$ roughly a factor 5 above
$\nu_b$ (Wijnands \& van der Klis 1999a).  Both components sometimes
feature QPOs located around $\nu_b$ and/or around (or instead of) the
hump near $\nu_h$.  The QPO near $\nu_h$ may show several harmonics
and is often called the LF QPO (at $\nu_{LF}$).  The BLN at $\nu_b$
goes under various names (LFN; \S\myref{s:nslfc}, LS noise;
\S\myref{s:bhlfc}, broken power law, flat-topped noise) and is often
just called 'the' BLN, or \Lb\ (L for 'Lorentzian', Belloni et
al.\ 2002a).  By analogy, other components go by names such as \Lh,
\LLF, \Ll, \Lu\ for the LF hump/QPO, lower/upper kHz QPO, etc; see
Table\,\myref{t:lfc}. \editnote{still add a better lf complex illustration?} 
\begin{figure*}[htbp]
$$
\psfig{figure=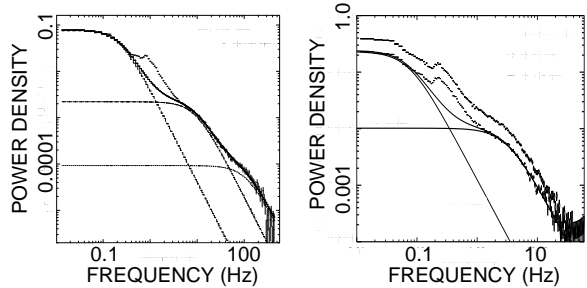,height=1.5in}
$$
\caption{The power spectra of neutron stars ({\it left}: 1E\,1724--3045) and 
black holes ({\it right}: GRO\,J0422+32) in the low state can be strikingly
similar. Note, however, the subtle {\it differences} in characteristic
frequencies, power levels and number of detectable components, which
may in fact be part of a systematic neutron-star black-hole
distinction; cf., Fig.\,\myref{f:psa} LS and EIS. After Olive et
al.\ (1998).
\editnote{could remove entire figure}
\mylabel{f:olive}}
\end{figure*}
\paragraph{Low states} 
When \Lb\ and \Lh\ are at low frequency ($\nu_b\approxlt1$\,Hz;
black-hole LS and atoll EIS, cf., Table\,\myref{t:states}), a third
BLN component is often present above the frequency of the LF hump/QPO.
The combination of these three noise components leads to the
characteristic power-spectral shapes displayed in
Fig.\,\myref{f:olive} that can be remarkably similar between neutron
stars and black holes (van der Klis 1994a and references therein,
Olive et al.\ 1998, Wijnands \& van der Klis 1999a, Belloni et
al.\ 2002a), and also show similar time lags (Ford et al.\ 1999).
This third component has been suggested to be the lower kHz QPO, \Ll,
but at frequencies as low as \about10\,Hz (\S\myref{s:freqcorr}), and
is designated \Llow\ here.  In neutron stars a fourth, $>$100\,Hz, BLN
component occurs above \Llow\ whose frequency connects smoothly with
that of the upper kHz QPO and, like it, is designated \Lu.

\begin{table}[htbp]
\begin{center}
\caption{High frequency and low frequency complex component names \& symbols} 
\mylabel{t:lfc}
\begin{tabular}{llcc}
\hline
\hline
          & Name           & Symbol    & Frequency \\
          &                &           & (Hz)      \\     
\hline                                  
\hline                                  
High-     & Upper kHz      & \Lu       & 500--1200 \\   
frequency & Lower kHz      & \Ll       & 300--1000   \\ 
          & High frequency & \LHF      & 100--500  \\                     
          & hectoHz        & \LhHz     & 100--200  \\   
\hline                                   
Low-      & BLN            & \Lb, \Lbt & 0.01--50  \\   
frequency & LF hump, LF QPO& \Lh,\LLF  & 0.1--50   \\   
complex   & 3d Lorentzian  & \Llow     & 1--100    \\   
          & 4th Lorentzian & \Lu       & 10--500$^a$   \\   
\hline                    
\hline                                  
\end{tabular}
\begin{minipage}{8cm}
$^a$ \,In neutron stars 140--500\,Hz, probably the upper kHz QPO
at low Q; in black holes 10--100\,Hz.
\end{minipage}
\end{center}
\end{table}
{\null\ \ \ \ \it Low to intermediate states\ \ \ \ } When atoll
sources and black holes move out of these low states
($\nu_b\approxgt$1\,Hz) towards the IMS/VHS or via the island state
(IS) to the lower-left end of the banana (LLB), respectively, all
components increase in frequency and become weaker and, often, more
coherent.  By the time that $\nu_b\sim10$\,Hz, \Lb\ and \Lh\ are much
weaker and \Llow\ is below detection.  Black holes in this state often
show narrow 1--30\,Hz LF QPO peaks with sometimes several harmonics
that differ in strength, phase lags and coherence between odd and even
ones, as well as  energy dependencies in \Lb\ (and hence, $\nu_b$,
\S\myref{s:bhlfc}); sometimes no BLN is detected but only
power-law noise (\S\myref{s:statesbh}, \S\myref{s:bhpl}).  In the
atoll sources \Lb\ attains substructure and often needs to be
described by two components, a peaked noise/QPO plus a BLN a factor
2--3 below it which is sometimes called \Lbt.
\begin{figure*}[htbp]
\begin{center}
$$\psfig{figure=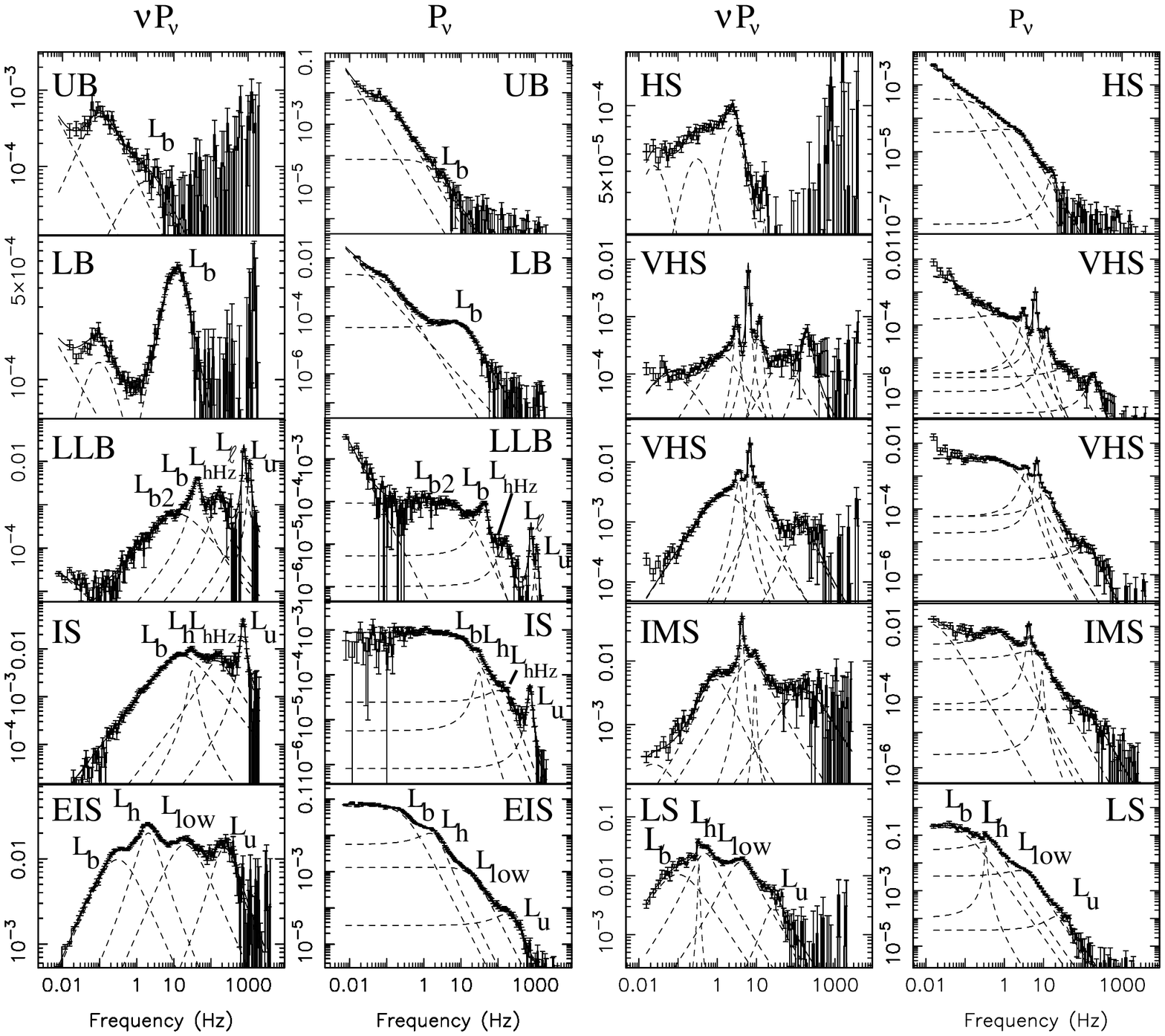,angle=-90,width=\linewidth}$$
\caption{Power spectra in {\it left:} atoll sources and {\it right:} black holes 
in their various states (\S\myref{s:states}) in $P_\nu$ and $\nu
P_\nu$ representations.  Two examples of VHS power spectra are shown,
with and without strong BLN.  Atoll source variability component
names are indicated (except the lowest-frequency, VLFN, components in
UB, LB and LLB; cf., Reerink et al.\ 2004).  The analogous broad-band
components in the black hole LS are also identified; the narrow
component is \LLF.
\mylabel{f:psa}}
\end{center}
\end{figure*}

Z sources, where $\nu_b$ is always $>1$\,Hz, are similar to this in
the HB and upper NB.  \Lb\ is sometimes called 'low-frequency noise'
(LFN) and has $\nu_b$ varying between 2 and 20 Hz.  A 15--60\,Hz LF
QPO superimposed on this called 'horizontal-branch oscillation' (HBO)
sometimes has several observable harmonics, with alternating high- and
low-Q harmonics somewhat reminiscent of those in black-hole LF QPOs;
warped disk geometries have been suggested as a cause of this (e.g.,
Jonker et al.\ 2000a, 2002a).

{\null\ \ \ \ \it Intermediate to high states\ \ \ \ } When atoll
sources move further onto the banana, the frequencies of the \Lb\
subcomponents rise (the QPO to 30-80\,Hz and the BLN to 20\,Hz) and
then, in the LB, decrease again (e.g., di Salvo et al.\ 2003, Reerink
et al.\ 2004).  The low-frequency complex components weaken until in
the UB they become undetectable.  A similar evolution (including
sometimes frequency reversals, e.g., Wijnands et al.\ 1996) is
observed in Z sources on the NB, and for black holes moving into the
HS, where dominant $\alpha$=0.7--1.4 power-law noise is also regularly
observed at this point (\S\myref{s:bhpl}), but no clear evidence for
frequency reversals has been reported (photon energy dependencies in
$\nu_b$ in the IMS further complicate the issue). Several more
components may occur in the range of the low-frequency complex
(\S\myref{s:nslfc}, \S\myref{s:bhlfc}) and disentangling them is
sometimes difficult.

\medskip
The $<$100\,Hz variability of black-hole LS and neutron-star EIS are
very similar and there is little doubt that they are physically
related.  There also obviously are close relations with, and between,
variability in black-hole IMS/VHS, atoll source IS/(L)LB, and Z source
HB/upper NB, but even empirically the exact nature of these relations
is not yet fully established (but see Klein-Wolt et al.\ 2004b for
recent progress).  An important clue is provided by the correlations
between the component frequencies (and strengths) which helps to
identify components across sources.  These are discussed next.
\section{Frequency correlations}\mylabel{s:freqcorr}
\subsection{Frequency correlations in neutron stars}\mylabel{s:nsfreqcorr}
Fig.\,\myref{f:termscheme}{\it a} displays the frequency correlations
of four well-studied atoll sources and four weak LMXBs (faint burst
sources, \S\myref{s:lmfnstypes}).  The frequencies of all components
described in \S\myref{s:components} are plotted vs.\ $\nu_u$. Source
state designations corresponding to the colour-colour diagrams in
Fig.\,\myref{f:cds} are also indicated.  As \Lu\ becomes undetectable
$\approxgt$1200\,Hz, the frequency evolution beyond that point is not
covered in Fig.\,\myref{f:termscheme}{\it a}.  The figure includes
data from sources covering an order of magnitude in luminosity when in
the same state, yet they display very similar power spectra and
essentially the same frequency correlations. The tracks of $\nu_b$
(and subcomponent $\nu_{b2}$), $\nu_h$, $\nu_{\ell ow}$, $\nu_{_{hHz}}$
and $\nu_\ell $ vs.\ $\nu_u$ are clearly recognizable.  This plot can
usefully serve as a template against which to match the
\rxrv\ of other objects.
\begin{figure*}[htbp]
\begin{center}
$$\psfig{figure=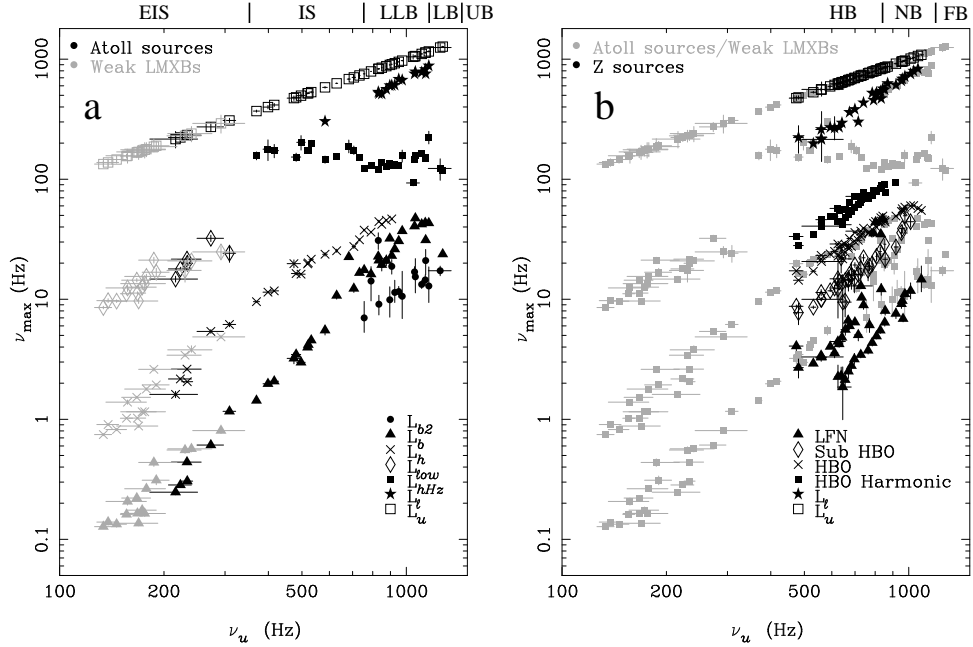,width=\linewidth}$$
\caption{Frequency correlations. {\it (a)} Atoll sources and weak
LMXBs, {\it (b)} Z sources compared with these objects.  The
characteristic frequencies ($\nu_{max}$, \S\myref{s:timing}) of the
components are plotted as indicated; approximate source state ranges
are indicated at the top.
\mylabel{f:termscheme}}
\end{center}
\end{figure*}

Fig.\,\myref{f:termscheme}{\it b} compares measurements for the Z sources
with the atoll source data.  The kHz QPOs match, as well as do the HBO
and \Lh.  Differences are that in Z sources no hectohertz QPO has been
reported so far, and that an additional HBO harmonic and
'sub-harmonic' occur.  How precisely these, and the LFN (the Z source
variant of \Lb) which varies even among the Z sources themselves,
relate to atoll source components is undecided.  The dependence of
$\nu_h$ and related QPOs on $\nu_u$ is approximately quadratic in both
Z and atoll sources (Stella \& Vietri 1998 and, e.g., Psaltis et
al.\ 1999b, Jonker et al.\ 1998, Homan et al.\ 2002, van Straaten et
al.\ 2003; Fig.\,\myref{f:quadratic}) suggesting that Lense-Thirring
precession of an orbit with frequency $\nu_u$ might be causing the LF
QPO (\S\myref{s:rpm}), but in GX\,17+1 at high frequency the HBO
frequency (and X-ray flux) start decreasing while $\nu_u$ continues
increasing (Homan et al.\ 2002).

An interesting discrepancy occurs in the frequencies of some
millisecond pulsars: in SAX\,J1808.4--3658 a pattern of correlated
frequencies occurs very similar to Fig.\,\myref{f:termscheme}{\it a},
but with relations that are offset.  At low frequencies, the match can
be restored by multiplying the $\nu_u$ values and the single measured
$\nu_\ell$ with \about1.45, which suggests a link with the 2:3
frequency ratios in black holes (\S\myref{s:hf}); at higher
frequencies the identification of the \Lb\ subcomponents is uncertain
(van Straaten et al.\ 2004).  Of the other millisecond pulsars,
XTE\,J0929--314 and XTE\,J1807--294 behave in the same way, but
XTE\,J1751--305 and XTE\,J1814--338 are like ordinary atoll sources in
this respect.  This suggests $\nu_u$ and $\nu_\ell$ form one group of
correlated frequencies and the LF complex another, independent one;
the case of GX\,17+2 above supports this.
\begin{figure*}[htbp]
\begin{center}
$$\psfig{figure=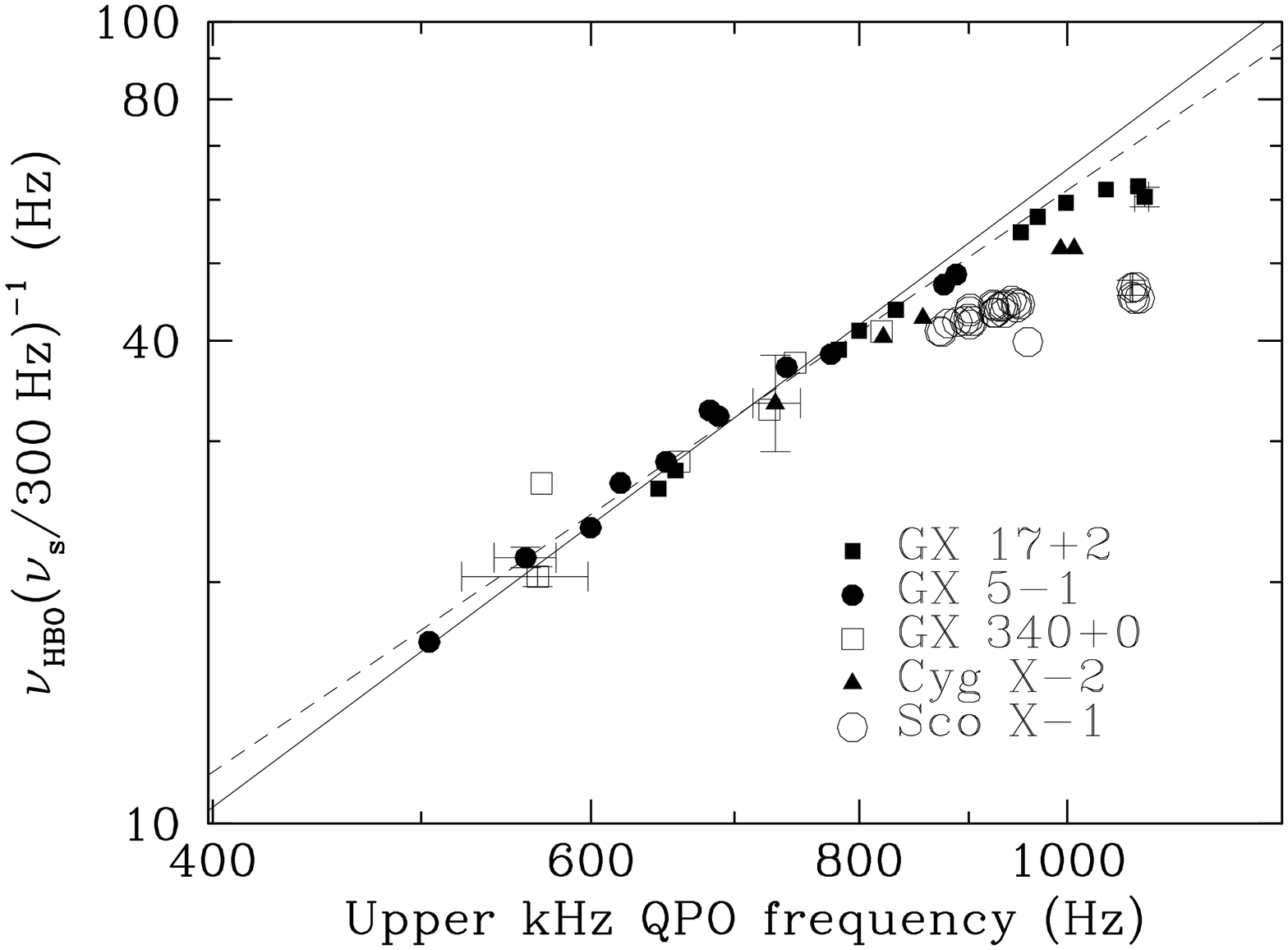,angle=-90,height=5cm}\qquad
\psfig{figure=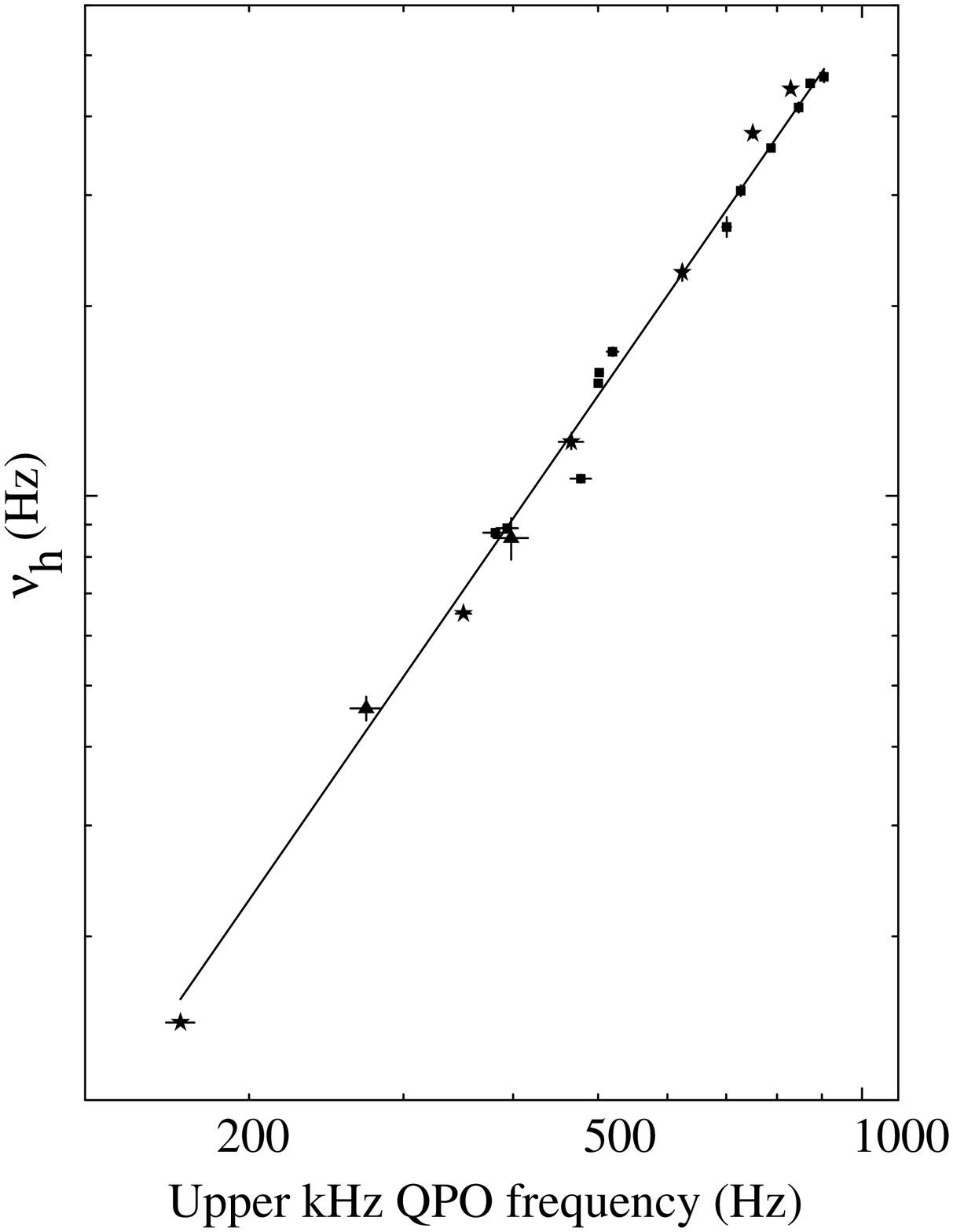,height=5cm}$$
\caption{The relation between upper kHz QPO frequency and, {\it
left}: 
HBO frequency in Z sources (scaled between sources by an 
inferred spin frequency), drawn line is for a quadratic
relationship, dashed line is best power-law fit, {\it right}: 
$\nu_h$ in atoll sources, line is power law with index 2.01.  In a
Lense-Thirring interpretation (\S\myref{s:grorbits}) a value of
$I_{45}/m$ of \about4 would be implied in both cases (see also
\S\myref{s:rpm}).  From Psaltis et al.\ (1999b) and van Straaten et
al.\ (2003). \mylabel{f:quadratic}}
\end{center}
\end{figure*}
\subsection{Frequency correlations of black holes compared to neutron stars}
\mylabel{s:bhnsfreqcorr}
Black holes can not be directly compared to Fig.\,\myref{f:termscheme}
as \Lu\ is not reliably detected.  However, correlations between other
frequencies in Fig.\,\myref{f:termscheme}, and one between frequency
and power, provide intriguing links between neutron stars and black
holes.

{\null\ \ \ \ \it BH relation\ \ \ \ }Belloni \& Hasinger (1990a)
noticed that in Cyg\,X-1 $\nu_b$ and the power-density level $P_{\nu
flat}$ of the BLN flat top anti-correlate (Fig.\,\myref{f:wkpbk}{\it
a}), an effect later also seen in other black holes in the LS and
neutron stars in the (E)IS (e.g., M\'endez \& van der Klis 1997, van
Straaten et al.\ 2000, Belloni et al.\ 2002a).  For $\nu_b$$<$1\,Hz
the relation is consistent with a BLN rms amplitude that remains
constant while $\nu_b$ varies; this constant amplitude is less for
neutron stars than for black holes and might even depend on mass
(Belloni et al.\ 2002a), but note that photon-energy dependencies can
affect $P_{\nu flat}$ as well.  Above 1\,Hz the BLN clearly weakens,
but the relation appears to extend to the black-hole I/VHS (van der
Klis 1994b) at $\nu_b$ up to \about10\,Hz, and to $\nu_b>10$\,Hz in
neutron stars (van Straaten et al.\ 2000).
\begin{figure*}[htbp]
$$\psfig{figure=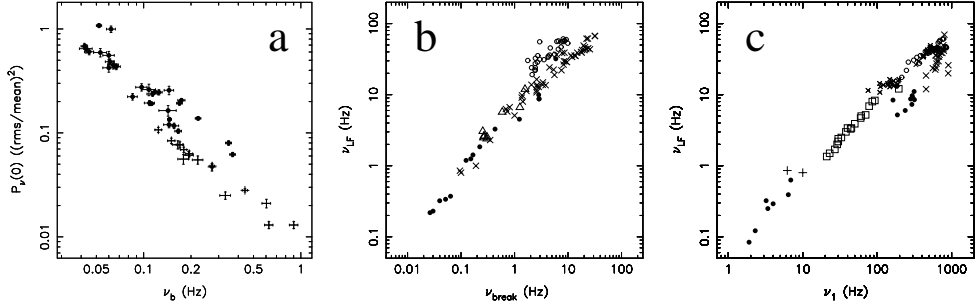,width=\linewidth}$$
\caption{Frequency and power correlations across neutron stars and 
black holes. {\it (a)} BLN flat-top power level vs.\ break frequency
(after Belloni et al.\ 2002a), {\it (b)} low-frequency hump/QPO vs.\
noise break frequency (after Wijnands \& van der Klis 1999a), {\it
(c)} low-frequency hump/QPO vs.\ $\nu_\ell$ or $\nu_{\ell\rm ow}$
(after Psaltis et al.\ 1999a). Filled circles represent black-hole
candidates, open circles Z sources, crosses atoll sources, triangles
the millisecond pulsar SAX\,J1808.4$-$3658, pluses faint burst sources
and squares Cir\,X-1. \mylabel{f:wkpbk}.}
\end{figure*}

{\null\ \ \ \ \it WK relation\ \ \ \ } Wijnands \& van der Klis
(1999a; WK) noted that in atoll sources (including weak LMXBs) and black
holes $\nu_b$ and $\nu_h$ are correlated over 3 orders of magnitude
(Fig.\,\myref{f:wkpbk}{\it b} , with the hump at $\nu_h$ a factor 8 to
2 above the break, decreasing with $\nu_b$.  This relation is that
between the two lower traces in Fig.\,\myref{f:termscheme}.  Z sources
are slightly above the main relation (cf., \S\myref{s:nsfreqcorr}),
with $\nu_h$ an also decreasing factor 10 to 3 above $\nu_b$. Of
course, the QPOs on the break of band-limited noise components
mentioned in \S\myref{s:lfc} hug the line $\nu_{BLN} = \nu_{QPO}$;
these may be physically the same components and are not shown in
Fig.\,\myref{f:wkpbk}. Cases of QPOs below the noise break have also
been reported (e.g., Brocksopp et al.\ 2001), and while in some
instances in the fits they could be reinterpreted as the reverse, this
may indicate that the description of the low-frequency complex is not
yet complete.

{\null\ \ \ \ \it PBK relation\ \ \ \ } Psaltis et al.\ (1999a; PBK)
found a rather good correlation between the frequencies of \Ll\ and
\Llow\ on the one hand and \Lh\ and \LLF\ on the other spanning nearly
three decades in frequency, with the Z and atoll sources populating
the $\nu_\ell$ ($>100$\,Hz) range and the weak LMXBs and black holes
in the LS the $\nu_{\ell ow}$ ($<10$\,Hz) one, and Cir X-1
(\S\myref{s:nsother}) filling in the gap between \Ll\ and \Llow\
(Figs.\,\myref{f:wkpbk}{\it c}, \myref{f:rpm}{\it b}).  The
correlation combines features from different sources with very
different Q values with relatively little overlap, and, as Psaltis et
al (1999a) note, although the data are suggestive, they are not
conclusive.  Whether the black-hole HF QPOs contributing to the small
branch below the main relation in Fig.\,\myref{f:wkpbk}{\it c} are
part of these correlations is questionable in view of their presumed
constant frequencies (\S\myref{s:hf}) --- on the other hand, in
XTE\,J1550--564 a correlation does occur between {\it observed} HF and
LF QPO frequencies (and spectral hardness; see \S\myref{s:bhhf}).

\bigskip
Further work (e.g., van der Klis 1994b, Crary et al.\ 1996, van der
Hooft et al.\ 1996, Belloni et al.\ 1996, 2002a,b, M\'endez and van
der Klis 1997, Ford \& van der Klis 1998, M\'endez et al.\ 1998d, van
der Hooft et al.\ 1999a, Wijnands \& van der Klis 1999b, Nowak 2000,
Remillard et al.\ 2002c, Revnivtsev et al.\ 2000c, van Straaten et
al.\ 2000, 2002, 2003, 2004, di Salvo et al.\ 2001a, Kalemci et al.\
2001, 2003 Wijnands \& Miller 2002, Yu et al.\ 2003, Olive et al.\
2003, Reig et al.\ 2004, Klein-Wolt et al.\ 2004b) produced many
examples of power spectra confirming these correlations, with in
particular the weak LMXBs and some ordinary atoll sources, as well as
some transient black holes in their decay bridging the gap between
neutron stars and black holes in the PBK relation.  In black holes a
feature perhaps similar to \Lu\ in neutron stars in the 10--100\,Hz
range was reported (Nowak 2000, Belloni et al.\ 2002a).  However, some
possible discrepancies also turned up in some of these same works
(e.g., Belloni et al.\ 2002a, van Straaten et al.\ 2002, 2004,
Pottschmidt et al.\ 2003) and as there is no direct observation of a
gradual transition, the identification of the high-Q lower kHz QPO in
Z and atoll sources with the low-Q \Llow\ component in the
neutron-star and black-hole low states remains conjectural.
\paragraph{Consequences of the correlations}
The relations of Fig.\,\myref{f:wkpbk} suggest that physically similar
phenomena cause the frequencies plotted there.  If so, then these
phenomena are extremely tunable, in some cases over nearly three
orders of magnitude in frequency, and occur in neutron stars as well
as black holes, which probably means they arise in the disk.  In
\S\myref{s:generic} we look at some of the models for this.

Warner \& Woudt (2002) and Mauche (2002), noted that the 'PBK'
relation may even extend to white dwarf systems (\S10\editnote{check})
down to frequencies a factor 10$^2$ below those in X-ray binaries,
which would mean that strong field gravity {\it per se} is not a
requirement for the accretion disk to produce the correlated
frequencies $\nu_\ell$ and $\nu_b$. Note that even then orbital motion
in the strong-field region of neutron stars and black holes is
implicated by the high frequencies observed, and that producing the
second kHz QPO may require strong-field gravity.  As remarked above,
the identification of the variability phenomena across source types is
difficult, so these results should be interpreted with caution.
\section{Orbital and epicyclic frequency models}
\mylabel{s:generic} 
Because of their direct link with physical time scales, interpretation
of observed characteristic frequencies dominates the discussion about
the nature of \rxrv.  What modulates the X-ray flux
(\S\myref{s:modulation}) and why the phenomenon is quasi-periodic
rather than periodic (\S\myref{s:decoherence}) often gets less
attention. Many QPO models are essentially models for a periodic
phenomenon (e.g., orbital motion) supplemented with a decohering
mechanism (e.g., damping).  Some broad-band noise models also work in
this way, but many (and some QPO models as well) are instead
intrinsically aperiodic (e.g., shots), with characteristic frequencies
arising through, e.g., correlations in the signal (cf.,
\S\myref{s:shots}).  Another important distinction is that between
constant variability frequencies, which may be expressible in
parameters of the compact object ($M$, $R$, $J$) alone, and variable
ones, which involve flow, or radiation-field parameters.

Phenomena that occur in both neutron stars and black holes can not
rely on physical properties unique to either object, such as either a
material surface or a horizon, a magnetic field not aligned with the
spin or extreme values of $J/M$. Hence, these essentially require an
origin in the accretion flow. Instead, in phenomena unique to either
neutron stars or black holes a role for unique compact-object
properties is likely.

For accretion-flow phenomena, the most obvious source of variability
is the disk, which with Keplerian orbital motion at each radius and
various oscillation modes provides a multitude of variability
frequencies.  Other structures, e.g., the 'corona' from X-ray spectral
models (\S\myref{s:statesbh}), a magnetosphere, a neutron-star/disk
boundary layer, or a jet (\S9\editnote{check}) can contribute as well.
Nevertheless, orbital motion (including general-relativistic epicyclic
motions, \S\myref{s:grorbits}) and disk oscillations
(\S\myref{s:diskoscillations}) are the mechanisms most often considered
for QPO phenomena.

In the current section we look at the interpretations of variability
frequencies that involve orbital and epicyclic motions in the
accretion disk.  Some interpretations recur in various models and can
be considered model 'building blocks'.  Models based on flow
instabilities, and interpretations of other aspects of the variability
(amplitude, coherence, phase and their photon-energy dependencies) are
discussed in \S\myref{s:othermodels}. Some specific models proposed
for specific variability components are mentioned in
\S\S\,\myref{s:ns}--\myref{s:bh}.
\subsection{General-relativistic orbital motion}
\mylabel{s:grorbits}
In classical physics, free-particle orbits around a spherically
symmetric mass $M$ are closed, and occur with Keplerian frequency
$$\nu_K=\sqrt{GM/r^3}/2\pi\approx
1184\,\hbox{Hz}\,\left(r\over15\,\hbox{km}\right)^{-3/2}m_{1.4}^{1/2}
\approx
184\,\hbox{Hz}\,\left(r\over100\,\hbox{km}\right)^{-3/2}m_{10}^{1/2},$$
where $m_{1.4}$ and $m_{10}$ are the compact object's mass in units of
1.4 and 10\,\msun, respectively, and $r$ is the orbital radius. In
general relativity, orbits are not closed, as the frequencies of
azimuthal, radial and vertical motion differ (Fig.\,\myref{f:rrm},
e.g., Merloni et al.\ 1999): in addition to the azimuthal motion at
the {\it general-relativistic orbital frequency} $\nu_\phi$, there are
the {\it radial} and {\it vertical epicyclic} frequencies $\nu_r$ and
$\nu_\theta$.\footnote[4]{Various other terms are used for these
motions: e.g., 'longitudinal' for $\nu_\phi$; 'latitudinal' or
'meridional' for $\nu_\theta$.} Due to this, eccentric orbits waltz at
the {\it periastron precession frequency} $\nu_{peri}=\nu_\phi-\nu_r$
and orbits tilted relative to the equatorial plane of a spinning
central mass wobble at the {\it nodal precession frequency}
$\nu_{nodal}=\nu_\phi-\nu_\theta$.
\editnote{would be nice to have a figure about the precession geometries, 
and another one like bardeen et al of rms rmb etc}
\begin{figure*}[htbp]
\begin{center}
\begin{tabular}{c}
\psfig{figure=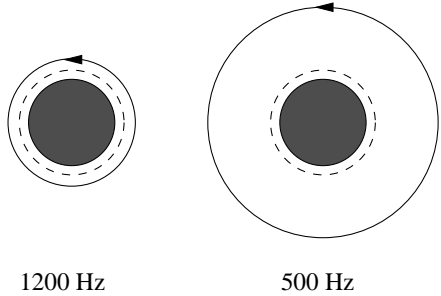,height=1.5in,angle=-90}
\end{tabular}
\hspace{1cm}
\begin{minipage}{5cm}
\caption{\small A 10-km radius, 1.4\msun\ neutron star with the
corresponding innermost stable circular orbit (ISCO; dashed circles)
and orbits (drawn circles) corresponding to orbital frequencies of
1200 and 500\,Hz, drawn to scale. \editnote{could put two NS orbits 
together, add Sch and j=0.9 BH}} \mylabel{f:circles}
\end{minipage}
\end{center}
\end{figure*} 

For equatorial circular orbits in Kerr spacetime (i.e., around a
spinning point mass $M$ with angular momentum $J$) the orbital
frequency is given by
\def\rg{(r_g/r)}
$$\nu_\phi={\sqrt{GM/r^3}/2\pi\over{1+j\rg^{3/2}}} =
\nu_K(1+j\rg^{3/2})^{-1},$$
where $j\equiv Jc/GM^2$ is the Kerr angular-momentum
parameter\footnote[5]{The quantity $j$ is also sometimes called $a_*$
or, exceptionally, $a$. Most authors let $a$ denote $Jc/GM$, some use
$a=J/M$, e.g., Stella \& Vietri (1998).  $r_g$ sometimes denotes
$2GM/c^2$.}; $0<j<1$ for prograde orbits, $-1<j<0$ for retrograde
ones, and $r_g\equiv GM/c^2$.  For the Schwarzschild geometry, $j=0$,
we have $\nu_\phi=\nu_K$. Infinitesimally tilted and eccentric orbits
will have radial and vertical epicyclic frequencies of
$$\nu_r=\nu_\phi\left(1-6\rg+8j\rg^{3/2}-3j^2\rg^2\right)^{1/2},
\hbox{\rm\ and}$$
$$\nu_\theta=\nu_\phi\left(1-4j\rg^{3/2}+3j^2\rg^2\right)^{1/2}$$
(Fig.\,\myref{f:rrm}).  For arbitrary-inclination orbits, see
Sibgatullin (2002).  All these frequencies are as observed by a static
observer in asymptotically flat spacetime, i.e., at infinity.  In a
Schwarzschild geometry a local observer, due to gravitational time
dilation, sees a frequency higher by $(1-2\rg)^{-1/2}$; in the general
case, the locally observed frequency also depends on the observer's
angular motion.

General relativity predicts (e.g., Bardeen et al.\ 1972) that in a
region close to a compact object no stable orbital motion is possible;
hence the above expressions are directly useful only outside that
region.  In a Schwarzschild geometry the {\it innermost stable
circular orbit} (ISCO) or {\it marginally stable orbit} has a radius
$$r_{ms} = 6r_g = 6GM/c^2\approx 12.5m_{1.4}\,\hbox{km}\approx
89m_{10}\,\hbox{km},$$
and the corresponding orbital frequency, the highest stable orbital
frequency, is
$$\nu_{ms} = c^3/2\pi6^{3/2}GM \approx
(1566/m_{1.4})\,\hbox{Hz}\approx (219/m_{10})\,\hbox{Hz}.$$
For prograde orbital motion in the equatorial plane of a Kerr geometry
the ISCO is smaller; as $j\rightarrow1$, $r_{ms}\rightarrow r_g$ and
$$\nu_{ms} = c^3/4\pi GM \approx (1611/m_{10})\,\hbox{Hz}.$$
\editnote{check w/ cole}
This is still outside the hole, because the horizon radius
$$r_{horizon}=r_g(1+(1-j^2)^{1/2})$$
shrinks from $2r_g$ at $j=0$ to $r_g$ as $j\rightarrow1$ as well.  The
corresponding frequency increases with $j$, and hence for a black hole
of known mass a high (test-particle) orbital frequency can be used to
argue for its spin (Sunyaev 1973; 
see also \S\myref{s:radii}). The full expression for the ISCO radius
(Bardeen et al.\ 1972) follows from the condition
$r^2-6r_gr+8r_g^{3/2}r^{1/2}j-3r_g^2j^2\ge0$ for stable orbits and is
somewhat tedious; substitution into the expression for $\nu_\phi$
yields the Kerr geometry ISCO frequency in closed form. To first order
in $j$ (Klu\'zniak et al.\ 1990, Miller et al.\ 1998a,b)
$$r_{ms}\approx(6GM/c^2)(1-0.54j)\qquad \hbox{and} \qquad \nu_{ms}\approx
(c^3/2\pi6^{3/2}GM)(1+0.75j).$$ 
Some disk flows can penetrate down to inside the ISCO before the
matter plunges in (e.g., Abramowicz et al.\ 2004, cf., \S\myref{s:radii}), but
not beyond the marginally bound orbit at $r_{mb} =
r_g(2-j)+2r_g(1-j)^{1/2}$, which is inside the ISCO (at $4r_g$ in the
Schwarzschild geometry).

Spacetime outside a spherically symmetric non-rotating star is
Schwarzschild. For {\it spinning} stars the exterior spacetime is Kerr
to first order in $j$; to higher order the metric, and hence the
precise frequencies, depend on the mass distribution (Hartle \& Thorne
1968; see Miller et al.\ 1998a, Shibata \& Sasaki 1998, Morsink \& Stella
1999, Markovi\'c 2000, Sibgatullin 2002, Abramowicz et al.\ 2003a).
Depending on mass and EOS (\S\myref{s:intro}), spinning neutron stars
could have appreciable angular momentum (e.g., $j$\about0.2 and
\about0.5 for 500 and 1000 Hz spins, respectively, Miller et al.\ 1998a). For
small-$j$ holes and slowly spinning neutron stars the first-order
expressions above apply.  In all cases $\nu_{ms}$ scales as $1/M$.

As $\nu_r$ and $\nu_\theta$ are both $<\nu_\phi$, periastron and nodal
precession are both prograde. Periastron precession is a consequence
of the non-$1/r^2$ nature of gravity in general relativity; the
classic example is Mercury's general-relativistic perihelion
precession. Nodal or 'Lense-Thirring' precession (Lense \& Thirring
1918) is due to the 'frame dragging' caused by the central object's
spin and does not occur if $j=0$; this
'gravito-magnetic'\footnote[6]{This term does not refer to any
magnetic field, but to an analogy with electromagnetism.} effect has
not yet been detected with certainty in any system.  In the weak-field
($r_g/r\ll1$) slow-rotation ($j\ll1$) limit the prediction is that
$\nu_{nodal}=(GM)^2j/\pi c^3r^3=8\pi^2\nu_\phi^2I\nu_{spin}/Mc^2$,
\editnote{how derived? do again} where $I$ is the neutron-star moment of
inertia and $\nu_{spin}$ its spin frequency: by measuring
$\nu_{nodal}$, $\nu_\phi$ and $\nu_{spin}$ the neutron-star
structure-dependent quantity $I/M$ is constrained. In terms of
$I_{45}/m$, where $I_{45}$ and $m$ are $I$ and $M$ in units of
\E45;\,g\,cm$^2$ and \msun, respectively,
$$\nu_{nodal}=13.2\ {\rm Hz}\ \left(I_{45}\over
m\right)\left(\nu_\phi\over1000\ {\rm Hz}\right)^2\left(\nu_{spin}\over300\
{\rm Hz}\right);$$
values of $I_{45}/m$ between 0.5 and 2 are expected (Stella \& Vietri
1998).
%
\begin{figure*}[htbp]
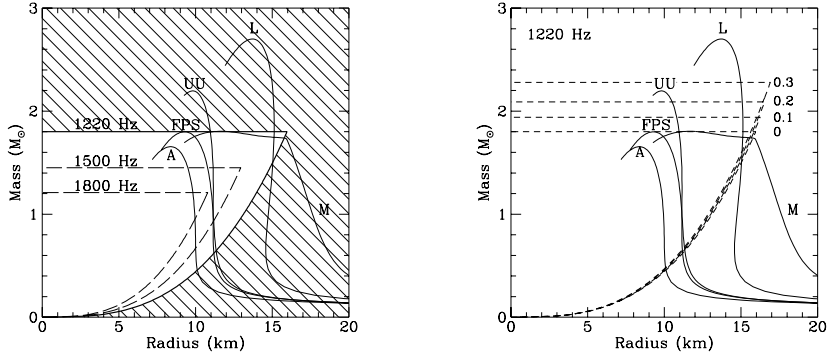

$$\psfig{figure=millerlambpsaltis_fig14.postscript,height=2.25in}
\psfig{figure=millerlambpsaltis_fig15.postscript,height=2.25in}$$
\caption{Constraints on neutron-star $M$ and $R$ from
orbital motion. {\it Left:} for $j=0$, orbital frequencies as indicated; the
hatched area is excluded if $\nu_\phi=1220$\,Hz. {\it Right:} with first order
corrections for frame dragging for the values of $j$ indicated.
Mass-radius relations for representative EOS are shown. (from Miller
et al.\ 1998a)
\mylabel{f:wedge}}
\end{figure*}

The detection of stable orbital motion at frequency $\nu$ around a
neutron star constrains both its mass $M$ and radius $R$ (Miller et
al.\ 1998a): (i) $R$ must be smaller than the orbital radius $r$, and
(ii) the ISCO must {\it also} be smaller than $r$ (assuming the orbit
is circular).  Condition (i) through the expression for $\nu_\phi$
translates into a mass-dependent upper limit on $R$, and condition
(ii) is an upper limit on $M$, in Schwarzschild geometry:
$6GM/c^2<(GM/4\pi^2\nu^2)^{1/3} \Rightarrow M<c^3/(2\pi6^{3/2}G\nu)$.
Fig.\,\myref{f:wedge} shows these limits in the neutron-star
mass-radius diagram. \hide{isn't there the problem that these radii
are BL but should be locally observed circumferential? no, says cole,
that's a higher order issue}

Detection of the ISCO would constitute direct proof of a qualitative
strong-field general-relativistic effect and at the same time
demonstrate that the neutron star is smaller than the ISCO. This
possibility was discussed since early on (e.g., Klu\'zniak \& Wagoner
1985, Paczy\'nski 1987, Biehle \& Blandford 1993).  Miller et
al.\ (1998a) pointed out that when the inner edge of the accretion
disk reaches the ISCO, the associated kHz QPO frequency might reach a
ceiling while \mdot\ continues rising (see also
\S\myref{s:khzinterp}).
\subsection{Relativistic precession models}\mylabel{s:rpm}
The term {\it relativistic precession model} (Stella \& Vietri 1998)
is used for a class of models in which observed frequencies are
directly identified with orbital, epicyclic, and precession
frequencies.  These models need additional physics to pick out one or
more particular radii in the disk whose frequencies correspond to those observed
(\S\myref{s:radii}).  Stella \& Vietri (1998, 1999) identify the upper
kHz QPO frequency $\nu_u$ (\S\myref{s:hf}) with the orbital frequency
$\nu_\phi$ at the inner edge of the disk (\S\myref{s:radii}), and
relate $\nu_\ell$ and $\nu_h$ (or related QPOs, \S\myref{s:lfc}) with,
respectively, periastron precession ($\nu_{peri}$) and nodal
precession ($\nu_{nodal}$) of this orbit.
\begin{figure*}[htbp]
$$\psfig{figure=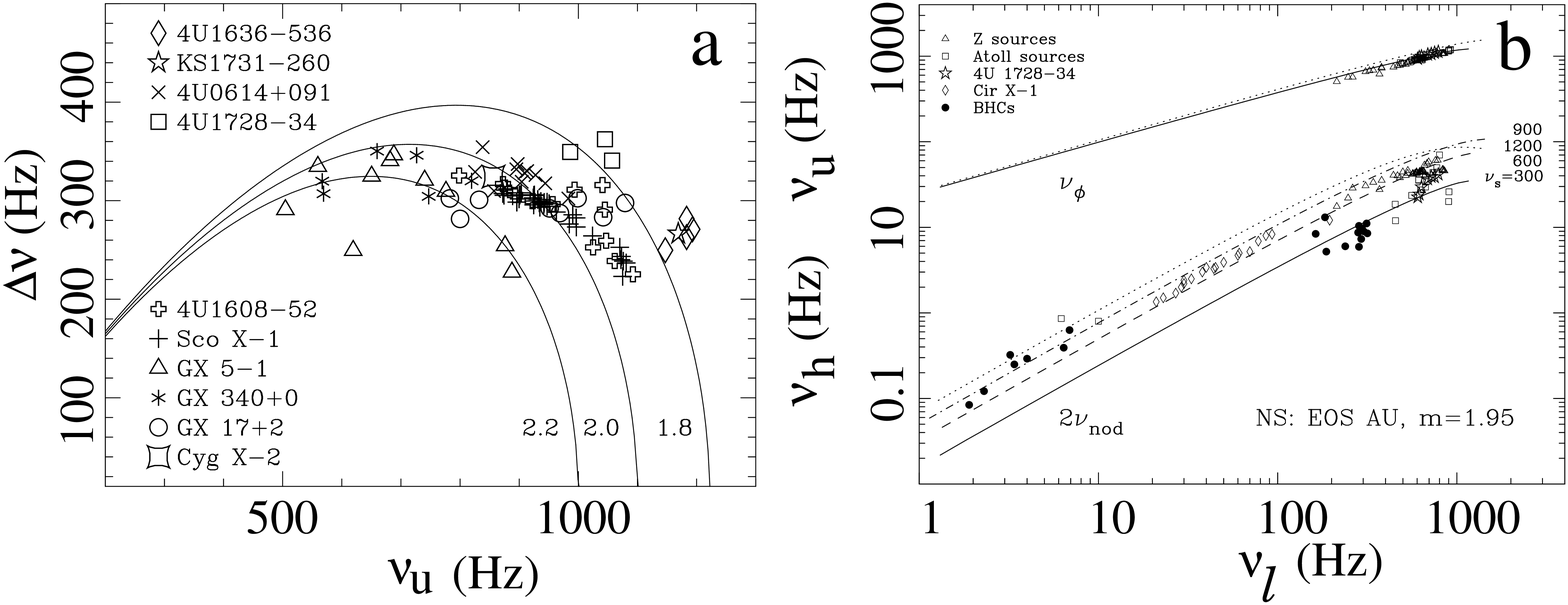,width=\textwidth}$$
\caption{Predicted relations between {\it (a)} $\nu_u$ and $\Delta\nu$ 
and {\it (b)} $\nu_\ell$ and $\nu_h$ as well as $\nu_u$ for different
values of the spin frequency (300\,Hz, 600\,Hz, etc.) in the
relativistic precession model, compared with observed values. (Stella
\& Vietri 1999, Stella et al.\ 1999). See \S\myref{s:bhnsfreqcorr}
for a discussion of the data in the right-hand frame. \mylabel{f:rpm}}
\end{figure*}

So, $\nu_h$ is predicted to be proportional to ${\nu_u}^2$
(\S\myref{s:grorbits}), which is indeed as observed
(\S\myref{s:nsfreqcorr}), and kHz QPO peak separation
$\Delta\nu\equiv\nu_r$.  Stellar oblateness affects the precession
rates and must be corrected for (Morsink \& Stella 1999, Stella et
al.\ 1999).  This model does not explain why $\Delta\nu$ is
commensurate with the spin frequency, nor how neutron stars with
different spins can have the same $\nu_u$--$\nu_h$ relation (van
Straaten et al.\ 2004).  A clear prediction is that $\Delta\nu$ should
decrease not only when $\nu_u$ increases (as observed) but also when
it sufficiently decreases (Fig.\,\myref{f:rpm}, see also
\S\myref{s:deltanu}).  For acceptable neutron-star parameters ($I/M$,
see \S\myref{s:grorbits}), $\nu_h$ is several times larger than the
$\nu_{nodal}$ predicted.  In a warped disk geometry $\nu_h$ could be
$2\nu_{nodal}$ or $4\nu_{nodal}$ (Morsink \& Stella 1999).  Cui et
al.\ (1998b) propose identifications of some black-hole frequencies
with $\nu_{nodal}$ (see also Merloni et al.\ 1999).

A precise match between model and observations requires additional
free parameters.  Stella \& Vietri (1999) propose that orbital
eccentricity systematically varies with orbital frequency.  Stella et
al.\ (1999), identifying $\nu_\ell$ and $\nu_{low}$ with
$\nu_{peri}$, and $\nu_h$ and $\nu_{LF}$ with $2\nu_{nodal}$ (cf.,
\S\myref{s:components}), produce an approximate match to the PBK
relation (\S\myref{s:freqcorr}) across neutron stars and black holes
(Fig.\,\myref{f:rpm}{\it b}) for reasonable black-hole masses and
$j$=0.1--0.3, but requiring neutron-star spin rates higher than
measured and masses of $\sim2$\msun.  For critical discussions of
these models see Psaltis et al.\ (1999b) and Markovi\'c \& Lamb (1998,
2000).  Vietri \& Stella (1998) and Armitage \& Natarajan (1999) have
performed calculations relevant to the problem of sustaining the
tilted orbits required for Lense-Thirring precession in a viscous
disk.  Miller (1999) calculated the effects of radiation forces on
Lense-Thirring precession.

Relativistic precession models are very predictive, as the frequency
relations are set by little more than compact-object parameters and
general relativity, and in unmodified form most are contradicted by
observations (e.g., Homan et al.\ 2002, van Straaten et
al.\ 2004). Yet the observed quadratic dependencies between $\nu_u$
and $\nu_h$ (\S\myref{s:bhnsfreqcorr}) which suggest Lense-Thirring
precession (\S\myref{s:grorbits}), are striking.  Calculations of the
theoretically expected QPO sideband patterns produced by luminous
clumps in orbits with epicyclic motions (Karas 1999a, Schnittman \&
Bertschinger 2004), and observations of such patterns, can help to
further test models in this class.  Certain disk oscillation models
are predicted to produce frequencies close to the orbital and
epicyclic ones and may be considered one way to implement the models
described here; hydrodynamic effects are expected to somewhat modify
the frequencies and to produce combinations between them, which may
allow to better fit the data.  These models are discussed in
\S\myref{s:diskoscillations}.  The general idea that an observed
frequency is the orbital frequency $\nu_\phi$ at the appropriate
radius in the disk is applied very often, and papers identifying
observed frequencies with orbital frequencies of clumps, vortices,
etc., are too numerous to cite (see e.g., Chagelishvili et al.\ 1989,
Abramowicz et al.\ 1992, Bao \& \O stgaard 1994, 1995, Karas 1999b,
Wang et al.\ 2003 and \S\myref{s:khzinterp}).
\subsection{Relativistic resonance models}\mylabel{s:rrm}
Relativistic resonance models (Klu\'zniak \& Abramowicz 2001,
Abramowicz \& Klu\'zniak 2001) make use of the fact that at
particular radii in the disk the orbital and epicyclic frequencies
have simple integer ratios or other commensurabilities with each other
or with the spin frequency.  At these radii resonances may occur which
show up in the observations: general relativity itself picks out
the frequencies from the disk.  Various different physical effects can
produce resonance.  A type of resonance invoked in some of these models
is 'parametric resonance': resonance in a system whose eigenfrequency
$\nu_0$ is itself perturbed at a frequency $\nu_1$ commensurate with
$\nu_0$; resonances occur when $\nu_0/\nu_1=2/n,\ n = 1,2,3,...$.
\begin{figure*}[htbp]
\begin{center}
\begin{tabular}{c}
\psfig{figure=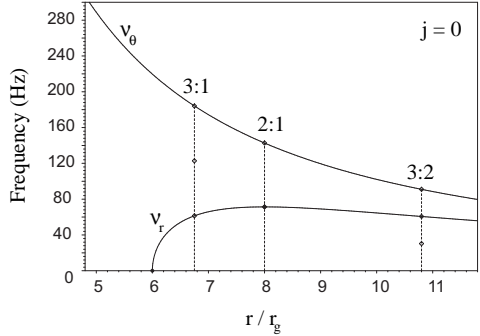,width=8cm}
\end{tabular}
\hspace{-1cm}
\begin{minipage}{5cm}
\caption{Radial and vertical epicyclic frequencies vs.\ $r/r_g$ in 
Schwarzschild geometry.  Three resonant radii are indicated.  After
Abramowicz et al.\ (2004).
\editnote{still to put in j and rg} 
\mylabel{f:rrm}}
\end{minipage}
\end{center}
\end{figure*}

Various resonant radii have been discussed
(Fig.~\myref{f:rrm}): the radii where $\nu_r/\nu_\phi$ equals $1/2$ or
$1/3$ (Klu\'zniak \& Abramowicz 2001), that where
$\nu_r/\nu_\theta=2/3$ (Klu\'zniak \& Abramowicz 2002; 
parametric resonance with $n=3$, the lowest value allowed as
$\nu_r<\nu_\theta$), and those where $\nu_\theta=2\nu_r$ or
$\nu_\theta=3\nu_r$ (Abramowicz \& Klu\'zniak 2004) can all be used to
explain the observed 2:3 ratios for HF QPOs in black holes
(\S\myref{s:bhhf}), with other resonant frequencies (1:2:3:5) also
being predicted (Klu\'zniak \& Abramowicz 2003). Like the ISCO
frequencies, all these resonant frequencies are predicted to scale
with $1/M$ (Abramowicz et al.\ 2004), and if the observed resonance is
identified they can be used to constrain both $M$ and $J$.

In the interaction between the neutron-star spin and the disk,
resonances could arise as well (e.g., Psaltis 2001, van der Klis
2002).  Klu\'zniak et al.\ (2004) suggest several ways in which the
epicyclic frequencies can resonate with spin (either $\nu_{spin}$ or
$\nu_{spin}/2$; see also Wijnands et al.\ 2003 who suggest
$\nu_\phi-\nu_r$ equals $\nu_{spin}$ or $\nu_{spin}/2$ and Lee et al.\
2004, who consider $\nu_\theta-\nu_r=\nu_{spin}/2$), which is relevant
to neutron-star kHz QPOs and their commensurabilities with the spin
frequency (\S\myref{s:deltanu}), and also point out that disk g-modes
(\S\myref{s:diskoscillations}) may resonate in a frequency ratio
tending to $\sqrt{2}$, all of which may be relevant to the twin kHz
QPOs observed in SAX\,J1808.4--3658 (\S\myref{s:deltanu}).  Lamb \&
Miller (2003) propose a resonance at the 'spin-resonance' radius where
the spin-orbit beat frequency equals the vertical epicyclic one:
$\nu_{spin} - \nu_\phi = \nu_\theta$; at this radius the pulsar beam
stays pointed at particles in the same {\it phase} (e.g., an antinode)
of their vertical epicyclic motion.  This radius is sufficiently far
out in the disk that $\nu_\phi \approx \nu_\theta$, i.e., $\nu_\phi
\approx \nu_{spin}/2$, which may explain why the observed kHz QPO 
separation is sometimes the spin frequency, sometimes half that
(\S\myref{s:deltanu}).

The radii at which a resonance condition applies are fixed, so these
models in principle produce constant frequencies, which is appropriate
to black-hole HF QPOs, but analytic (Rebusco 2004) and numerical
(Abramowicz et al.\ 2003b) work on 'tunable' versions which might be
applicable to kHz QPOs has been done.  Tunable frequencies can also be
produced by beating constant resonant frequencies with a variable one
(Lamb and Miller 2003; see \S\myref{s:bfm}).
Observationally, the 2:3 ratios in black holes (\S\myref{s:bhhf}), the
half-spin separation of some kHz QPOs (\S\myref{s:deltanu}) and the
factor \about1.45 offsets of some of the frequency correlations in
accreting millisecond pulsars (\S\myref{s:nsfreqcorr}) all suggest
that resonances play a role, but it is not yet entirely clear exactly
which are the frequency ratios that can occur, when they occur, and
what is the incidence of each ratio.  Clarifying this will allow to
test the various proposals for frequency commensurabilities that have
been put forward.
\subsection{Beat-frequency models}
\mylabel{s:bfm}
By 'beating' orbital frequencies with the spin frequency of a central
neutron star more frequencies can be produced (e.g., Alpar \& Shaham
1985, Lamb et al.\ 1985, Miller et al.\ 1998a).  This requires that
some azimuthally non-uniform structure co-rotating with the spin, e.g.,
a non-aligned magnetic field or radiation pattern, reaches out to the
relevant orbital radius. These models do not work for black holes, for
which in view of the no-hair theorem no such structures are expected.
Additionally there must be some azimuthal structure in orbit to
interact with the spin; the mechanism will not work if the flow is
completely axi-symmetric.
\begin{figure*}[htbp]
\begin{center}
\begin{tabular}{c}
\psfig{figure=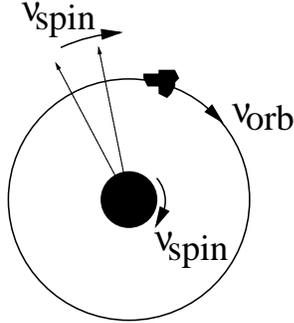,height=1.9in} 
\end{tabular}
\hspace{1cm}
\begin{minipage}{5cm}
\caption{Example of a beat-frequency interaction. The clump orbiting
with $\nu_{orb}$ periodically overtakes the magnetic-field lines or
the pulsar beam that is sweeping around with $\nu_{spin}$.  This
happens $\nu_{beat}$ times per unit of time, where
$\nu_{beat}=\nu_{orb} -
\nu_{spin}$ is the beat frequency. \mylabel{f:bfm}}
\end{minipage}
\end{center}
\end{figure*}

Spin-orbit interaction then takes place at the difference frequency
between the orbital and spin frequencies: the {\it beat frequency}
$\nu_{beat} = \nu_{orb} - \nu_{spin}$.  This is the natural disk/star
interaction frequency: the frequency at which a particle orbiting in
the disk periodically overtakes a given point on the spinning star.
An example of an interaction making $\nu_{beat}$ observable would be
the periodic illumination of an orbiting blob by a pulsar beam
sweeping around at $\nu_{spin}$ (Fig.\,\myref{f:bfm}).  Only {\it one}
beat frequency (a 'single sideband') is produced: for prograde orbital
motion no signal occurs at $\nu_{orb}+\nu_{spin}$ while for retrograde
motion {\it only} this sum frequency is generated. If the beat
interaction occurs at the magnetospheric radius (see
\S1\editnote{check}) of an accreting magnetic star one expects
$\nu_{spin}<\nu_{orb}$, but in other settings, if the relevant orbit
is sufficiently far out, $\nu_{spin}>\nu_{orb}$ is possible, and for a
variable orbital radius $\nu_{beat}$ could go through zero.

If there is an n-fold symmetric azimuthal pattern associated with spin
and/or orbit, frequency multiples occur (e.g., for 2 pulsar beams
$\nu_{beat} = 2(\nu_{orb} - \nu_{spin})$).  If spin and orbit have an
$n_{spin}$ and $n_{orb}$-fold symmetry, respectively, then a
degeneracy occurs and $\nu_{beat} = \mu(n_{spin},n_{orb})( \nu_{orb} -
\nu_{spin})$, where $\mu(x,y)$ denotes the least common multiple
(Wijnands et al.\ 2003; additionally non-sinusoidal signals include
harmonics to the fundamental frequency, and if these also interact a
complex spectrum results, Miller et al.\ 1998a).  Usually $\nu_{orb}$
is the orbital frequency $\nu_\phi$, but other azimuthal motions,
e.g., periastron and nodal precession, could beat with the spin as
well.

In the magnetospheric beat-frequency model, the spin interacts with
clumps orbiting just outside the magnetosphere (\S\myref{s:radii};
Alpar \& Shaham 1985, see also Alpar \& Y\i lmaz 1997).  The model was
originally applied to CVs (Patterson 1979) and to the HBO
(\S\myref{s:nslfc}); similar models have been proposed for kHz QPOs
(\S\myref{s:nskhz}) by Cui (2000) and Campana (2000).  In the
sonic-point beat-frequency model for kHz QPOs (Miller et al.\ 1998a)
$\nu_u$ and $\nu_\ell$ are identified with respectively $\nu_\phi$ and
$\nu_{beat}$ at the sonic radius $r_{sonic}$ (\S\myref{s:radii}).  The
beat occurs as an X-ray pulsar beam irradiates clumps orbiting at
$r_{sonic}$ once per beat period, modulating the rate at which the
clumps' material accretes.  In its pure form the model predicts
$\Delta\nu$ to be constant at $\nu_{spin}$, contrary to observations
(\S\myref{s:deltanu}).  However, if the clumps spiral down while their
material accretes, $\Delta\nu$ can be less than this (Lamb \& Miller
2001).  The observation of twin kHz QPOs separated by half the
neutron-star spin frequency is inconsistent with this model (see
\S\myref{s:deltanu}).  Lamb \& Miller (2003) have proposed that a
beat-frequency interaction between orbital motion at $r_{sonic}$ and
an azimuthal structure at the spin-resonance radius (\S\myref{s:rrm})
explains $\nu_\ell$ (and maintain that $\nu_u=\nu_\phi(r_{sonic})$).
So, in this scenario the beat is between two different radii in the
disk.  With the spin frequencies of an increasing number of low
magnetic-field neutron stars in LMXBs becoming known
(\S\myref{s:deltanu}), further testing of the predictions of
beat-frequency models for the QPO frequencies in these systems can be
expected to put further rigorous constraints on the theoretical
possibilities.

\bigskip
This completes the summary of the 'orbital motion' QPO models.  How
can observations discriminate between these various models?  As it
turns out, every physical frequency model predicts its own
'fingerprint' set of parasitic, weaker frequencies in addition to the
strong ones it set out to explain in the first place (e.g., Miller et
al.\ 1998a).  These predicted patterns of weaker frequencies provide a
strong test of each model.  Searches for such weaker power-spectral
features are very difficult because this is really work at (or beyond)
the limit of the sensitivity of current instrumentation, and the
distinct impression of the observers is that there is still much
hiding below the formal detection levels.  However, a small number of
weak sidebands have been detected in neutron stars (Jonker et al.\
2000b, Wijnands et al.\ 2003; \S\myref{s:nskhz}) and the 2:3 ratio HF
QPOs in black holes (Strohmayer 2001a, Miller et al.\ 2001, Remillard
et al.\ 2003; \S\myref{s:bhhf}) provide similar constraints on possible
models.  These detections have led to only a very limited number of
proposed precise theoretical explanations in each case (e.g., Psaltis
2000, Klu\'zniak and Abramowicz 2001), an unusual situation that
testifies to the discriminating power of such additional frequencies.
However, too few of these frequencies have been detected yet for the
full power of this method to be applied; more sensitivity to weak
timing features is required for this.  To some extent these
fingerprint patterns depend on the way in which the frequencies
physically modulate the X-ray flux; models for this are discussed
next.
\subsection{Preferred radii}\mylabel{s:radii}
In models involving frequencies depending strongly on radius (e.g., of
orbital and epicyclic motion), {\it preferred radii} in the disk are
required to produce specific frequencies.  Constant and variable radii
produce corresponding frequencies.  Variable radii usually depend on
\mdot\ through disk physics (e.g., $r_{mag}$, below), whereas constant 
radii follow from just the compact object's parameters, perhaps
through strong-field gravity effects (e.g., the ISCO radius $r_{ms}$,
\S\myref{s:grorbits} or resonant radii, \S\myref{s:rrm}).

The 'inner edge of the disk' at radius $r_{in}$, where the
near-Keplerian flow ends and the radial velocity becomes appreciable
compared to the azimuthal one, provides a natural preferred radius.
The density contrast at $r_{in}$ can be sharp, as the radial velocity
of the flow can change suddenly (e.g., Miller et al.\ 1998a); this is
a strong-field gravity effect related to the small difference in
orbital angular momentum between different radii (see Paczy\'nski
1987).  In the absence of magnetic stresses, $r_{ms}$ (for standard
thin disks) or $r_{mb}$ (for low radiative efficiency disks such as
ADAFs, slim disks and super-Eddington flows; \S\myref{s:grorbits}) set
a lower limit to the flow, but for magnetized flows predictions are
uncertain; e.g., Abramowicz et al.\ 1978, Liang \& Thompson 1980, Lai
1998, Krolik \& Hawley 2002, Watarai \& Mineshige 2003a).  In neutron
stars the surface provides a fixed radius $R$ limiting $r_{in}$;
depending on the EOS may be smaller or larger than $r_{ms}$, but note
that some modulation models (\S\myref{s:modulation}) do not work if
the disk actually extends down to the star.

Radiation drag by photons emitted from within $r_{in}$ removing
angular momentum from the disk flow through the Poynting-Robertson
effect can truncate the flow at radius $r_{rad}$ (Miller et
al.\ 1998a).  As the radiation providing the drag is produced by the
same accretion flow that it interacts with on the way out, to first
order this radius is fixed, but details of flow and scattering
geometries can make it variable.  According to Miller et al.\ (1998a)
$r_{rad}$ can not be larger than \about15$GM/c^2$, and shrinks when
\mdot\ increases.  For a sufficiently magnetic neutron star,
electromagnetic stresses truncate the disk well outside the ISCO and
$r_{rad}$, at the magnetospheric radius (\S1\editnote{check;
ghosh lamb}) $r_{mag}$ which shrinks when $\dot M$ increases; note that
according to some authors orbital motion of part of the accreting
matter down to $r_{rad}$ still occurs within $r_{mag}$ (e.g., Miller
et al.\ 1998a).  Both these disk truncation mechanisms may only work
for neutron stars: black holes are not expected to have a sufficiently
strong magnetic field (\S1\editnote{check, see commented text on this
above}), and may also have difficulty to produce a sufficiently
strong, and low specific angular-momentum, radiation field from within
$r_{in}$.

Some authors treat the radius of maximum flux from the disk (e.g.,
Gruzinov 1999) or the radius of maximum pressure in a toroidal flow
(Klu\'zniak et al.\ 2004) as preferred; in such cases one needs to
investigate if the function whose extremum is used to define the
radius is sufficiently narrow to produce the observed QPO.  Laurent \&
Titarchuk (2001) proposed an interesting model where just the presence
of a Comptonizing converging flow makes one particular disk orbital
frequency stand out.
\subsection{Decoherence}\mylabel{s:decoherence}
The lack of coherence (aperiodicity) of particular observed
frequencies attributed to orbital and epicyclic phenomena is usually
taken to more or less naturally follow from the sheared and turbulent
nature of the flow, where any feature such as a density enhancement
('clump') or a vortex will have a finite lifetime (see
\S\myref{s:shots}).  Although suggested by the Lorentzians used to
describe QPO peaks (\S\myref{s:timing}), and sometimes implicitly
assumed in calculations (e.g., Titarchuk 2002), there is not much
evidence that observed signals are produced by a (superposition of)
exponentially damped harmonic oscillators.  Rapid frequency
fluctuations, for example related to a varying preferred radius
(e.g., a clump spiralling in, Stoeger 1980), or superposition of
multiple frequencies, e.g., generated from a finite-width disk annulus
are alternatives (e.g., Abramowicz et al.\ 1992, Nowak 1994, Bao \&
\O stgaard 1994).  As noted by Belloni et al.\ (2002a), the distinction
(\S\myref{s:timing}) between a QPO (with two frequencies, $\nu_0$ and
$\lambda$) and a BLN (with just one break frequency) becomes much less
fundamental when interpreted in terms of the width of a disk annulus
than when described in terms of damped harmonic oscillators.  This may
be why the use of $\nu_{max}$ (\S\myref{s:timing}) to summarize both
cases seems to work empirically.
\subsection{Modulation}\mylabel{s:modulation}
Orbital and epicyclic motions must modulate the X-ray flux for them to
be observable.  A classic mechanism (Cunningham \& Bardeen 1972,
Sunyaev 1973; see the end of \S\myref{s:rpm} for further references)
is that of a self-luminous blob orbiting a black hole in
general-relativistic spacetime.  Doppler boosting and gravitational
light bending produce the modulation.  Related models that are
considered are centrally-illuminated scattering blobs (e.g., Lamb \&
Miller 2003) and self-luminous turbulent disk flows (e.g., Armitage \&
Reynolds 2003).  Variable obscuration of central emitting regions by
matter orbiting further out is often invoked, but is attractive only
if the predicted inclination dependencies occur.  This is for example
the case for the \about1\,Hz dipper QPOs (\S\myref{s:nsotherrxrv}),
where a similar modulation of all emission (persistent and bursts,
different photon energies) further strengthens the case for this
mechanism (Jonker et al.\ 1999).
\begin{figure*}[htbp]
\begin{center}
\begin{tabular}{c}
\psfig{figure=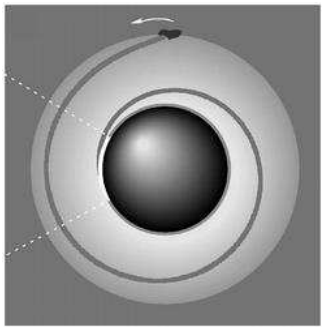,height=1.7in} 
\end{tabular}
\hspace{1cm}
\begin{minipage}{5cm}
\caption{%
Example of a modulation mechanism (see \S\myref{s:modulation}). The
clump with its spiral flow and the emission from the flow's footpoint
(dashed lines) in the Miller et al.\ (1998a) model all rotate at the
same angular velocity (that of the clump's orbit), irrespective of the
stellar spin, and produce a beaming modulation at that frequency. From
Miller et al.\ (1998a). \editnote{get rid of b}
\mylabel{f:modul}}
\end{minipage}
\end{center}
\end{figure*}

A fundamentally different mechanism is modulation of the accretion
rate into the inner regions, either taking the form of an actual
modulation of total \mdot\ (e.g., Lamb et al.\ 1985) or affecting only
the pattern by which the matter accretes (e.g., Miller et al.\ 1998a).
Likewise, either the total emitted flux can be modulated ({\it
luminosity modulation}), or only its angular distribution ({\it
beaming modulation}).  'Beaming' here is any anisotropy in the
emission.  Scattering material around the source suppresses a beamed
modulation much more effectively than a luminosity modulation (Wang \&
Schlickeiser 1987, Brainerd \& Lamb 1987, Kylafis \& Klimis 1987,
Bussard et al.\ 1988, Asaoka \& Hoshi 1989, Miller 2000).  In
magnetospheric beat-frequency models the accretion of matter from
orbiting clumps (and hence the luminosity) is modulated at
$\nu_{beat}$ as accretion is easier near the magnetic poles (Lamb et
al.\ 1985), leading to oscillating shots (\S\myref{s:timing}) in which
the luminosity oscillates at the beat frequency and whose lifetime is
that of a clump.  In the sonic point model of Miller et al.\ (1998a),
material from clumps orbiting at $r_{in}$ gradually accretes following
a fixed spiral-shaped trajectory in the frame co-rotating with the
clumps (Fig.\,\myref{f:modul}) at whose ``footpoint'' accretion, and
hence emission, is enhanced.  The resulting hot spot moves around the
surface at $\nu_\phi(r_{in})$ irrespective of the star's spin, and
produces a beaming modulation.  Both of these models predict surface
hot spots (revolving at $\nu_{spin}$ or $\nu_\phi(r_{in})$,
respectively) whose emission is itself modulated at $\nu_{beat}$,
leading to the potential for sideband formation.

Fully specified orbital and epicyclic flux-modulation models make
specific predictions for fractional amplitude, harmonic content and
sideband pattern, and energy dependencies (see also \S\myref{s:Edep}) of the
variability, so there is no lack of stringent tests to distinguish
between such models, and hence, the frequency models to which they are
appropriate.  As the modulation models shape up from the currently
still common simple qualitative ideas to quantitatively described
mechanisms, it can be expected that many simple models will turn out
to need revision.
%
%
\section{Low magnetic-field neutron stars}\mylabel{s:ns}
The main variability components and the correlations between their
frequencies have already been summarized in \S\S\myref{s:components}
and \myref{s:freqcorr}.  Here (\S\S\myref{s:ns}--\myref{s:bh}) a more
detailed overview of the observed timing properties of neutron stars
and black holes, respectively, is provided.  Tables\,\myref{t:spin}
and \myref{t:ns} summarize the literature on neutron stars and
reference codes in the current section are linked to those tables;
likewise the codes in \S\myref{s:bh} refer to Table\,\myref{t:bh}.
\subsection{Kilohertz quasi-periodic oscillations}\mylabel{s:nskhz}
\begin{figure*}[htbp]
$$\psfig{figure=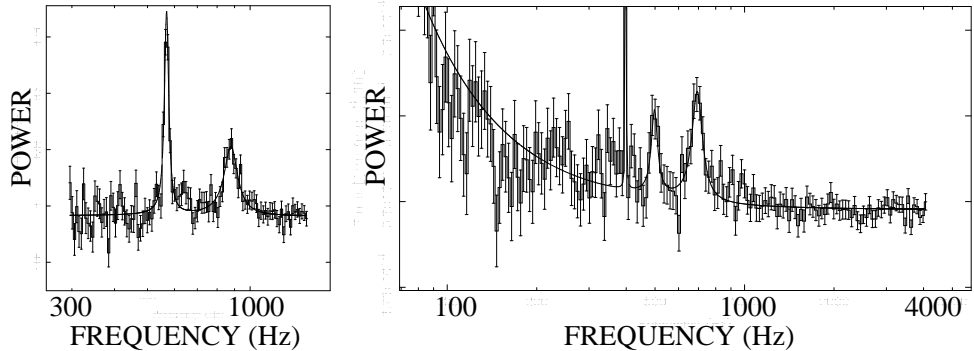,width=\textwidth}$$
\caption{Twin kHz QPOs in {\it left:} 4U\,1608$-$52 (M\'endez et al.\ 1998b) 
and {\it right:} the 401\,Hz accreting pulsar SAX\,J1808.4$-$3658 
(Wijnands et al.\ 2003; the 401~Hz pulsar spike is seen as well).
\mylabel{f:khz}}
\end{figure*}
kHz QPOs (\S\myref{s:hf}) are seen in a wide range of low
magnetic-field neutron-star sources, including all Z sources, most
atoll sources, several transients and two millisecond pulsars
(14c,13b\hide{Wijnands et al.\ 2003, Markwardt 2004 priv. comm.}), but
not the GX atoll sources (\S\myref{s:lmfnstypes}), which do not
usually reach the relevant states (\S\myref{s:statesa}).
Table\,\myref{t:spin} lists all sources with kHz QPOs, as well as
those with burst oscillations (\S3) or millisecond pulsations
(\S\myref{s:deltanu}).

kHz QPO frequencies increase with source state ($S_a$, $S_z$) in all
cases.  Table\,\myref{t:khzfreq} summarizes the typical frequencies
seen in well-covered sources; see \S\myref{s:freqcorr} for the
corresponding states.  Below \about500\,Hz \Lu\ turns into a BLN
component (\S\myref{s:lfc}) which is seen (also in weak LMXBs) down to
frequencies as low as 120\,Hz (\hide{van Straaten et al.\ 2004}14e).
A 10--20\,Hz BLN (\Llow, \S\myref{s:lfc}) seen in the EIS is sometimes
interpreted as a low-frequency version of \Ll\ (4z, but see 14e).

kHz QPO amplitudes increase with photon energy by typically a factor 4
between 3 and $>$10\,keV (8a,b,g,j,9b,d,11a,19a,g,21a,27a\hide{Berger
et al.\ 1996, Zhang et al.\ 1996a,b, Wijnands et al.\ 1997a,b,
1998c, M\'endez et al.\ 1997, 2001, Yu et al.\ 1997, Ford et al.\
1997b, 1998a, Homan \& van der Klis 2000, Gilfanov et al.\ 2003}).
In similar bands, the QPOs are weaker in the more luminous sources,
with 2--60\,keV rms amplitudes ranging from nearly 20\% in the weakest
atoll sources to typically 2--5\% in the Z sources (Jonker et
al.\ 2001).  They also weaken towards the extremes of their observable
range (Fig.\,\myref{f:mendezamplitudes}), where often only one peak is
detected.  At high energy, amplitudes are much higher (e.g., 40\% rms
$>$16\,keV in 4U\,0614+09; 9d).  This has been interpreted in
different ways, e.g., as an effect of Comptonization in a central
corona (Miller et al.\ 1998a) and as due to the combination of
variable hard flux from a boundary layer with a constant, softer, disk
flux (Gilfanov et al.\ 2003).
\begin{table}[htbp]
\begin{center}
\caption{kHz QPO frequencies (Hz)\mylabel{t:khzfreq}} 
\begin{tabular}{cccc}
\hline
\hline
\ms2{\hfil Atoll \& weak LMXBs\hfil}          & \ms2{\hfil Z\hfil }              \\
\hline
$\nu_\ell$ & $\nu_u$ & $\nu_\ell$ & $\nu_u$ \\
\hline
        & (120)      &         &            \\
        & (500)      & (200)   & (500)      \\
300     & 600        & (300)   & 600        \\      
500     & 800        & 500     & 800        \\
800     & 1000       & 800     & 1000       \\
1000    & 1200       &         &            \\
\hline
\hline
\end{tabular}
\hspace{1mm}
\begin{minipage}{5cm}\scriptsize
Typical observed kHz QPO frequency ranges.  Exceptions include
narrower ranges in GX\,5--1 and GX\,340+0 (maximum frequencies are
150--200\,Hz less; 7b,16b\hide{Jonker et al.\  2000a, 2002a}).
Parenthesized values refer to ranges in which usually $Q<2$.
\end{minipage}
\end{center}
\end{table}

Peak widths are affected by variations in centroid frequency during the
integration.  Typical values are several 10--100\,Hz, somewhat broader
in Z sources than in atoll sources and usually higher-Q at higher
frequency. \Ll\ in atoll sources can be very sharp (Q\about100), then
attains excellent signal-to-noise and exhibits soft lags of
10--60\,$\mu$s (9c,19c).  These lags are opposite to those expected from
inverse Compton scattering, so they may originate in the QPO production
mechanism rather than in propagation delays. \hide{was er niet een met
omgekeerde lag? wsl markwardt abstract, never published}
\begin{figure}[htbp]
\begin{center}
\begin{tabular}{c}
\psfig{figure=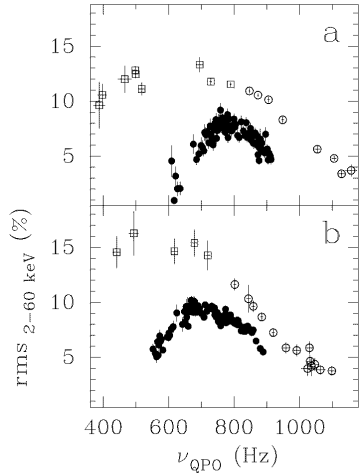,height=2.7in}
\end{tabular}
\hspace{1cm}
\begin{minipage}{5cm}
\caption{Lower ({\it filled symbols}) and upper ({\it open
symbols}) kHz QPO amplitudes as a function of their frequency for {\it
(a)} 4U\,1728--34 and {\it (b)} 4U\,1608--52 (M\'endez et al.\ 2001).
\mylabel{f:mendezamplitudes}}
\end{minipage}
\end{center}
\end{figure}

\def\mt#1{\ms{2}{\hfil#1\hfil}}
\def\dnns{${\Delta\nu\over\nu_{spin}}$}
\def\dn{$\Delta\nu$}
\def\ns{$\overline{\nu_{spin}}$}
\begin{table*}[tbp]
\caption{Kilohertz QPOs and neutron-star spin\mylabel{t:spin}}\scriptsize
\begin{center}
\begin{tabular}{lcccccll}
\hline                                             
\hline
Source              & kHz & $\Delta\nu$ & $\nu_{burst}$ & $\nu_{pulse}$ & \dn       & Rem.   & References  \\ 
                    & QPOs& (Hz)        & (Hz)          & (Hz)          & \ns       &        &       \\
\hline                                 
\hline                    
\ms{8}{\bf Millisecond pulsars\hfill}  \\            
\hline                                 
XTE\,J0929--314     & ---  & ---       & ---           & 185           & ---        &  T     & 10a;14e \\
XTE\,J1751--305     & ---  & ---       & ---           & 435           & ---        &  T     & 12a;14e \\
XTE\,J1807--294     & 2    & \about190 & ---           & 191           &\about0.99  &  T     & 13a,b  \\
SAX\,J1808.4--3658  & 3$^a$& 195       & 401           & 401           & 0.49       &  T     & 14a-e  \\
XTE\,J1814--338     & ---  & ---       & 314           & 314           & ---        &  T     & 15a,b;14e \\
\hline                                 
\ms{8}{\bf Z sources\hfill} \\                     
\hline                                 
Sco\,X-1(1617--155) & 2   & 245--310   & ---           & ---           & ---       &         & 4a-f,z,9g,h    \\
GX\,340+0 (1642--455)& 2   & 280--410   & ---           & ---           & ---       &         & 16a,b;4f,z;8j;9g,h  \\
GX\,349+2 (1702--363)& 2   & 265        & ---           & ---           & ---       &         & 17a-c       \\
GX\,5--1 (1758--250)& 2   & 240--350   & ---           & ---           & ---       &         & 7a,b;4f,z;9g,h     \\
GX\,17+2 (1813--140)& 2   & 240--300   & ---           & ---           & ---       &         & 11a-d;4f;9g,h      \\
Cyg\,X-2 (2142+380) & 2   & 345        & ---           & ---           & ---       &         & 3a,b;4f,z;9g,h      \\
\hline                                 
\ms{8}{\bf Atoll sources\hfill} \\                 
\hline                                 
4U\,0614+09         & 2   & 240--360   & ---           & ---           & ---        &        & 9a-h;4f;22h      \\  
2S\,0918--549       & 1   & ---        & ---           & ---           & ---        &        & 18a          \\
4U\,1608--52        & 3$^b$& 225--325  & 619           & ---           & 0.36--0.53 & T      & 8a-j,t;4f;9c,e,g,h \\
4U\,1636--53        & 3$^b$& 240--330  & 582           & ---           & 0.41--0.57 &        & 19a-g;4f;8f,t;9c,g,h  \\  
4U\,1702--43        & 2   & 320--340   & 330           & ---           & 0.97--1.03 &        & 20a;9h \\   
4U\,1705--44        & 1   & ---        & ---           & ---           & ---        & T      & 21a,b;9h         \\
4U\,1728--34        & 3$^b$& 275--350  & 363           & ---           & 0.76--0.96 &        & 22a-h;4f;8f,g,t;9g,h\\ 
KS\,1731--260       & 2   & 260        & 524           & ---           & 0.50       & T      & 23a;4f;9g,h \\  
4U\,1735--44        & 2   & 295--340   & ---           & ---           & ---        &        & 24a,b;4f;9h\\
SAX\,J1750.8--2900  &2$^c$& 317$^c$    & 601           & ---           & ---        & T      & 32a \\ 
4U\,1820--30        & 2   & 220--350   & ---           & ---           & ---        & G      & 25a-e;4f;9g,h\\  
Aql\,X-1 (1908+005) & 1   & ---        & 549           & ---           & ---        & T      & 26a-e;8g,t;9g,h\\     
4U\,1915--05        & 2   & 290,350$^g$& 272           & ---           & 1.07,1.29$^g$& D    & 33a,d  \\    
XTE\,J2123--058     & 2   & 255--275   & ---           & ---           & ---        & T      & 34a,b \\      
\hline                                 
\ms{8}{\bf Other sources$^d$\hfill} \\                         
\hline                    
EXO\,0748--676      & 1   & ---        & ---           & ---           & ---        & T,D    & 27a         \\
MXB\,1659--298      & --- & ---        & 567           &               & ---        & T,D    & 28a \\
XTE\,J1723--376     & 1$^e$& ---       & ---           & ---           & ---        & T      & 29a          \\
MXB\,1743--29       & --- & ---        & 589           & ---           & ---        & $^f$   & 30a;9g \\             
SAX\,J1748.9--2021  & --- & ---        & 410           & ---           & ---        & T,G    & 31a  \\
\hline                                             
\hline                                         
\end{tabular}
\end{center}
\begin{minipage}{\hsize}\scriptsize
References see Table\,\myref{t:ns}.  kHz QPOs: number of kHz
QPO peaks and sidebands with $Q\geq2$; $\nu_{burst}$: see also
\S3.4\editnote{check}.  Remarks: T: transient; D: dipper; G: in
globular cluster.  Notes: $^a$\,Sideband to pulsation, $^b$\,Sideband
to lower kHz QPO, $^c$\,Second QPO tentative, $^d$\,Faint, transient,
unidentified, etc., and not unambiguously classified, $^e$\,QPO not
confirmed, $^f$\,Source identification uncertain, $^g$ Two
incompatible values of $\Delta\nu$ reported, 33a.
\end{minipage}
\end{table*}
Weak sidebands 50--64\,Hz above $\nu_\ell$ with powers \about0.1 that of
\Ll\ have been detected in three objects (Jonker et al.\ 2000b), as well
as an additional high-frequency QPO peak 8--12\,Hz above the 401\,Hz
pulse frequency in SAX\,J1808.4--3658 (Wijnands et al.\ 2003).  In
both cases the separation between main peak and sideband increased
with QPO frequency, but was different from simultaneous LF QPO
frequencies, complicating interpretation.  These sidebands are the
first examples of the ``fingerprints'' of weaker parasistic
frequencies expected to accompany the kHz QPOs in all models
(\S\myref{s:generic}).  Lense-Thirring precession, orbital motion
within the inner disk edge (Jonker et al.\ 2000b) and relativistic
disk-oscillation modes (Psaltis 2000) have been considered as
explanations.

\subsubsection{Relation with luminosity}
\mylabel{s:parlines}
On time scales of hours and in a given source kHz QPO frequency $\nu$
typically correlates well with \Lx, but, remarkably, on longer time
scales, and across sources, this $\nu$-\Lx\ correlation is lost, with
the result that similar QPO frequencies are observed over two orders
of magnitude in \Lx\ (e.g., van der Klis 1997, 8b,9a\hide{Yu et al.\
1997, Ford et al.\ 1997a,}; \S\myref{s:states}).  Due to this, parallel
tracks form in $\nu$--\Lx\ diagrams covering either (i) several
sources (Fig.\,\myref{f:parlines}{\it a}), or (ii) several
observations of a single source separated by $\approxgt$1 day
(Fig.\,\myref{f:parlines}{\it b}).  In a given source, QPO amplitude
is affected much less by the \Lx\ shifts between tracks than predicted
if the shifts were caused by an extra source of X-rays unrelated to
the QPOs (8g\hide{M\'endez et al.\ 2001}).  This is also true across
sources.

That different sources have different $\nu$--\Lx\ tracks might be
because $\nu$ depends not only on \Lx, but also (inversely) on a
parameter related to average \Lx\ (e.g., van der Klis 1997, 1998,
Zhang et al.\ 1997b), such as magnetic-field strength $B$ (White \&
Zhang 1997) which previously, on other grounds, was hypothesized to
correlate to average \Lx\ (e.g., Hasinger \& van der Klis 1989).
Alternatively, the two parallel-track phenomena might be explained
{\it together} by noting that kHz QPO frequency seems to be governed
not by \Lx, but by how much \Lx\ deviates from its time average
$\langle$\Lx$\rangle$ over a day or so (van der Klis 1999).  This
could arise if there is an \Lx\ component (e.g., due to nuclear
burning or the accretion of a radial inflow) that is proportional to
some {\it time average} over the inner-disk accretion rate $\dot M_d$,
whose effect is to make the frequency lower (e.g., via the Miller et
al.\ 1998a radiative disk-truncation mechanism;
\S\myref{s:radii}).  Frequency would then scale with $\dot M_d/L_x$
where $L_x=\dot M_d+\alpha\langle\dot M_d\rangle$, with $\alpha$ the
relative efficiency of the additional energy release (cf.,
\S\myref{s:states}).  Simulations of this toy model (van der Klis
2001) reproduce the salient features of the observed parallel tracks
seen in Fig.\,\myref{f:parlines}; testing this requires sustained
monitoring of the $\nu$--\Lx\ correlation.
\begin{figure*}[htbp]
$$\psfig{figure=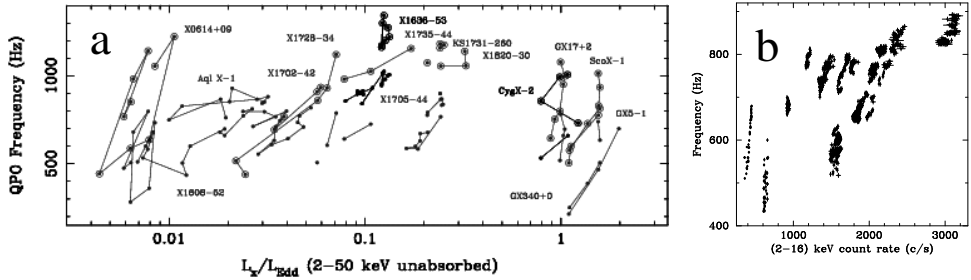,width=\linewidth}$$
\caption{The two parallel tracks phenomena. {\it (a)} Across sources 
(Ford et al.\ 2000; upper and lower kHz peaks are indicated with
different symbols). {\it (b)} In time, in the source
4U\,1608--52 (M\'endez et al.\ 1999), frequency plotted is
$\nu_\ell$. \mylabel{f:parlines}}
\end{figure*}

kHz QPOs are sensitive to very short-term \Lx\ variations as well.  In
Sco\,X-1 amplitudes and frequencies systematically vary on sub-second
time scales, in phase with the 6-Hz flux variations of the NBO
(\S\myref{s:nsotherrxrv}; 4d\hide{Yu et al.\ 2001}).  In 4U\,1608--52,
kHz QPO frequency $\nu$ has been observed to increase, as usual, with
\mdot-induced \Lx\ increases, but to {\it de}crease in response to
probable nuclear-burning ('mHz' QPO, \S\myref{s:nsotherrxrv}) induced
\Lx\ increases (8h\hide{Yu \& van der Klis 2002}), confirming a
prediction of the Miller et al.\ (1998a) model, where the QPO occurs
at $\nu_\phi$ of the inner disk radius $r_{in}$ (\S\myref{s:radii})
which increases in response to the extra, nuclear-burning generated
luminosity component.
\subsubsection{Relation with neutron-star spin} 
\mylabel{s:deltanu}
Peak separation $\Delta\nu$ usually decreases with increasing kHz QPO
frequency $\nu$ (Fig.\,\myref{f:deltanu}{\it a}; 4b,c,7b,8d,19f,22c).
In the two cases where a probable positive $\nu$--$\Delta\nu$
correlation ocurred, this was at the lowest detectable frequencies
(11b,22f\hide{Homan et al.\ 2002, Migliari et al.\ 2003b}).

$\Delta\nu$ is approximately commensurate with $\nu_{spin}$
(Table\,\myref{t:spin}), and this is the main motivation for
beat-frequency models (\S\myref{s:bfm}).  Of course, $\Delta\nu$
varies so any commensurability can not be exact.  Spin is measured
directly in millisecond pulsars; also, burst oscillations (as verified
in two msec pulsars; 14b,d,15b) likely occur very near $\nu_{spin}$
(\S3).  In the eight sources where both $\Delta\nu$ and spin were
measured by one of these two methods (Table\,\myref{t:spin}),
$\Delta\nu$ was between 0.7 and 1.3 times $\nu_{spin}$ for
$\nu_{spin}<400$\,Hz, and 0.36--0.57 times $\nu_{spin}$ for
$\nu_{spin}>400$\,Hz (Fig.\,\myref{f:deltanu}{\it b}); the largest offsets of the ratios from 1.0 and
0.5, respectively, are \about8$\sigma$ but usually the discrepancies
are much less.  In the two pulsars the ratios are 0.49 and 0.99.  So,
a commensurability with spin does indeed seem to exist where $\Delta\nu\approx\nu_{spin}$
for low, and $\Delta\nu\approx\nu_{spin}/2$ for high $\nu_{spin}$.

\begin{figure*}[htbp]
$$\psfig{figure=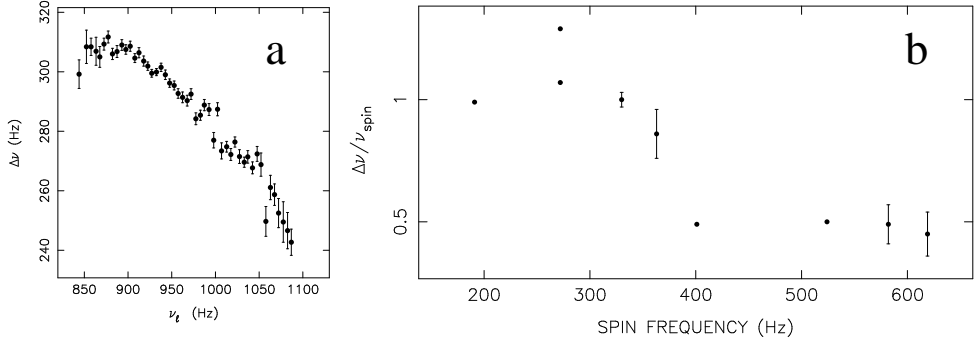,width=\linewidth}$$
\caption{{\it (a)} The variation in kHz QPO peak separation as a function of 
the lower kHz QPO frequency in Sco X-1, after M\'endez \&
van der Klis (1999); {\it (b)} $\Delta\nu/\nu_{spin}$ vs. $\nu_{spin}$, after
Table\,\myref{t:spin}.  Vertical bars indicate the range of variation in
$\Delta\nu$.  
\mylabel{f:deltanu}}
\end{figure*}
That the spin frequency can be twice $\Delta\nu$ was initially
suspected based on measurements of burst oscillations, where an
alternative explanation, namely that $\nu_{burst} = 2\nu_{spin}$ could
not be excluded (see \S3).  However, the case of the 401\,Hz pulsar
SAX\,J1808.4--3658 now leaves little doubt that $\Delta\nu$ {\it can}
indeed be half the spin frequency (Wijnands et al.\ 2003).  This
falsifies the direct spin-orbit beat-frequency interpretation
(\S\myref{s:bfm}) even when allowing for multiple neutron-star hot
spots or orbiting clumps, but may be explained by models involving
resonances, perhaps combined with a beat (\S\myref{s:rrm}; 14c).
Sixteen low magnetic-field neutron-star spins have now been measured
(thirteen burst oscillations and five msec pulsars, with an overlap of
two sources), and ten more spins are known up to a factor of two.
Table\,\myref{t:spin} lists these 26 objects.  The spin frequencies
are between \about200 and \about700\,Hz, suggesting a cut off well
below the limit set by observational constraints and indicating that a
braking mechanism limits $\nu_{spin}$ (see Chakrabarty et al.\ 2003).
If the stars spin at the magnetospheric equilibrium spin rates
(\S1\editnote{check}) corresponding to their current \Lx, this
predicts a tight correlation between \Lx\ and magnetic-field strength
$B$ (White \& Zhang 1997; a similar possibility came up to explain the
similar kHz QPO frequencies at very different \Lx,
\S\myref{s:parlines}).  Another possibility is that gravitational
radiation limits $\nu_{spin}$ by transporting angular momentum out as
fast as accretion is transporting it in; this predicts these sources
to be the brightest gravitational-wave sources, with a known
$\nu_{spin}$ facilitating their detection (Bildsten 1998, Andersson et
al.\ 1999, 2000, also Levin 1999, and e.g., Rezzolla et al.\ 2000,
Bildsten \& Ushomirsky 2000, Brown \& Ushomirsky 2000, Ushomirsky et
al.\ 2000, Yoshida \& Lee 2001, Wagoner 2002\hide{Kinney and Mendell
2002 PhRevD 67 4032, Reisenegger and Bonacic 2003 PhRevLett 91 1103}).

Some caution is advised in interpreting the SAX\,J1808.4--3658 kHz QPO
result, as the twin peaks were observed only once in this pulsar, its
$\Delta\nu$ (195\,Hz) is lower than in non-pulsars, and other
commensurabilities also exist between the observed frequencies (14c).
kHz QPOs in XTE\,J1807--294 have a $\Delta\nu$ close to its 191\,Hz
spin frequency (13b).  The frequency correlations of some pulsars are
a factor \about1.45 off the usual ones (\S\myref{s:freqcorr}).
Clearly, further detections of twin kHz QPOs in millisecond pulsars
would help to clarify the systematics in this phenomenology.  While
the kHz QPO frequency ratio $\nu_\ell/\nu_u$ is certainly not
constant, Abramowicz et al.\ (2003c) propose that ratios near 2:3
may occur more often than others, which might provide a link with
black-hole high-frequency QPOs.
\subsubsection{Interpretations}\mylabel{s:khzinterp}
kHz QPO models include beat-frequency, relativistic precession and
relativistic resonance models (\S\myref{s:generic}) as well as
disk-oscillations and other flow instabilities
(\S\myref{s:othermodels}).  All proposals involve plasma motion in the
strong-field region, and with one exception (photon bubbles,
\S\myref{s:nonflow}) in all models the QPOs originate in the disk.
Most identify $\nu_u$ with orbital motion at a preferred radius in the
disk (in models of Titarchuk and co-workers $\nu_\ell$ has this role),
Fig.\,\myref{f:wedge} illustrates the constraints the orbital-motion
hypothesis provides on neutron-star parameters.  The highest measured
$\nu_u$ of 1329\p4\,Hz in 4U\,0614+09, when interpreted as $\nu_\phi$
for $j=0$, implies $M_{NS}<1.65$\msun\ and $R_{NS}<12.4$\,km,
imperiling the hardest equations of state; for a 300\,Hz spin these
numbers become 1.9\msun\ and 15.2\,km (van Straaten et al.\ 2000).

The maximum kHz QPO frequencies in well-studied sources are
constrained to a relatively narrow range of $\nu_u=$\ 850--1330\,Hz.
If this is the ISCO frequency $\nu_{ms}$ (\S\myref{s:grorbits}), the
neutron-star masses are near 2\msun\ (Zhang et al.\ 1997b, see also
Kaaret et al.\ 1997).  An apparent leveling off of $\nu_u$ as a
function count rate, flux and $S_a$ at
\about1060\,Hz was found in 4U\,1820--30 (25b-d\hide{Zhang et al.\
1998b, Kaaret et al.\ 1999b, Bloser et al.\ 2000a}), but further
observations cast doubt on this (25e).  A tendency to level off (but
without a true 'ceiling') may be a general feature of the
parallel-tracks phenomenon (cf., Fig.\,\myref{f:parlines}); indeed,
the toy model described in \S\myref{s:parlines} predicts such a
pattern as a consequence of $\nu_u$ depending on $\dot M_d/\langle\dot
M_d\rangle$.

Calculations exploring to what extent kHz QPOs constrain the EOS,
usually assuming $\nu_u=\nu_\phi$, have further been performed by
e.g., Miller et al.\ (1998b), Datta et al.\ (1998, 2000), Akmal et
al.\ (1998), Klu\'zniak (1998), Bulik et al.\ (1999, 2000), Thampan et
al.\ (1999), Li et al.\ (1999), Schaab \& Weigel (1999), Kalogera \&
Psaltis (2000), Stergioulas et al.\ (1999), Heiselberg \&
Hjorth-Jensen (1999), Zdunik et al.\ (2000), Glendenning \& Weber
(2001), Gondek-Rosinska et al.\ (2001), Ouyed (2002) and Mukhopadhyay
et al.\ (2003).

The evidence for orbital motion as the cause of kHz QPOs is not yet
iron-clad.  Orbital motion could be empirically demonstrated by a
number of different possible measurements, such as independent
measurements of frequency and orbital radius, where radius could be
measured from continuum or line spectroscopy, demonstrating the
predicted frequency - radius relation, measurements of orbital and
epicyclic frequencies varying together in the way predicted by general
relativity, measurements of the Doppler effect in an orbiting hot spot
or measurements of the orbital frequency ceiling predicted at the
ISCO.  Indications for some of these possibilities have already been
obtained.
\subsection{Hectohertz QPOs}\mylabel{s:nshhzqpos}
Figs.\,\myref{f:hf} and \myref{f:psa} show examples of hHz QPOs, and
Fig.\,\myref{f:termscheme} illustrates their approximately constant
frequencies (\S\myref{s:hf}) in the 100--200\,Hz range.  hHz QPOs
occur in atoll sources (including the millisecond pulsar
SAX\,J1808.4--3658) when $\nu_b\approxgt1$\,Hz (8i,9f,14a,
e,21b,22b,e,f,g,h,26e\hide{Wijnands \& van der Klis 1998b, van
Straaten et al.\ 2000, di Salvo et al.\ 2001, van Straaten et al.\
2002, van Straaten et al.\ 2003, Olive et al.\ 2003, Migliari et al.\
2003a,b, van Straaten et al.\ 2004, Reig et al.\ 2004}, see also
Fragile et al.\ 2001), but have not been seen in most weak LMXBs
(which may not reach the relevant states) nor or black holes.  It may
also occur in Z sources (e.g., 11g).  At low $\nu_b$ (\about0.2\,Hz)
confusion occurs with \Lu.  The phenomenon is usually rather low-Q,
but occasionally peaks up to Q$>$2, has an amplitude between 2 and
20\% (rms) and becomes weaker and more coherent as $\nu_b$ rises.  The
near-constant frequency allows diskoseismic modes as a model
(\S\myref{s:diskoscillations}) and suggests a link with black-hole HF
QPOs (see \S\myref{s:hf}).  Fragile et al.\ (2001) suggest $\nu_{_{hHz}}$
could be the orbital frequency at the radius where a warped disk is
forced to the equatorial plane by the (strong-field gravity) Bardeen
\& Petterson (1975) effect.  Titarchuk (2003) proposed Rayleigh-Taylor
gravity waves at the disk-star boundary layer as the origin of hHz
QPOs (cf. Brugmans 1983).
\subsection{The low-frequency complex}\mylabel{s:nslfc}
Components of the low-frequency complex have been studied in Z and
atoll sources since the 1980's (e.g., 7c,3B; van der Klis 1995a for a
review).  The 15--60\,Hz LF QPO in the Z-source HB and upper NB (the
'horizontal branch oscillation' or HBO) and the 0.1--30\,Hz BLN
component (formerly called 'low-frequency noise' or LFN in Z sources
and confusingly, sometimes 'high-frequency noise' in atoll sources,
3B,8l) were first detected in GX\,5--1 (7c); HBO harmonics were
sometimes seen (e.g., 11g).  Frequencies were found to correlate with
spectral state (4i-l,n,7e,i,j,38a,b,3h,q,y,B, 11f-i\hide{Z Priedhorsky
et al.\ 1986 sco, van der Klis et al.\ 1987 5-1 state, van der Klis et
al.\ 1987 sco, Lewin et al.\ 1987 3+1, Hasinger \& van der Klis 1989,
Hasinger et al.\ 1989 sco, Makishima et al.\ 1989 3+1, Langmeier et
al.\ 1990 17+2, Hasinger et al.\ 1990 x-2, Penninx et al.\ 1990 17+2,
Hertz et al.\ 1992 sco, Lewin et al.\ 1992 5-1, Kuulkers et al.\ 1994
5-1, 1996 x-2, 1997 17+2, 1999 x-2, Wijnands et al.\ 1996 17+2, 1997
x-2, Dieters \& van der Klis\,2000 sco},19h\hide{atoll HK89, Prins \& van de
Klis 1997}).  The variations were spectrally hard (e.g.,
16d,7i,k\hide{ Penninx et al.\ 1991 340, Lewin et al.\ 1992, Vaughan
et al.\ 1994 5-1}), and the HBO showed hard lags of a few ms (e.g.,
7f,k\hide{van der Klis et al.\ 1987 5-1 lags, Vaughan et al.\ 1994
5-1}).  Similarities with black-hole LS and VHS variability were
pointed out (e.g., van der Klis 1994a,b, 8l,m\hide{Yoshida et
al.\ 1993, Berger \& van der Klis 1998}).  The BLN was sometimes
peaked (in atoll sources, e.g., 38a,b,25f,g\hide{ Lewin et al.\ 1987
3+1, Stella et al.\ 1987a 1820, Dotani et al.\ 1989 1820, Makishima et
al.\ 1989 3+1}, and in Z sources, e.g., 11b,17d,4k,n\hide{Stella et
al.\ 1987b, Ponman et al.\ 1988 349, Hasinger et al.\ 1989 sco,
Dieters \& van der Klis\,2000 sco}), but in atoll sources QPOs were only rarely
seen (25g,8l\hide{Stella et al.\ 1987a 1820, Yoshida et al.\ 1993
1608}).  The classic Z and atoll tracks in CDs and HIDs
(\S\myref{s:statesns}) were known (3B and e.g., 19h,7i,j\hide{HK89,
and e.g., Prins and van der Klis , Kuulkers et al.\ 1994 5-1, Lewin et
al.\ 1992 5-1}), but the fainter EIS was hardly explored (e.g.,
8k,21c,19i\hide{Mitsuda et al.\ 1989 1608, Langmeier et al.\ 1989
1705, van der Klis et al.\ 1990 1636}).

In recent work, with the exploration of the fainter sources and
states, similarities in the phenomenology across source types were
clarified (\S\myref{s:components}), with the frequency correlations
between, and among, kHz and LF phenomena (\S\myref{s:freqcorr})
particularly revealing
(e.g., 4a,b,e,24b,3a,20a,33a,16b,7b,11b,3b,s,22g,21b,26e\hide{van der
Klis et al.\ 1996 sco k, 1997 sco k, Ford \& van der Klis 1998,
Wijnands et al.\ 1998 x-2 k, Markwardt et al.\ 1999a 1702 k, Boirin et
al.\ 2000 1915 k, Jonker et al.\ 2000 340, 2002 5-1, Homan et
al.\ 2002 17+2, Kuznetsov et al.\ 2002 x-2, Kuznetsov et al.\ 2001
x-2, 2002 sco kh, di Salvo et al.\ 2001 1728, Olive et al.\ 2003 1705
k, Reig et al.\ 2004 Aql}), and the EIS--LS match confirmed
(\S\myref{s:states}).  This led to the synthesis summarized in
\S\S\myref{s:lfc} and \myref{s:freqcorr} 
(4f,y,z,9c,f,11c,22g,h,8i,14e\hide{wk, pbk, bpk, psal1998,1999
freqcor, vstr2000, 2002,2003,2004}).  Clear examples of LF QPOs were
found in atoll sources
(23a,39a,24a,20a,9f,33a,22g,h,21b,26e,11c,8i\hide{26 Hz Wijnands \&
van der Klis 1997 1731, 57-69 Hz Homan et al.\ 1998 13+1, 67 Hz
Wijnands et al.\ 1998 1735, Markwardt et al.\ 1999a 1702, 18-37 Hz van
Straaten et al.\ 2000 0614, Boirin et al.\ 2000 1915, di Salvo et
al.\ 2001 1728, Olive et al.\ 2003 1705, Reig et al.\ 2004 Aql,
Belloni et al.\ 2002a BPK, van Straaten et al.\ 2002 0614/1728, 2003
1608}) and millisecond pulsars (14e\hide{van Straaten et al.\ 2004})
and these were incorporated into the general picture, although the
phenomenology of these QPOs is not yet entirely clear, with apparently
both \Lb\ and \Lh\ often producing a narrow ($Q>2$) peak when they
move to high ($>30$\,Hz) frequency, and indications for two harmonically
related QPOs in the 0.2--2\,Hz range (and once near 80\,Hz) very
roughly a factor 2 below $\nu_h$ (8i,l,11c,14e,\hide{bpk, van Straaten
03, 04, yoshi}).

Models for LF QPO (particularly, the Z-source HBO) include the
magnetospheric beat-frequency model (\S\myref{s:bfm}),
Lense-Thirring precession (\S\myref{s:rpm}) and various disk
oscillations (\S\myref{s:diskoscillations}), e.g., the magnetically
warped precessing-disk model (Lai 1999, Shirakawa \& Lai 2002a).  The
frequency correlations among low-frequency complex phenomena and with
kHz QPOs, and in particular the factor \about1.5 shifts observed in
those relations in some millisecond pulsars (\S\myref{s:nsfreqcorr})
are among the main observational features to be accounted for in model
scenarios, but presently are too fresh to have affected the models
reported here.

\begin{table*}[tbp]
\caption{Neutron-star variability and states {\it (a.)}\mylabel{t:ns}}\scriptsize
\begin{center}
\begin{tabular}{lcccl}
\hline
\hline 
Source                      & HB/NB/FB     & LF    & HB/N-FB/kHz     &  References \\
                            &              & noise & \hfill QPOs\ \ \ \ \ & \\
\hline                                                              
\ms{3}{\bf Z sources\hfill} \\
\hline
Sco\,X-1 (1617--155)          & \c/\b/\b     & \b    & \b/\b/\b        & 4a-A;3A,B;9g,h \\
GX\,340+0 (1642--455)         & \b/\b/\b     & \b    & \b/\b/\b        & 16a-e;3A,B;4y,z;8j,r;9g,h     \\
GX\,349+2$^a$ (1702--363)     & \c/\b/\b     & \b    & \c/-/\b         & 17a-j;3A,B      \\
GX\,5--1 (1758--250)         & \b/\b/\b     & \b    & \b/\b/\b        & 7a-q;3A,B;4f,y,z;9g,h;8r \\
GX\,17+2 (1813--140)          & \b/\b/\b     & \b    & \b/\b/\b        & 11a-l;3A,B;4f,x,y;9g,h;8r         \\
Cyg\,X-2 (2142+380)           & \b/\b/\b     & \b    & \b/\b/\b        & 3a-B;4f,n,x-z;7d,f;8r;9g,h       \\
\hline 
\ms{3}{\bf Z source candidates$^a$\hfill} \\
\hline
RX\,J0042.6+4115            & \c/\c/\c     &  -    & -/-/-           & 35a  \\
LMC\,X-2 (0521--720)          & \c/\c/\c     &  -    & -/-/-           & 36a-c \\
Cir\,X-1 (1516--569)          & \c/\c/\c     & \b    & \b/-/-          & 37a-j;4f,x;11c\\
\hline
\hline
\end{tabular}
\end{center}
\begin{minipage}{\hsize}
HB: horizontal branch, NB: normal branch, FB: flaring branch. 
\b: variability of this type observed, \c: some doubt (uncertain,ambiguous, atypical, rare 
and not clearly seen, etc.), -: not reported. Note: $^a$\,Z/atoll character somewhat
ambiguous, or evidence incomplete.
\end{minipage}
\end{table*}
\addtocounter{table}{-1}
\begin{table*}[tbp]
\caption{Neutron-star variability and states {\it (b.)}}\scriptsize
\begin{center}
\begin{tabular}{lcccl}
\hline
\hline 
                            & EIS/IS/B     & Strong & LF/hHz/kHz     & References \\
                            &              & BLN &  \hfill QPOs\ \ \ \ \ &      \\
\hline                                                              
\ms{3}{\bf Millisecond pulsars\hfill}  \\            
\hline                                 
XTE\,J0929--314             & \b/-/-       & \b    & \b/-/-          & 10a;14e \\
XTE\,J1751--305             & \c/-/-       & \b    & -/-/-           & 12a;14e,j \\
XTE\,J1807--294             & -/-/\c       & -     & -/-/\b          & 13a,b;14j \\    
SAX\,J1808.4--3658          & \b/\b/\c     & \b    & \b/\c/\b        & 14a-k;4x,y;8q \\
XTE\,J1814--338             & \c/-/-       & \b    & \b/-/-          & 15a-c;14e,j \\
\hline                                 
\ms{3}{\bf Atoll sources\hfill} \\
\hline
4U\,0614+09                 & \b/\b/\b     & \b    & -/-/\b          & 9a-k;4f,y;8q;22h\\
2S\,0918--549               & \c/\b/\b     & \b    & -/-/\b          & 18a\\
4U\,1608--52                & \b/\b/\b     & \b    & \b/\b/\b        & 8a-t;4x,y;3B;9c,e,g,h \\
4U\,1636--53                & \c/\b/\b     & \b    & -/-/\b          & 19a-j;3B;8f,m,p,t;9c,g,h \\
4U\,1702--43                & -/-/\c       & \c    & \b/\c/\b        & 20a,b;8p;9h \\
4U\,1705--44                & \b/\b/\b     & \b    & -/\b /\b        & 21a-e;3B;4x,y;8m,q-s;9h,l \\ 
4U\,1728--34 (GX354--0)       & \b/\b/\b     & \b    & \b/\b/\b        & 22a-j;3B;4f,y;8f,g,p-r,t;9g,h\\ 
GX 9+9 (1758--169)            & -/-/\b       & -     & -/-/-           & 3A,B;38h\\ 
KS\,1731--260               & -/-/\b       & -     & \b/-/\b         & 23a,b;4f;8p,q;9g,h \\ 
4U\,1735--44                & -/\c/\b      & -     & \b/-/\b         & 24a-e;3B;4f,y;9h\\
GX\,3+1 (1744--265)           & -/-/\b       & -     & -/-/-           & 38a-h;3A,B \\
SAX\,J1750.8--2900          & -/\b/\b      & -     & \c/-/\b         & 32a \\
GX\,9+1 (1758--205)           & -/-/\b       & -     & -/-/-           & 3A,B;38h \\
GX\,13+1 (1811--171)$^a$      & -/\c/\b      & -     & \b/-/-          & 39a-d;3A,B;8r \\
4U\,1820--30 (NGC6624)        & -/\b/\b      & -     & -/-/\b          & 25a-i;3B;4f;8r;9g,h \\
Ser\,X-1 (1837+049)           & -/-/\b       & -     & \c/\c/-         & 38c;8r;25i \\
Aql\,X-1 (1908+00)            & \b/\b/\b     & \b    & \c/\c/\b        & 26a-f;4x;8g,p,r,s;9g,h \\
4U\,1915--05                & \c/\c/\c     & \c    & \b/-/b          & 33a-d \\
XTE\,J2123--058             & \c/\b/\b     & -     & -/-/\b          & 34a,b \\   
\hline                                                   
\ms{3}{\bf Weak LMXBs / atoll source candidates$^b$\hfill} \\              
\hline                                                   
EXO\,0748--676              &  -/\c/\c     & -     & -$^c$/-/\b      & 27a,b \\
4U\,1323--62                & -/-/-        & -     & -$^c$/-/-       & 41a \\
MXB\,1659--298              & \c/-/\c      & -     & -/-/-           & 28a,b \\ 
XTE\,J1709--267             & -/-/-        & \b    & -/-/-           & 42a \\
XTE\,J1723--376             & -/-/-        & -     & -/-/\c          & 29a \\
1E\,1724--3045 (Ter\,2)       & \b/-/-       & \b    & \b/-/-          & 43a;4f,y;8q;11c;14e;23b \\ 
MXB\,1730--335 (Lil 1; RB)   & -/-/-        & \b    & \b/-/-          & 44a-m \\
SLX\,1735--269              & \b/-/-       & \b    & -/-/-           & 45a;8q;11c;14e;23b \\
4U\,1746--37 (NGC6441)        & -/\c/\b      & -     & -$^c$/-/-       & 40a \\
GRS\,1747--312 (Ter\,6)       & -/\c/-       & -     & -/-/-           & 46a \\ 
SAX\,J1748.9--2021 (NGC6440)  &  -/-/\c      & -     & -/-/-           & 31a,b \\
XTE\,J1806--246             & \c/\c/\c     & \b    & \c/-/-          & 47a,b \\
4U\,1812--12                & -/-/-        & \b    & -/-/-           & 48a \\
GS\,1826--238               & \b/-/-       & \b    & \b/-/-          & 49a;4f;8q,r;11c;14e;23b \\
4U\,1850--08 (NGC6712)        & -/\c/-       & -     & -/-/-           & 50a \\
\hline
\hline
\end{tabular}
\end{center}
\begin{minipage}{\hsize}
EIS: extreme island state, IS: island state, B: banana state. 
\b: variability of this type observed, \c: some doubt (uncertain,ambiguous, 
atypical, rare and not clearly seen, etc.), -: not reported. Strong
BLN: $>15$\% rms. RB: rapid burster. Notes: $^a$\,Z/atoll character
somewhat ambiguous, $^b$\,Evidence incomplete, $^c$\,\about1\,Hz
'dipper' QPO, \S\myref{s:nsotherrxrv}.
\end{minipage}
\end{table*}
\begin{table*}[htb]
\scriptsize
\begin{tabular}{c}
\hline
\hline
\begin{minipage}{\hsize}
%
%
%
References in Table\,\myref{t:ns}a and b: 3a Wijnands et al.\ 1998a,
3b Kuznetsov 2002a, 3c Branduardi et al.\ 1980, 3d Hasinger et
al.\ 1986, 3e Norris \& Wood 1987, 3f Hasinger 1987, 3g Mitsuda \&
Dotani\,1989, 3h Hasinger et al.\ 1990, 3i Vrtilek et al.\ 1990, 3j
Chiapetti et al.\ 1990, 3k Hjellming et al.\ 1990a, 3l Hirano et
al.\ 1995, 3m Kuulkers et al.\ 1995, 3n Focke 1996, 3o Wijnands et
al.\ 1997c, 3p Smale 1998, 3q Kuulkers et al.\ 1999, 3r Wijnands \&
van der Klis 2001, 3s Kuznetsov 2001, 3t Piraino et al.\ 2002, 3u
O'Brien et al.\ 2004, 3v di Salvo et al.\ 2002, 3w Done et al.\ 2002,
3x Vrtilek et al.\ 2003, 3y Kuulkers et al.\ 1996, 3z Kuulkers \& van
der Klis 1995, 3A Schulz et al.\ 1989, 3B Hasinger \& van der Klis
1989, 3C Dubus et al.\ 2003; 4a van der Klis et al.\ 1996, 4b van der
Klis et al.\ 1997, 4c M\'endez \& van der Klis 2000, 4d Yu et
al.\ 2001, 4e Kuznetsov 2002b, 4f Psaltis et al.\ 1999a, 4g Ilovaisky
et al.\ 1980, 4h Middleditch \& Priedhorsky 1986, 4i Priedhorsky et
al.\ 1986, 4j van der Klis et al.\ 1987b, 4k Hasinger et al.\ 1989, 4l
Hertz et al.\ 1992, 4m Kallman et al.\ 1998, 4n Dieters \& van der
Klis 2000, 4o Dieters et al.\ 2000a, 4p White et al.\ 1985, 4q
Hjellming et al.\ 1990b, 4r Vrtilek et al.\ 1991, 4s Augusteijn et
al.\ 1992, 4t Barnard et al.\ 2003a, 4u McNamara et al.\ 2003, 4v
Santolamazza et al.\ 2003, 4w Bradshaw et al.\ 2003, 4x Done \&
Gierli\'nski 2003, 4y Wijnands \& van der Klis 1999a, 4z Psaltis et
al.\ 1999b, 4A Canizares et al.\ 1975, 4B Robinson \& Warner 1972; 7a
Wijnands et al.\ 1998c, 7b Jonker et al.\ 2002a, 7c van der Klis et
al.\ 1985, 7d Elsner et al.\ 1986, 7e van der Klis et al.\ 1987a, 7f
van der Klis et al.\ 1987c, 7g Norris et al.\ 1990, 7h Mitsuda et
al.\ 1991, 7i Lewin et al.\ 1992, 7j Kuulkers et al.\ 1994, 7k Vaughan
et al.\ 1994, 7l Kamado et al.\ 1997, 7m Vaughan et al.\ 1999, 7n van
der Klis et al.\ 1991, 7o Tan et al.\ 1992, 7p Blom et al.\ 1993, 7q
Asai et al.\ 1994; 8a Berger et al.\ 1996, 8b Yu et al.\ 1997, 8c
M\'endez et al.\ 1998a, 8d M\'endez et al.\ 1998b, 8e M\'endez et al.\ 1999, 8f
Jonker et al.\ 2000b, 8g M\'endez et al.\ 2001, 8h Yu \& van der Klis
2002, 8i van Straaten et al.\ 2003, 8j Gilfanov et al.\ 2003, 8k
Mitsuda et al.\ 1989, 8l Yoshida et al.\ 1993, 8m Berger \& van der
Klis 1998, 8n Revnivtsev et al.\ 2001a, 8o Gierli\'nski \& Done 2002a,
8p Muno et al.\ 2004, 8q Sunyaev \& Revnivtsev 2000, 8r Muno et
al.\ 2002, 8s Gierli\'nski \& Done 2002b, 8t M\'endez 1999; 9a Ford et
al.\ 1997a, 9b Ford et al.\ 1997b, 9c Vaughan et al.\ 1997, 1998, 9d
M\'endez et al.\ 1997, 9e Kaaret et al.\ 1998, 9f van Straaten et
al.\ 2000, 9g Psaltis et al.\ 1998, 9h Ford et al.\ 2000, 9i Ford et
al.\ 1999, 9j M\'endez et al.\ 2002, 9k Singh \& Apparao 1994; 10a
Galloway et al.\ 2002; 11a Wijnands et al.\ 1997b, 11b Homan et
al.\ 2002, 11c Belloni et al.\ 2002a, 11d Kuulkers et al.\ 2002, 11e
Stella et al.\ 1987b, 11f Langmeier et al.\ 1990, 11g Penninx et
al.\ 1990, 11h Wijnands et al.\ 1996, 11i Kuulkers et al.\ 1997a, 11j
Penninx et al.\ 1988, 11k di Salvo et al.\ 2000; 12a Markwardt et
al.\ 2002; 13a Markwardt et al.\ 2003, 13b Markwardt 2004 priv. comm;
14a Wijnands \& van der Klis 1998b, 14b in 't Zand et al.\ 2001, 14c
Wijnands et al.\ 2003, 14d Chakrabarty et al.\ 2003, 14e van Straaten
et al.\ 2004, 14f Uttley 2004, 14g Wijnands \& van der Klis 1998a, 14h
Menna et al.\ 2003, 14i Gierli\'nski et al.\ 2002, 14j Wijnands 2004,
14k van der Klis et al.\ 2000; 15a Markwardt \& Swank 2003, 15b
Strohmayer et al.\ 2003, 15c Wijnands \& Homan 2003; 16a Jonker et
al.\ 1998, 16b Jonker et al.\ 2000a, 16c van Paradijs et al.\ 1988a,
16d Penninx et al.\ 1991, 16e Oosterbroek et al.\ 1994; 17a Zhang et
al.\ 1998a, 17b Kuulkers \& van der Klis 1998, 17c O'Neill et
al.\ 2002, 17d Ponman et al.\ 1988, 17e O'Neill et al.\ 2001, 17f
Agrawal \& Bhattacharyya 2003, 17g di Salvo et al.\ 2001b, 17h Agrawal
\& Sreekumar 2003, 17i Iaria et al.\ 2004; 18a Jonker et al.\ 2001;
19a Zhang et al.\ 1996a,b; 19b M\'endez et al.\ 1998c, 19c Kaaret et
al.\ 1999a, 19d di Salvo et al.\ 2003, 19e M\'endez 2002a, 19f Jonker
et al.\ 2002b, 19g Wijnands et al.\ 1997a, 19h Prins \& van der Klis
1997, 19i van der Klis et al.\ 1990, 19j Damen et al.\ 1990; 20a
Markwardt et al.\ 1999a, 20b Oosterbroek et al.\ 1991; 21a Ford et
al.\ 1998a, 21b Olive et al.\ 2003, 21c Langmeier et al.\ 1989, 21d
Langmeier et al.\ 1987, 21e Barret \& Olive 2002; 22a Strohmayer et
al.\ 1996, 22b Ford \& van der Klis 1998, 22c M\'endez \& van der
Klis 1999, 22d Piraino et al.\ 2000, 22e Migliari et al.\ 2003a, 22f
Migliari et al.\ 2003b, 22g di Salvo et al.\ 2001a, 22h van Straaten
et al.\ 2002, 22i van Straaten et al.\ 2001, 22j Franco 2001; 23a
Wijnands \& van der Klis 1997, 23b Barret et al.\ 2000; 24a Wijnands
et al.\ 1998b, 24b Ford et al.\ 1998b, 24c Penninx et al.\ 1989, 24d
van Paradijs et al.\ 1988b; 24e Corbet et al.\ 1989; 25a Smale et
al.\ 1997, 25b Zhang et al.\ 1998b, 25c Kaaret et al.\ 1999b, 25d
Bloser et al.\ 2000a, 25e M\'endez 2002b, 25f Dotani et al.\ 1989, 25g
Stella et al.\ 1987a, 25h Wijnands et al.\ 1999a, 25i Migliari et
al.\ 2004; 26a Zhang et al.\ 1998c, 26b Cui et al.\ 1998a, 26c Yu et
al.\ 1999, 26d Reig et al.\ 2000a, 26e Reig et al.\ 2004, 26f Yu et
al.\ 2003; 27a Homan \& van der Klis 2000, 27b Homan et al.\ 1999b;
28a Wijnands et al.\ 2001b, 28b Wijnands et al.\ 2002b; ; 29a Marshall
\& Markwardt 1999; 30a Strohmayer et al.\ 1997; 31a Kaaret et
al.\ 2003, 31b in 't Zand et al.\ 1999a; 32a Kaaret et al.\ 2002; 33a
Boirin et al.\ 2000, 33b Bloser et al.\ 2000b, 33c Narita et al.\
2003, 33d Galloway et al.\ 2001; 34a Homan et al.\ 1999a, 34b Tomsick
et al.\ 1999; 35a Barnard et al.\ 2003b; 36a Smale \& Kuulkers 2000,
36b Smale et al.\ 2003, 36c McGowan et al.\ 2003; 37a Tennant 1987,
37b Tennant 1988, 37c Igekami 1986, 37d, Makino 1993, 37e Oosterbroek
et al.\ 1995, 37f Shirey et al.\ 1996, 37g Shirey et al.\ 1998, 37h
Shirey et al.\ 1999, 37i Qu 2001, 37j Ding et al.\ 2003; 38a Lewin et
al.\ 1987, 38b Makishima et al.\ 1989, 38c Oosterbroek et al.\ 2001,
38d Makishima et al.\ 1983, 38e dal Fiume et al.\ 1990, 38f Asai et
al.\ 1993, 38g den Hartog et al.\ 2003, 38h Reerink et al.\ 2004; 39a
Homan et al.\ 1998, 39b Schnerr et al.\ 2003, 39c Matsuba et al.\
1995, 39d Homan et al.\ 2003c; 40a Jonker et al.\ 2000c; 41a Jonker et
al.\ 1999; 42a Jonker et al.\ 2003; 43a Olive et al.\ 1998; 44a
Tawara et al.\ 1982, 44b Stella et al.\ 1988a, 44c Stella et al.\
1988b, 44d Dotani et al.\ 1990, 44e Kawai et al.\ 1990, 44f Lubin et
al.\ 1991, 44g Tan et al.\ 1991, 44h Lubin et al.\ 1992a, 44i Lubin et
al.\ 1992b, 44j Lubin et al.\ 1993, 44k Rutledge et al.\ 1995, 44l
Kommers et al.\ 1997, 44m Fox et al.\ 2001; 45a Wijnands \& van der
Klis 1999b; 46a in 't Zand et al.\ 2000; 47a Wijnands \& van der Klis
1999c, 47b Revnivtsev et al.\ 1999a; 48a Barret et al.\ 2003; 49a in
't Zand et al.\ 1999b; 50a Kitamoto et al.\ 1992.
\end{minipage}\\
\hline
\hline
\end{tabular}
\end{table*}
\subsection{Other phenomena}\mylabel{s:nsotherrxrv}
A 1--3\% rms Q\about2 QPO near 6 Hz occurs in Z sources in the NB
(e.g., 4h,i,j,n,11b,g,i,3h,r,7b,i,j,l,3o,16b\hide{Midleditch \&
Priedhorsky 1986, Priedhorsky et al.\ 1986, van der Klis et al.\ 1987
sco, Penninx et al.\ 1990 17+2, Hasinger et al.\ 1990 Cyg x-2, Lewin et
al.\ 1992, Kuulkers et al.\ 1994 5-1, 1997 17+2, Kamado et al.\ 1997
5-1, Wijnands et al.\ 1997 x-2, Jonker et al.\ 2000 340, 2002 5-1,
Wijnands \& van der Klis 2001 x-2, Homan et al.\ 2002 17+2}).
\about180$^\circ$ energy-dependent phase lags occur in some of these
QPOs which can be interpreted in terms of a quasi-periodically
pivoting X-ray spectrum with a pivot point between 3 and 7\,keV
(3g,r,4o,7m\hide{Mitsuda \& Dotani 1989 x-2, Vaughan et al.\ 1999 5-1,
Wijnands \& van der Klis \,2001 x-2}).  In some sources, this 'normal
branch oscillation' or NBO seems to jump to 10--20 Hz when the source
moves into the FB (e.g., 4j,l,n,11g,i) while in others it
disappears.  The 6--14\,Hz QPOs rarely seen at the tip of the UB in
some atoll sources may be related to this N/FBO
(25h,45a,47a,26e\hide{Wijnands et al.\ 1999a 1820, 1999 1806,
Revnivtsev et al.\ 1999a 1806, Reig et al.\  2003 aql}).  In Z sources
the phenomenon has been modeled as a radiation-force feedback
instability in spherical near-Eddington accretion (Fortner et
al.\ 1989), but detections in atoll sources occur at luminosities
well below Eddington.  Disk oscillation models for NBO have also been
proposed (e.g., Alpar et al.\ 1992, Wallinder 1995).

An $\alpha=$ 1.2--2 power-law component called very-low frequency
noise (VLFN) with an rms amplitude of usually a few percent often
dominates the variability $\approxlt$1\,Hz.  This component is
detected in all Z (e.g., 7b,f,i,j,4j,l,n,11b,g,i,3h,o,17c,e,f\hide{van
der Klis et al.\ 1987 5-1, 1987 sco, Hasinger and van der Klis 1989,
Penninx et al.\ 1990 17+2, Hasinger et al.\ 1990 x-2, Hertz et
al.\ 1992 sco, Lewin et al.\ 1992 5-1, Kuulkers et al.\ 1994 5-1,
Wijnands et al.\ 1997 Cyg X-2, Kuulkers et al.\ 1997a 17+2, Dieters and
van der Klis \,2000 sco, Agrawal \& Bhattacharya 2001 349, O'Neill et
al.\ 2001 349, Jonker et al.\ 2002 5-1 22\% tot 0.001Hz, Homan et
al.\ 2002 17+2, O'Neill et al.\ 2002 349}) and atoll sources (e.g.,
9f,19d,22g,25f,8i\hide{van Straaten et al.\ 2000 0614, Di Salvo et
al.\ 2003 1636, di Salvo et al.\ 2001 1728, Dotani et al.\ 1989 1820,
van Straaten et al.\ 2003 1608}), including some faint and/or
transient sources (31a,47a,b\hide{Kaaret et al.\ 2003 1748.9, Wijnands
\& van der Klis 1999 1806, Revnivtsev et al.\ 1999a 1806}) as well as
the Rapid Burster and Cir X-1 (\S\myref{s:nsother}).  It is rarely
detected in the EIS \hide{(see Campana et al.\ 2004 601 474 for a
report on low-frequency variability in the transient Cen X-4 in
quiescence} and usually becomes stronger (up to typically 6\% rms but
sometimes much stronger, 7b) and often also steeper (up to $\alpha$=2)
towards higher states.  It has been variously ascribed to
accretion-rate variations and unsteady nuclear burning
(\S\myref{s:nonflow}).  In recent work, breaks and broad $<$1\,Hz
Lorenzians have sometimes been detected in the VLFN range (e.g.,
39b,26e,38h\hide{Schnerr et al.\ 2003 13+1, Reig et al.\ 2004 aql.,
Reerink et al.\  2004}), which provides some support for models that
produce the VLFN from a superposition of finite events such as nuclear
'fires' (Bildsten 1995).

In X-ray dip sources (\S1), a 0.6--2.4\,Hz QPO occurs
(41a,27b,40a\hide{Jonker et al.\ 1999 1323, Homan et al.\ 1999 0748,
Jonker et al.\ 2000 1746}) which, contrary to nearly all other
variability has an amplitude (5--10\% rms) that hardly depends on
photon energy.  Its occurrence in dippers suggests a link with high
system inclination; the QPO persists at near-constant fractional
amplitude right through X-ray bursts and dips, so quasi-periodic
obscuration of central emitting regions by structure above the plane
of the disk is an attractive model (cf. 3z).  Titarchuk \& Osherovich
(2000) proposed a specific global normal disk mode that might
accomplish this.  The 401\,Hz pulsar SAX\,J1808.4--3658 in the late
decay of its outbursts sometimes exhibits violent \about1\,Hz highly
non-sinusoidal flaring (14e,j,k\hide{van der Klis et al.\ 2000 iauc
7358, Wijnands et al.\ 2003 review, van Straaten et al.\ 2004 puls})
the nature of which is unclear.

Revnivtsev et al.\ (2001a) discovered 'mHz' (0.007--0.009\,Hz) QPOs
that only occurred in a very particular \Lx\ range and at energies
$<$5\,keV in 4U\,1608--52 and 4U\,1636--53 and interpreted them as
possible variations in nuclear burning on the neutron-star surface
(cf., \S3\editnote{check}).  These QPOs have been found to affect the
kHz QPOs (8h, \S\myref{s:parlines}).
\subsection{Peculiar low magnetic field neutron stars}
\mylabel{s:nsother}
{\it Cir\,X-1}, apart from behaviour attributed to periodic surges of
mass transfer at periastron in its 17-d highly eccentric orbit and
ill-understood long-term changes (Murdin et al.\ 1980 and references
therein), is also peculiar in that it sometimes seems to exhibit the
correlated spectral and $<$100\,Hz timing behaviour of an ordinary
atoll source (37e\hide{Oosterbroek et al.\ 1995}) and at other times,
at higher \Lx, that of a (somewhat atypical) Z source (37h\hide{Shirey
et al.\ 1999}).  However, no kHz QPOs have been reported, which sets
it apart from both classes.  BLN is seen and LF QPOs with frequencies
from 30 down to \about1\,Hz, as well as broad noise out to several
100\,Hz (37a,b,f-h\hide{Tennant 1987, 1988, Shirey et al.\ 1996, 1998,
1999}).  Both the timing properties and the hard X-ray spectrum
actually make the source resemble a black hole, but the detection of
X-ray bursts from the field, nearly certainly from the source itself
(Tennant et al.\ 1986), strongly argues in favour of a neutron star.

The {\it Rapid Burster} (MXB\,1730--335) is a transient source with
thermonuclear bursts (\S3) as well as repetitive 'type II' bursts
attributed to a recurrent accretion instability (e.g., Lewin et
al.\ 1993) possibly related to similarly-interpreted phenomena seen in
the high magnetic-field neutron star GRO\,J1744--28
(\S\myref{s:hmfns}, cf., Lewin et al.\ 1996) and the black hole
GRS\,1915+105 (\S\myref{s:bhotherrxrv}).  In addition to
\about0.05\,Hz wave trains with gradually increasing frequency
apparently excited by the type II bursts (44i,l\hide{Lubin et al.\
1992b, see also Kommers et al.\ 1997}), complex 0.5--7\,Hz QPOs are
observed both in- and outside the type II bursts (references see
Table\,\myref{t:ns}) which can be quite strong (up to 35\% rms,
44b\hide{ Stella et al.\ 1988}) with sometimes (usually weak)
harmonics.  Apart from the similar frequencies and the fact that
correlations exist with flux and spectral parameters, no obvious
similarities exist between the correlated spectral and timing
behaviour of the Rapid Burster and other LMXBs.  A link with Z-source
NBOs was suggested (44d\hide{Dotani et al.\ 1990}), but the properties
of the source do not fit in well with those of either Z or atoll
sources, or Cir\,X-1 (44k\hide{Rutledge et al.\ 1995}).  No
significant kHz variability has been detected (44m\hide{Fox et al.\
2001}).  Models for type II bursts in which the neutron-star magnetic
field temporarily interrupts accretion (e.g., Hanawa et al.\ 1989)
provide a setting where QPOs can be produced through magnetic
interactions with the accretion flow (e.g., Hanami 1988); non-magnetic
disk oscillations are a possibility as well (e.g., Cannizzo 1997).

Other sources which do not fit seamlessly within the framework
sketched here are GX\,13+1 and GX\,349+2 which show somewhat ambiguous
Z/atoll behaviour (cf., 3B and other references in
Table\,\myref{t:ns}), and the millisecond pulsar XTE\,J1751--305
(14e).  The other millisecond pulsars have properties similar to those
of other weak LMXBs (14e,j).
%
%
\section{Black holes}\mylabel{s:bh}
The reference codes in this section refer to those listed with
Table\,\myref{t:bh}.
\subsection{High-frequency QPOs}\mylabel{s:bhhf}
The 7 black holes from which HF QPOs (\S\myref{s:hf}) $>$100\,Hz have
been reported (2x,z,A,C,6b,f,h,i,m,59c,g,60b,61c,e,68a,74a,b) are
listed in Table\,\myref{t:bh} (some possibly related $<$100\,Hz
oscillations are discussed at the end of this section).  HF QPOs are
weak, transient, and energy dependent, and are detected only at high
count rate.  In many cases they hover around formal detection levels,
so there is considerable uncertainty about their exact properties, but
typically Q ranges between less than 2 and 10, and amplitudes are
between 0.5 and 2\% rms (2--60\,keV).  HF QPOs usually occur in the
VHS, but in XTE\,J1550--564 a 250-Hz QPO was seen in an IMS at a count
rate 70--85\% below the VHS (6f\hide{Homan et al.\ 2001}).  

In two objects two peaks at an approximate 2:3 frequency ratio were
detected together (GRO\,J1655--40: \about300 and \about450\,Hz,
61e\hide{Strohmayer 2001a}, see also 6m,61c\hide{h5 Remillard et
al.\ 1999c, h13 Remillard et al.\ 2002a}; XTE\,J1550--564: 188$\pm$3
and 268$\pm$3\,Hz, 6h\hide{Miller et al.\ 2001}; see also
6m\hide{Remillard et al.\ 2002a apj580,1030} and 6b,f,i,\hide{h2
Remillard et al.\ 1999a, h6 Homan et al.\ 2001, h9 Remillard et
al.\ 2002}).  A similar case (but not simultaneous) has been mentioned
but not yet fully reported for GRS\,1915+105 (113$\pm$3\,Hz and
165$\pm$3\,Hz, 2C\hide{Remillard et al.\ 2003 A27};
\S4.4.3\editnote{check}).  The frequencies of the 2:3 pairs may scale
with inverse black-hole mass (6m\hide{Remillard et al.\ 2002a}), but
there is some choice in frequencies and the systematic uncertainties
in the masses are considerable.  Marginal cases involving 1:2
frequency ratios have also been discussed (92\,Hz in XTE\,J1550--564,
6m\hide{Remillard et al.\ 2002a 580 1030}; 328\,Hz in GRS\,1915+105, 
2z\hide{Remillard et al.\ 2002b A25}).  Usually HF QPO peaks are seen
alone, but with a tendency to occur near fixed frequencies, in
addition to sources mentioned above: 184$\pm$5\,Hz in 4U\,1630--47
(59c,g\hide{Remillard \& Morgan 1999 AAS}), 250$\pm$5\,Hz in
XTE\,J1650--500 (60b\hide{Homan et al.\ 2003b 586 1262}), 240\,Hz in
H1743--322 (68a\hide{Homan et al.\ 2003a Atel 162}) and, marginally, at
150$^{+17}_{-28}$ and 187$^{+14}_{-11}$\,Hz in XTE\,J1859+226
(74a,b\hide{Cui et al.\ 2000b 535 L123, Markwardt et al.\ 20..}).
However, values well off the nominal values 184 and 276\,Hz occur in
XTE\,J1550--564 (e.g., 123$\pm$2\,Hz, 6f\hide{Homan et al.\ 2001 apjs
132 377}, but see 6m\hide{Remillard et al.\ 2002a 580 1030};
141$\pm$3\,Hz, 6i\hide{Remillard et al.\ 2002 564 962}).  

Below 6\,keV the HF QPOs are usually not detected, while at higher
energies amplitudes up to 5\% rms have been measured.  The
higher-frequency peak is more evident at energies $>$13\,keV in
GRO\,J1655--40 (61e\hide{Strohmayer 2001a}) but not in XTE\,J1550--564
(6h\hide{Miller et al.\ 2001}). In XTE\,J1550--564 and GRO\,J1655--40
the higher harmonic becomes stronger when the spectrum becomes softer
(6f,m\hide{Homan et al.\ 2001, Remillard et al.\ 2002a 580,1030}),
i.e, while all other variability frequencies increase
(\S\myref{s:bhlfc}).  In XTE\,J1550--564 this, together with the QPOs
off the fixed frequencies, produces a correlation between the
frequencies of actually observed HF and LF QPOs (6f\hide{Homan et
al.\ 2001}).  An observed correlation with LF QPO type and coherence
(6i,m\hide{Remillard et al.\ 2002 both 1550HFQPO}) is related to this.

In GRS\,1915+105 a 67-Hz QPO occurs (2b,x\hide{Morgan et al.\ 1997,
Remillard and Morgan 1998}) whose frequency varies by only a few
percent in no apparent correlation to factor-several X-ray flux
changes.  Q is usually around 20 (but sometimes drops to 6;
2A\hide{Remillard et al.\ 1999b AAS}), the amplitude is \about1\% (rms), and
hard lags up to 2.3\,rad occur (2y\hide{Cui 1999}).  The spectral
characteristics of this QPO depend strongly on the peculiar rapid
state changes of this source (\S\myref{s:bhotherrxrv}, 2p\hide{Belloni
et al.\ 2001 HF}).  QPOs at 40 and possibly 56\,Hz sometimes occur
with the 67-Hz one (2q\hide{Strohmayer 2001b HF}), and a 27-Hz QPO,
again strongly dependent on the rapid state changes, occurs on other
occasions (2p\hide{Belloni et al.\ 2001 HF}).  Note that 27:40:56:67
is close to 2:3:4:5.  The relation of these QPOs to the
$\approxgt$100\,Hz HF QPOs is unclear, but of course the reported
113/165\,Hz pair in GRS\,1915+105 (above) complicates their
interpretation.  In XTE\,J1550--564, a possibly related Q\about4.4 QPO
near 65\,Hz has been reported (6g\hide{Kalemci et al.\ 2001}).

The high frequencies of the QPOs discussed in this section, similar to
ISCO frequencies in stellar mass black holes, strongly suggest an
origin in the strong-field region.  In view of the reported constant
frequencies, diskoseismic modes (\S\myref{s:diskoscillations}) are a
candidate explanation, but these models do not predict harmonic sets
of frequencies such as seen in HF QPOs (Nowak et al.\ 1997, P\'erez et
al.\ 1997, Ortega-Rodriguez \& Wagoner 2000, Silbergleit et al.\ 2001).
Rezzolla et al.\ (2003) calculated p-modes in a toroidal accretion
geometry which can explain some of the key HF QPO properties such as
the integer frequency ratios.  Cui et al.\ (1998b) proposed that HF
QPOs occur at $\nu_{nodal}$ (\S\myref{s:rpm}).  Relativistic resonance
models (\S\myref{s:rrm}) predicted constant frequencies in small
integer ratios which were then found in GRO\,J1655--40 (Strohmayer
2001a, Abramowicz \& Klu\'zniak 2001).  The previously mentioned
models of Schnittman \& Bertschinger (2004, see \S\myref{s:grorbits})
and Laurent \& Titarchuk (2001, see \S\myref{s:radii}) both aim at
explaining HF QPOs.

HF QPO properties, to the extent that it has been possible to
establish them for this somewhat elusive phenomenon, seem quite
different from those of neutron-star kHz QPOs: the neutron-star QPOs
come in pairs with a separation related to the star's spin, have
strongly tunable frequencies and low harmonic content, while the
black-hole ones are single, but with high harmonic content, and have a
much more stable frequency.  Yet, the two phenomena might still be
reconciled within a single theoretical description.  For example, the
variable frequencies in neutron stars might occur because the
phenomenon occurs at a variable (e.g., inner disk) radius, set by
interaction of the disk flow with either a magnetic field or radiation
from the stellar surface, while in black holes, in the absence of
these influences, the same phenomenon occurs at a more constant radius
(perhaps close to the ISCO); the second QPO may occur only in neutron
stars because it is due to an interaction with the spin; the high
harmonic content in black-hole QPOs may be due to relativistic effects
(e.g, as a simple example, extreme Doppler boosting near the ISCO of a
spinning black hole) on the flow and its emission that become
important only near the ISCO.  Future observations at higher
sensitivity will help much in understanding this phenomenon and
elucidating the relation with kHz QPOs.
\subsection{The low-frequency complex}\mylabel{s:bhlfc}
\begin{table*}[htbp]
\caption{Black-hole rapid X-ray variability \mylabel{t:bh}}\scriptsize
\begin{tabular}{lcccccl}
\hline 
\hline
Source               & LS/IMS/HS & Strong & LF    & HF  & References \\
                     &           & BLN    & QPO   & QPO &            \\
\hline                           
GRO J0422+32         & \b/--/--  & \b     & \c    & --- & 51a-d;1v;5s;64d \\
LMC X-1 (0538--641)  & --/\b/\b  & ---    & \b    & --- & 52a-d \\
LMC X-3 (0540--679)  & \b/--/\b  & \b     & \b    & --- & 53a-e;52c,d \\
A0620--00            & \c/\c/\c  & \c     & ---   & --- & 54a-b,56d,f \\
XTE J1118+480        & \b/--/--  & \b     & \b    & --- & 55a-j;64d \\
GS 1124--68          & \b/\b/\b  & \b     & \b    & --- & 56a-f;1t \\
GS 1354--64          & \b/--/--  & \b     & \b    & --- & 57a,b;1y;64d \\
4U 1543--47          & \b/\b/\b  & \b     & \b    & --- & 58a \\ 
XTE\,J1550--564      & \b/\b/\b  & \b     & \b    & \b  & 6a-r;1A,H;5s,x;61g \\
4U\,1630--47         & \b/\b/--  & \b     & \b    & \b  & 59a-g;1y;5x,6l;61g \\
XTE\,J1650--500      & \b/\b/\b  & \b     & \b    & \b  & 60a-d;5u;61g \\
GRO J1655--40        & \b/\b/\b  & \b     & \b    & \b  & 61a-h;1v,y,A;5s,x;6e,i,l \\
GX 339--4 (1659--487)& \b/\b/\b  & \b     & \b    & --- & 5a-w;1l,p,t-v,y,z,A,H,K;56b,e\\ 
IGR J17091--3624     & \b/--/--  & \b     & ---   & --- & 62a \\
SAX J1711.6--3808$^a$& \b/\b/--  & \b     & \b    & --- & 63a,b \\
GRO J1719--24 (1716--249)& \c/--/--  & ---    & \b    & --- & 64a-c \\  
GRS 1737--31         & \b/--/--  & \b     & ---   & --- & 65a \\
GRS 1739--278        & --/\b/\b  & ---    & \b    & --- & 66a,b \\
1E 1740.7--2942      & \b/--/--  & \b     & \b    & --- & 67a,b;1v,y,K;5v \\
H\,1743--322         & --/--/\c  & ---    & \b    & \b  & 68a \\
XTE J1748--288       & \b/\b/\b  & \b     & \b    & --- & 69a;1v,y;5x;6l;61g \\
XTE J1755--324       & \b/--/\c  & \b     & ---   & --- & 70a;1v,y \\
GRS 1758--258        & \b/--/\b  & \b     & \b    & --- & 71a,b;1v,A,K;5v;67a,b\\
XTE J1819--254       & \c/--/--  & \b     & ---   & --- & 72a-d \\ 
EXO 1846--031        & --/--/\b  & ---    & ---   & --- & 73a,b \\
IGR J18539+0727      & \b/--/--  & \b     & ---   & --- & 62a \\
XTE\,J1859+226       & --/\b/--  & ---    & \b    & \b  & 74a-c;1H \\
XTE J1908+094        & \b/--/--  & ---    & \c    & --- & 75a \\
GRS\,1915+105        & \c/\c/\c  & \b     & \b    & \b  & 2a-C;1y,A;5s\\
4U 1957+11           & --/--/\b  & ---    & ---   & --- & 76a-c \\
Cyg\,X-1 (1956+35)   & \b/\b/\c  & \b     & \b    & --- & 1a-L;5p,t,v\\
GS 2000+25           & \b/\b/\b  & \b     & \c    & --- & 77a;1t \\
XTE J2012+381        & --/--/\b  & ---    & ---   & --- & 78a-c \\
GS 2023+338          & \b/\c/\c  & \b     & ---   & --- & 79a-c;1u\\
\hide{C->1 A->2 l->5 h->6}
\hline 
\hline
\end{tabular}
\bigskip
\begin{minipage}{\hsize}
This is not a list of certified black holes; relevant low
magnetic-field compact-object timing behaviour plus an absence of
bursts and pulsations suffice for inclusion. LS: low state, IMS:
intermediate state (includes VHS), HS: high state.  \b: variability of
this type observed; \c: some doubt (uncertain, ambiguous, atypical,
rare and not clearly seen, etc.); ---: not reported. 
Strong BLN: $>$15\% rms.
Note: $^a$\,Contamination by SAX J1712.6-3739, 63b.
\end{minipage}
\end{table*}
\begin{table*}
\scriptsize
\begin{tabular}{c}
\hline
\hline
\begin{minipage}{\hsize}
References in Table\,\myref{t:bh}: 1a Oda et al.\ 1971, 1b Terrell
1972, 1c Nolan et al.\ 1981, 1d Miyamoto \& Kitamoto 1989, 1e Belloni
\& Hasinger 1990a, 1f Negoro et al.\ 1994, 1g Vikhlinin et al.\ 1994,
1h Crary et al.\ 1996, 1i Belloni et al.\ 1996, 1j Cui et
al.\ 1997a,b, 1k Revnivtsev et al.\ 1999b, 1l Ford et al.\ 1999, 1m
Nowak et al.\ 1999a,b, 1n Gilfanov et al.\ 1999, 1o Gilfanov et
al.\ 2000, 1p Nowak 2000, 1q Uttley \& McHardy 2001, 1r Churazov et
al.\ 2001, 1s Pottschmidt et al.\ 2003, 1t Rutledge et al.\ 1999, 1u
Miyamoto et al.\ 1992, 1v Wijnands \& van der Klis 1999a, 1w Reig et
al.\ 2002, 1x Berger \& van der Klis 1998, 1y Sunyaev \& Revnivtsev
2000, 1z Miyamoto et al.\ 1988, 1A Li \& Muraki 2002, 1B Zdziarski et
al.\ 2002, 1C Wen et al.\ 2001, 1D Frontera et al.\ 2001, 1E
Gierli\'nski et al.\ 1999, 1F Zhang et al.\ 1997a, 1G Ling et al.\ 1983,
1H Done \& Gierli\'nski 2003, 1I Kotov et al\,2001, 1J Maccarone et
al.\ 2000, 1K Lin et al.\ 2000a, 1L Maccarone \& Coppi 2002a; 2a Chen
et al.\ 1997, 2b Morgan et al.\ 1997, 2c Paul et al.\ 1997, 2d Belloni
1998, 2e Swank et al.\ 1998, 2f Reig et al.\ 2003b, 2g Trudolyubov et
al.\ 1999a,b, 2h Markwardt et al.\ 1999b, 2i Feroci et al.\ 1999, 2j
Muno et al.\ 1999, 2k Naik et al.\ 2000, 2l Rao et al.\ 2000a,b, 2m
Chakrabarti et al.\ 2000, 2n Reig et al.\ 2000b, 2o Lin et al.\ 2000c,
2p Belloni et al.\ 2001, 2q Strohmayer 2001b, 2r Nandi et al.\ 2001,
2s Tomsick \& Kaaret 2001, 2t Muno et al.\ 2001, 2u Trudolyubov 2001,
2v Rodriguez et al.\ 2002a, 2w Ji et al.\ 2003, 2x Remillard \& Morgan
1998, 2y Cui 1999, 2z Remillard et al.\ 2002b, 2A Remillard et
al.\ 1999b, 2B Belloni et al.\ 2000, 2C Remillard et al.\ 2003; 5a
Samimi et al.\ 1979, 5b Motch et al.\ 1982, 5c Motch et al.\ 1983, 5d
Maejima et al.\ 1984, 5e Makishima et al.\ 1986, 5f Imamura et
al.\ 1990, 5g Grebenev et al.\ 1991, 5h Miyamoto et al.\ 1991, 5i
M\'endez \& van der Klis 1997, 5j Steiman-Cameron et al.\ 1997, 5k
Kong et al.\ 2002, 5l Belloni et al.\ 1999a, 5m Smith \& Liang 1999, 5n
Nowak et al.\ 1999c, 5o Revnivtsev et al.\ 2001b, 5p van Straaten et
al.\ 2003, 5q Nowak et al.\ 2002, 5r Nespoli et al.\ 2003, 5s Psaltis
et al.\ 1999a, 5t Belloni \& Hasinger 1990b, 5u Belloni 2004, 5v Lin
et al.\ 2000a, 5w Vaughan \& Nowak 1997; 6a Cui et al.\ 1999, 6b
Remillard et al.\ 1999a, 6c Wijnands et al.\ 1999b, 6d Cui et
al.\ 2000a, 6e Sobczak et al.\ 2000a, 6f Homan et al.\ 2001, 6g
Kalemci et al.\ 2001, 6h Miller et al.\ 2001, 6i Remillard et
al.\ 2002c, 6j Rodriguez et al.\ 2002b, 6k Belloni et al.\ 2002b, 6l
Vignarca et al.\ 2003, 6m Remillard et al.\ 2002a, 6n Kubota \& Done
2004, 6o Kubota \& Makishima 2004, 6p Gierli\'nski \& Done 2003, 6q
Rodriguez et al.\ 2003, 6r Sobczak et al.\ 2000b, 6s Reilly et
al.\ 2001; 51a Denis et al.\ 1994, 51b Vikhlinin et al.\ 1995, 51c
Grove et al.\ 1998, 51d van der Hooft et al.\ 1999a; 52a Ebisawa et
al.\ 1989, 52b Schmidtke et al.\ 1999, 52c Haardt et al.\ 2001, 52d
Nowak et al.\ 2001; 53a Treves et al.\ 1990, 53b Boyd et al.\ 2000,
53c Boyd et al.\ 2001, 53d Wilms et al.\ 2001, 53e Cowley et
al.\ 1991; 54a Carpenter et al.\ 1976, 54b Kuulkers 1998; 55a
Revnivtsev et al.\ 2000b, 55b Wood et al.\ 2000, 55c Malzac et
al.\ 2003, 55d Frontera et al.\ 2003, 55e Hynes et al.\ 2003b, 55f
Belloni et al.\ 2002a, 55g Kanbach et al.\ 2001, 55h Spruit \& Kanbach
2002, 55i McClintock et al.\ 2001, 55j Esin et al.\ 2001; 56a Miyamoto
et al.\ 1994, 56b Belloni et al.\ 1997, 56c Takizawa et al.\ 1997, 56d
Hynes et al.\ 2003a, 56e Miyamoto et al.\ 1993, 56f Esin et al.\ 2000,
e7 Esin et al.\ 1997, e8 Ebisawa et al.\ 1994; 57a Revnivtsev et
al.\ 2000a, 57b Brocksopp et al.\ 2001; 58a Park et al.\ 2003; 59a
Parmar et al.\ 1986, 59b Kuulkers et al.\ 1997b, 59c Remillard \&
Morgan 1999, 59d Dieters et al.\ 2000b, 59e Tomsick \& Kaaret 2000, 59f
Trudolyubov et al.\ 2001, 59g Klein-Wolt et al.\ 2004a; 60a Kalemci et
al.\ 2003, 60b Homan et al.\ 2003b, 60c Tomsick et al.\ 2004, 60d Rossi
et al.\ 2004; 61a Zhang et al.\ 1997c, 61b M\'endez et al.\ 1998d, 61c
Remillard et al.\ 1999c, 61d Yamaoka et al.\ 2001, 61e Strohmayer
2001a, 61f Sobczak et al.\ 1999, 61g Kalemci et al.\ 2004, 61h Hynes
et al.\ 1998; 62a Lutovinov \& Revnivtsev 2003; 63a Wijnands \& Miller
2002, 63b in 't Zand et al.\ 2002; 64a van der Hooft et al.\ 1996, 64b
van der Hooft et al.\ 1999b, 64c Revnivtsev et al.\ 1998a, 64d
Brocksopp et al.\ 2004; 65a Cui et al.\ 1997c; 66a Borozdin \&
Trudolyubov 2000, 66b Wijnands et al.\ 2001a; 67a Smith et al.\ 1997,
67b Main et al.\ 1999; 68a Homan et al.\ 2003a; 69a Revnivtsev et
al.\ 2000c; 70a Revnivtsev et al.\ 1998b; 71a Lin et al.\ 2000b, 71b
Smith et al.\ 2001; 72a Wijnands \& van der Klis 2000, 72b Revnivtsev
et al.\ 2002, 72c Uemura et al.\ 2002, 72d Uemura et al.\ 2004;
\hide{aanvullen V4641!}  73a Parmar et al.\ 1993, 73b Sellmeijer et
al.\ 1999; 74a Cui et al.\ 2000b, 74b Markwardt 2001, 74c Brocksopp et
al.\ 2002; 75a Woods et al.\ 2002; 76a Ricci et al.\ 1995, 76b Nowak
\& Wilms 1999, 76c Wijnands et al.\ 2002a; 77a Terada et al.\ 2002;
78a Vasiliev et al.\ 2000, 78b Naik et al.\ 2000, 78c Campana et
al.\ 2002; 79a Oosterbroek et al.\ 1996, 79b Oosterbroek et
al.\ 1997, 79c $\dot{\rm Z}$ycki et al.\ 1999a,b.
\end{minipage}\\
\hline
\hline
\end{tabular}
\end{table*}
The strong BLN in the LS was first noticed in Cyg\,X-1 (Oda et al.\
1971) and initially interpreted in terms of a 73~msec periodicity.  Later
shot-noise models (\S\myref{s:timing}) were applied to this noise (see
also \S\myref{s:shots}).  Nolan et al.\ (1981) presented the first
power spectrum unambiguously showing the characteristic
$\nu_b$=0.01--1\,Hz flat-topped noise shape (Fig.\,\myref{f:psa},
LS). The energy dependence of the noise amplitude is generally small
in the 2--40\,keV range, above that the rms may decrease (e.g.,
1m,I,5n,v,51a,b,c,57a,\hide{Denis et al.\ 1994 0422, Vikhlinin et al.\
1995 0422, Grove et al.\ 1998 0422, Nowak et al.\ 1999 cygx1, 1999
339, Lin et al.\ 2000a surv. , Revnivtsev et al.\ 2000a 1354, Kotov et
al.\ 2001 Cygx1}), although background subtraction issues may have
affected some of these results.  While the noise is well correlated
between energy bands (0.1-10 Hz cross-coherence,
\S\myref{s:timing}, is \about1), hard \about0.1\,rad lags are seen
between $<$6 and $>$ 15\,keV, increasing logarithmically with energy.
Contrary to predictions from basic Comptonization light travel-time
models (\S\myref{s:Edep}), time lag is not constant but decreases with
Fourier frequency as $\nu^{-0.7}$, with some structure possibly
related to the characteristic noise frequencies ($\nu_b, \nu_h$,
\S\myref{s:lfc}; 1d,l,m,u,z,5n,w,64b,67a,71a\hide{Miyamoto et
al.\ (1988), 1992 surv, Miyamoto \& Kitamoto (1989), Vaughan and Nowak
1997, Smith et al.\ 1997 1758, van der Hooft et al.\ 1999b 1719, Ford
et al.\ 1999 nsbh, Nowak et al.\ 1999 cygx1, 1999 339, Lin et
al.\ 2000b 1758}).  While in some scattering geometries this could be
understood (Kazanas et al.\ 1997, Hua et al.\ 1997), lack of the
expected smearing of the variability towards higher photon energies by
light travel time effects makes models of this type unlikely
(1J,K\hide{Lin et al.\ 2000a, Maccarone et al.\ 2000}) and favours
propagation models for the lags instead (\S\myref{s:Edep}).

In the IMS/VHS, $\nu_b$ is higher, from 0.1--1\,Hz to $>$10\,Hz
(2--60\,keV).  Energy dependencies in noise amplitude, time lags and
coherencies tend to be larger and more complex and variable here than
in the LS (1s,t,5h,6a,c,d,52d,56c,e,60a,72b,79c\hide{Miyamoto et al.\
1991 339, Miyamoto et al.\ 1993 surv, Rutledge et al.\ 1995 survey,
Takizawa et al.\ 1997 1124, Cui 2000a 1550, Wijnands et al.\ 1999b
1550, {\. Z}ycki et al.\ 1999a or b 2023, Cui 1999 1550, Pottschmidt et al
2000 Cygx1, Nowak et al.\ 2001 LMCx1, Revnivtsev et al.\ 2002 1819,
Kalemci 2003 1650}), possibly due to competition between the
contribution of hard and soft spectral components to these quantities,
and characteristic frequencies depend on energy as well (e.g.,
1i,j,t,6a,f,55d,56b,60a,61c\hide{Belloni et. al.\ 1996 Cyg x-1, 1997
1124, Cui et al.\ 1997a or b??  Cygx1, 1999 1550, Homan et al.\ 2001
1550, Remillard et al.\ 1999c 1655, Rutledge et al.\ 1999, Kalemci et
al.\ 2003 1650, Frontera et al.\ 2003 1118}).  GRS\,1915+105 in its
hard state (C state; 2B\hide{Belloni et al.\ 2000}) has noise similar
to this (i.e., this state probably corresponds to a VHS/IMS, 2f\hide{
Reig et al.\ 2003b}, see also, e.g., 2n,o,t\hide{Muno et al.\ 2001,
Lin et al.\ 2000c, Reig et al.\ 2000b}).

Noise amplitude is less in the 6--7-keV (Fe-line) region
(1k,o,L,72b,79a,\hide{Oosterbroek et al.\ 1996 2023, Revnivtsev et
al.\ 1999b cygx1, Gilfanov et al.\ 2000 cygx1, Revnivtsev et al.\ 2002
1819, MaccaroneCoppi 2002a 0204160 cygx1}). This effect is seen in the
IMS, but is more pronounced in the LS, particularly $>$5\,Hz. Perhaps
this is due to a larger line forming ('reflection') region in that
state (1k,o\hide{Revn ea 99b aa 347 L23, Gilfanov et al.\ 2000}, see
also 79a\hide{Oosterbroek et al.\ 1996}).

QPOs ($Q>2$) with amplitudes of a few to more than 10\% rms are
sometimes observed superimposed on the noise at frequencies
(\S\myref{s:freqcorr}) a factor 2--8 above, or around, $\nu_b$. This
is seen in the LS (0.01--2\,Hz, sometimes including a harmonic:
1s,5c,g,m,o,q,s, 6i,53b,55a,61b,67a,71a\hide{ Motch et al.\ 1983 339 0.1
Hz, Grebenev et al.\ 1991 339 0.8 Hz, Smith et al.\ 1997 1 and
1758-258 2 en 0.4 Hz, M\'endez et al.\ 1998d 1655 0.77, 1.3, Smith \&
Liang 1999 339 0.8 Hz + harmonics, Remillard et al.\ 1999a 1550 0.2
0.8 Hz, Lin et al.\ 2000b 1758-258 1.4 Hz + harm, Boyd et al.\ 2000
lmcx3 0.46Hz, Revnivtsev et al.\ 2000b 1118, Revnivtsev et al.\ 2001b
339 0.1-0.65 Hz Q unclear, Nowak et al.\ 2002 339 0.2-0.8, Pottschmidt
et al.\ 2003 cygx-1 0.01-1 Hz}) where frequency often gradually
increases through the initial LS rise of a transient outburst (in the
0.015$\rightarrow$0.3\,Hz range: 55b,d,57b,64a,d\hide{ van der Hooft
et al.\ 1996 1719 0.04-0.3 Hz steady rise, Brocksopp et al.\ 2001 1354
0.014-0.05 Hz steady rise below break, Hynes et al.\ 2003b 1118
0.07-0.16Hz steady rise, Frontera et al.\ 2003 1118 same, Brocksopp et
al.\ 2004 survey 0.04-0.3, 0.015-0.047, 0.08-0.16 steady rises and
0.1-0.28 both rise and fall}).  It is also often observed in the IMS
(1--33\,Hz, often with rich harmonic structure:
1t,5h,r,6c,d,f,56a-c,58a,59d-f,60a,b,d,61b,c,g,63a,b,66a,b,69a\hide{
Miyamoto et al.\ 1991 339 5.9-13.8 harmonics, Takizawa et al.\ 1997
1124 3-16 Hz harmonics, Belloni et al 1997 same, Miyamoto et al 1994
same, M\'endez et al.\ 1998d 1655 6.5, 9.3, Wijnands et al.\ 1999b
1550 3-12 Hz harmonics, Rutledge et al.\ 1999 survey 3-8.5, Tomsick \&
Kaaret 2000 1630 5.3-11 harmonics/3.4-1.3, Dieters et al.\ 2000b 1630
0.8-14 harmonics?, Cui et al.\ 2000a? 6-12 harmonics, Borozdin \&
Trudolyubov 2000 1739 5, 9.8 harmonics, Revnivtsev et al.\ 2000c 1748
18-33, Homan et al.\ 2001 1550 1-18, Remillard et al.\ 1999c 1655 HF
8.7-22, Trudolyubov et al.\ 2001 1630 2.7-14 (also mHz), Wijnands et
al.\ 2001a 1739 1.7-7.5 harmonics, Wijnands \& Miller 2002 1711
4.5-13.4 harmonics, in 't Zand 2002 1711 0.5-7.3 harmonics, Rossi et
al.\ 2004 1650 1-10 Hz steady rise, Kalemci et al.\ 2003 1650 8.6-3.8
decay, Park et al.\ 2003 1543 7-10 Hz, Nespoli et al.\ 2003 339-4
5-7,12 harmonic, Homan et al.\ 2003b 1650 1.3-25 Q unclear, Kalemci et
al.\ 2003 survey 1.3-9.3}). Transitions in frequency occur in the
0.08$\rightarrow$13\,Hz range between LS and IMS both in the rise
(0.08$\rightarrow$13\,Hz range; 6a,e,i,s\hide{ Cui et al.\ 1999 1550
0.1-10, Sobczack et al.\ 2000a 1550 0.084-13.1, Reilly et al.\ 2001
1550 0.24-7.19, Remillard et al.\ 2002c 1550 HF 0.08-9.8 steady rise
0.08-19.6 harmonics}) and in the decay (6$\rightarrow$0.2\,Hz range;
6g,s,59e\hide{ Reilly et al.\ 2001 6.34-0.64, Kalemci et al.\ 2001
1550 HF 4-0.35 steady fall, Tomsick \& Kaaret 2000 1630 3-0.2 steady
fall}) of some transient outbursts.  Transitions that are smooth in
frequency may nevertheless be rapid in time and, depending on
coverage, seem abrupt.  IMS/VHS LF-QPO harmonic structure is
intricate, with subsets of 1:2:3:4:6:8 frequency ratios all observed
(references above).  Time lags of opposite sign and different Q values
are sometimes seen between alternating harmonics
(56b,c,6a,c,74a\hide{Takizawa 1997 1124, Belloni 1997 1124, Cui 1999
1550, Wijnands et al.1999b 1550, Cui et al.\ 2000b 1859 HF})
reminiscent of what is observed in Z-source HBOs (\S\myref{s:lfc}).
Source-state dependent QPO subtypes \rem{more about this?, ABC eg
Homan 1743 QPO paper geeft aardig overzicht} are distinguished based
on differences in peak width, time lags and accompanying noise
(6c,f,i,m\hide{ Wijnands et al.\ 1999b, Homan et al.\ 2001, Remillard
et al.\ 2002}, Ch.\,4).  In GX\,339--4 a 6-Hz LF QPO fluctuated by
about 1\,Hz on time scales down to 5\,sec in correlation with X-ray
flux variations (5r\hide{Nespoli et al.\ 2003}).
%
%
QPOs with similar harmonic structure and alternating time lags are
also observed in the 0.5--12\,Hz range in GRS\,1915+105 in its hard
state (e.g., 2a,b,g,h,j,l,n,o,s,v\hide{Chen et al.\ 1997, Morgan et
al.\ 1997 HF, Trudolyubov et al.\ 1999a,b, Markwardt et al.\ 1999b,
Muno et al.\ 1999, Rao et al.\ 2000, Reig et al.\ 2000b, Lin et
al.\ 2000c, Tomsick \& Kaaret 2001, Rodriguez et al.\ 2002a}).  
In the HS, weak ($\approxlt$1\% rms) 0.02--0.08 and 14--27\,Hz QPOs
are occasionally seen whose relation to the other LF QPOs is unclear
(6e,f,52a,61c,69a\hide{Ebisawa et al.\ 1989 LMC X-1 0.08 Hz, Sobczack et
al.\ 2000a 1550 18.1, Sobczack et al.\ 2000a 1655 14-23 really HS?,
Revnivtsev et al.\ 2000c 1748 0.029, Homan et al.\ 2001 1550 16-18}).

Correlations between the LF complex and X-ray spectral (usually,
Comptonization and reflection) model parameters are mostly an
expression of the source state dependencies already discussed above
and in \S\myref{s:lfc}: as the spectrum hardens (power-law fraction
increases, blackbody flux decreases, coronal compactness increases,
photon index decreases, etc.), amplitudes increase and frequencies
decrease (e.g., 1n,s,5o,q,6g, l,56a,59e,60c,61g,77a\hide{ Miyamoto et
al.\ 1994 1124, di Matteo \& Psaltis 1999,Gilfanov et al.\ 1999 cygx1,
Tomsick \& Kaaret 2000 1630, Kalemci et al.\ 2001 1550, Revnivtsev et
al.\ 2001b 339, Nowak et al.\ 2002 339, Terada et al.\ 2002 2000+25,
Vignarca et al.\ 2003 surv, Pottschmidt et al.\ 2003 Cygx1, Tomsick et
al.\ 2003 1650, Kalemci et al.\ 2004 survey}, also di Matteo and
Psaltis 1999). Gilfanov et al.\ (2004) present a tentative correlation
between QPO frequency and Fe line width (wider line for higher
frequency) which is roughy in accordance with expectations if the QPO
is an orbital frequency and the line width is determined by Doppler
shifts due to motion in that orbit.  Again, in the IMS correlations
are more complex, perhaps due to competition between spectral
components. In both XTE\,J1550--564 and GRO\,J1655--40 the QPO
frequency increases with blackbody flux, but in XTE\,J1550--564 the
photon index and total flux then both rise, whereas in GRO\,J1655--40
both fall (6e,i\hide{Sobczak et al 2000a, Remillard et al.\ 2002}, who
conclude that QPO frequency correlates with $\dot M_d$,
\S\myref{s:states}, and that the power-law fraction should be $>$20\%
for QPOs to be observed).

In XTE\,J1650--500, in decay towards the off state at a luminosity a
factor \about500 below that early in the outburst, a $\nu_b$ of
0.0035\,Hz was measured (60c\hide{Tomsick et al.\ 2003 1650}), so on
the way to the off state $\nu_b$ apparently continues to decrease.
Optical and UV power spectra in the low and off states are similar to
those in X-rays, with $\nu_b$ as low as 0.0001-0.001\,Hz in the off
state (55e,56d\hide{Hynes et al.\ 2003a,b}).  0.02-Hz BLN and 0.1-Hz
QPO were observed simultaneously in UV and X-rays in the LS in
XTE\,J1118+480, with 1-2\,sec UV lags possibly due to light travel
time delays in reprocessing, but the detailed correlation is complex
and not well understood (55c,e,g,h\hide{Kanbach et al.\ 2001, Spruit
\& Kanbach 2002, Malzac et al.\ 2003, Hynes et al.\ 2003b}).  In
GRO\,J1655--40 in the VHS, 10-20 sec UV lags were observed
(61h\hide{Hynes et al.\ 1998}), while in XTE\,J1819--254 7-min {\it
X-ray} lags were seen which were attributed to disk propagation
effects (72c,d\hide{Uemura et al.\ 2002, 2003}). In GX\,339--4 optical
BLN and QPOs occur in the 0.005--3 Hz range (e.g., 5b,c,f,j\hide{Motch
et al.\ 1982, 1983, Imamura et al.\ 1990, Steiman-Cameron et al.\
1997}).

The BLN in black holes has been extensively modeled in
terms of shot noise, often used as model for magnetic flares, and
chaos-, and certain MHD-simulations of disk flows also
produce power spectra reminiscent of LS noise (see
\S\myref{s:shots}).  For the LF QPOs a variety of, nearly exclusively
disk-based, models has been considered (orbiting hot spots,
\S\myref{s:grorbits}, Lense-Thirring precession, \S\myref{s:rpm}, disk
oscillations, \S\myref{s:diskoscillations}).  Superposition of various
oscillation frequencies can also produce BLN (Nowak 1994).  Churazov
et al.\ (2001) have proposed that $\nu_b$ is the orbital frequency of
the inner edge of the optically thick disk, and explain the LS and IMS
power spectra of Cyg\,X-1 by supposing that the variability is
generated at the local orbital time scale in the hot optically thin
regions of the flow located both within the inner edge, and sandwiching,
the disk.  
\subsection{Power-law noise}\mylabel{s:bhpl}
Power-law noise (in this section defined as noise with no clear flat
top in $P_\nu$ down to several 0.01\,Hz) is not normally seen in the
LS (strong BLN could mask it). The HS is characterized by weak
(0.5--3\%) power-law noise with index typically
0.7--1.2 (e.g., 5l,6f,56c,59d,61b,d,69a\hide{ Takizawa et al.\ 1997
1124, M\'endez et al.\ 1998d 1655, Belloni et al.\ 1999a 339, Dieters
et al.\ 2000b, Revnivtsev et al.\ 2000c 1748, Homan et al.\ 2001 1550,
Yamaoka et al.\ 2001 1655}) and sometimes a break to $\alpha=1.5-2$
occurs around 3\,Hz (Fig.\,\myref{f:psa}, HS; 6f,61b\hide{ M\'endez et
al.\ 1998d 1655, Homan et al.\ 2001 1550}). The most varied power-law
noise is observed in the IMS.

Rapid (minutes to days) variations between \about1-Hz BLN-dominated
and few-\% rms $\alpha$\about1 power-law dominated power spectra in
GS\,1124--68 and GX\,339--4 were an early defining characteristic of
the VHS (5h,56c\hide{1124: Takizawa et al.\ 1997 0.3-1.4\% a=1 + BLN
2-6 Hz, 339: Miyamoto et al.\ 1991}). In more recent work
(2f,6f,52c,d,59b,61a-c\hide{4U\,1630--47: Kuulkers et al.\ 1997b 7\%
a=1.5 1655 in the VHS: Zhang et al.\ 1997c 5.5\% a=1.2, M\'endez et
al.\ 1998d 5\%, Remillard et al.\ 1999c 5\% a=1 but .1-10 Hz curve
peak at 1 Hz, 1915: Reig et al.\ 2003b 1915, LMC\,X-1: Nowak et
al.\ 2001 7\% a=1, some structure; Haardt et al.\ 2001 6\% a=0.8 and
XTE\,J1550--564: Homan et al.\ 2001}), a 4--8\% rms $\alpha=$0.8--1.5
power-law component is commonly found, sometimes together with BLN,
usually at the 'soft side' of the VHS/IMS, where the BLN weakens.  While
the noise in these various examples is clearly not flat-topped, it
sometimes has a discernable curvature or break in the 1-10\,Hz range,
and steepens towards higher frequencies.  When XTE\,J1550--564 moves
from HS to gradually harder intermediate states, $\alpha$=0.7--1.2
power-law noise, which can be as weak as 0.5\% in the HS proper,
becomes stronger, up to 5.6\% in the VHS, 2.8\% in lower-luminosity
IMS and up to 8.9\% on the way to the LS, with no BLN detected (6f).
When the BLN appears, the power-law noise can still be detectable at
low frequencies at levels of 1.5--4\% rms, but as the BLN gets
stronger and $\nu_b$ drops, it disappears below the detection level.
In what may be a similar phenomenon, Cyg\,X-1 displays a strong (up to
33\% rms, 1j,r\hide{Cui et al.\ 1997Jan a or b?,Churazov et al.\ 2001})
$\alpha$=1 power law, usually with a clear break to $\alpha$=2 around
10\,Hz at the top of soft flares in what is usually called the 'high
state' in this source, but is actually more similar to an IMS
(1s,61g\hide{Pottschm, Kalemci et al.\ 2004}); when the spectrum gets
slightly harder in this IMS, BLN with $\nu_b$=1--3.5 Hz is seen
together with a flat ($\alpha$=0.3--0.8) power law at low frequency
(1i,j).  Because it dominates in the disk-dominated 'soft'
state, Cui et al.\ (1997b) interpret this noise to originate in the
disk, but Churazov et al.\ (2001) argue instead that it originates in
the corona. 

Like VLFN in neutron stars (\S\myref{s:nsotherrxrv}), power-law noise
in black holes may be due to \mdot\ variations, perhaps arising in the
outer disk.  Numerical simulations can produce power spectra that are
red over several decades in frequency (Mineshige et al.\ 1994b,
Kawaguchi et al.\ 2000, Hawley \& Krolik 2001) but as noted in
\S\myref{s:bhlfc}, flat tops are usually discernable in these cases.
Clearly, models involving surface phenomena such as nuclear burning do
not apply in the black-hole case, so such models for neutron-star
power-law noise (\S\myref{s:nsotherrxrv}) predict differences
between neutron stars and black holes which should show up in
comparative studies.
\subsection{Other phenomena and peculiar objects}\mylabel{s:bhotherrxrv}\mylabel{s:otherbh}
The number of black-hole X-ray binaries that is peculiar with respect
to the timing characteristics and source states discussed here is
small; mostly what makes systems peculiar is unusual variations on
longer time scales, likely originating outside the strong-field
region, or {\it lack} of variability (in possible black holes such as
SS433, Cyg\,X-3 and CI\,Cam; Grindlay et al.\ 1984, Berger \& van
der Klis 1994, Belloni et al.\ 1999b, but see van der
Klis \& Jansen 1985), usually plausibly attributable to
external circumstances.  This suggests that the innermost accretion
flow, where the rapid variability arises, is dominated by the black-hole
properties.  XTE\,J1819--254 (V4641 Sgr) in its brief outburst showed
unusually strong (46\% rms) power-law noise with rapidly varying
$\alpha$ (72a\hide{Wijnands \& van der Klis 2000}) and rapid optical
fluctuations (72c,d\hide{Uemura et al.\ 2002}). It may be an
SS433-like super-Eddington accretor (72b\hide{Revnivtsev et
al.\ 2002}).  Objects showing properties not clearly characteristic
for either low magnetic-field neutron stars or black holes include
4U\,1957+11 (76a-c\hide{Ricci et al.\ 1995, Nowak \& Wilms 1999,
Wijnands et al.\ 2002a}), possibly a black hole always in the HS, and
4U\,1543--62 and 4U\,1556--60, which might be atoll sources
(Table\,\myref{t:ns}; Farinelli et al.\ 2003).

A small number of rapid variability phenomena observed in black-hole
systems do not fit the above categories. Transient oscillations in the
1-1000\,mHz range are regularly observed in the very high state ('dips
and flip-flops', 5h; e.g., 56c,59d,f) which may just be relatively
rapid source-state transitions (e.g., Belloni et al.\ 2000) due to
disk instabilities; the peculiar nature of GRS\,1915+105 may be due to
the source showing such behaviour most of the time.  Of course, faster
and lower-luminosity phenomena could similarly be due to rapid state
transitions (cf., Yu et al.\ 2001), but this is harder to check.
\section{High-magnetic-field neutron stars}\mylabel{s:hmfns}
In the high magnetic-field neutron stars the typically 10$^{12}$\,G
$B$ field disrupts the disk flow at the magnetospheric radius
$r_{mag}$\about10$^8$\,cm and channels the plasma to the magnetic
poles.  The resulting strong periodic pulsations complicate the
analysis of the stochastic variability.  The dynamical time scale at
$r_{mag}$ is of order seconds, hence variability much faster than this
is not apriori expected from the accretion disk (but see e.g.,
Orlandini \& Morfill 1992, Orlandini and Boldt 1993, Klein et
al.\ 1996a,b for possible rapid magnetospheric accretion-flow
phenomena).  Indeed, aperiodic variability tends to be dominated by
strong (several 10\% rms) flat-topped BLN with $\nu_b$ often (but not
always) \about$\nu_{spin}$ (e.g., Frontera et al.\ 1987, Belloni \&
Hasinger 1990b, Angelini et al.\ 1989, Takeshima et al.\ 1991,
Takeshima 1992), suggesting that it arises in the disk outside
$r_{mag}$.  Possibly, such a BLN component is a
characteristic of any disk flow truncated at 10$^8$\,cm, and similar
physics underlies this component in high magnetic-field neutron stars
and in low magnetic-field compact objects in low states (e.g., Hoshino
\& Takeshima 1993, van der Klis et al.\ 1994b).  However, as noted by
Lazzati \& Stella (1997), if the pulse amplitude is itself modulated,
the strong periodic component can give rise to these noise components,
complicating interpretation.

A broad QPO often seen at factors up to \about10$^2$ below
$\nu_{spin}$ (see Table\,1.1 in \S1.2.1.2;  references above and in Shirakawa \& Lai 2002b, see also
Li et al.\ 1980) is sometimes interpreted as a magnetospheric beat
frequency (\S\myref{s:bfm}), implying that orbital motion at $r_{mag}$
occurs at approximately the spin frequency.  On occasion, when a QPO occurred above
$\nu_{spin}$ (e.g., Angelini et al.\ 1989, Finger et al.\ 1996) its
luminosity- or spin-up-rate related changes in frequency were
consistent with either the beat frequency or the orbital frequency at
$r_{mag}$, allowing to constrain neutron-star parameters\editnote{how,
if either/or?}.  Upper and lower sidebands were seen separated from
$\nu_{spin}$ by the QPO frequency (Kommers et al.\ 1998, Moon \&
Eikenberry 2001a,b), implying the pulse amplitude is modulated by the
QPO; although this is what a beat-frequency model predicts, asymmetry
between the two sidebands suggests that an orbital frequency in the
disk contributes to the QPO signal.  Disk oscillations (Titarchuk \&
Osherovich 2000) and warped disk modes (Shirakawa \& Lai 2002b) have
also been proposed for these QPOs.  The peculiar 0.5-sec pulsar
GRO\,J1744--28 which showed type II bursts (cf., \S\myref{s:nsother})
also produced 20, 40 and 60\,Hz QPOs (Zhang et al.\ 1996c).

Jernigan et al.\ (2000) reported QPO features near 330 and 760\,Hz in
the 4.8\,s accreting pulsar Cen\,X-3.  They interpreted this in terms
of the photon bubble model (\S\myref{s:nonflow}).  This is the only
report of millisecond oscillations from a high-magnetic-field neutron
star, but note that the QPO features are quite weak, and instrumental
effects are a concern at these low power levels.
\section{Flow-instability and non-flow models}\mylabel{s:othermodels}
In \S\myref{s:generic} we looked at models for observed variability
frequencies based on orbital and epicyclic motions supplemented with
decoherence and modulation mechanisms in order to fit the properties
of observed QPO and noise components.  In the current section, we
discuss \rxrv\ models in which the variability arises in instabilities
in the accretion flow.  Models of this type often include the
modulation mechanism as part of their physics, although the
consequences of this have not always been fully explored yet in terms
of predictions of observable phenomena. At the end of this section we
look at neutron-star boundary-layer and surface phenomena.
\subsection{Disk-oscillation models}\mylabel{s:diskoscillations}
The expressions in \S\myref{s:grorbits} are for free particle orbits.
An accretion disk is a hydrodynamic flow, so the particles are not
free but exert forces upon each other.  This leads to the potential
for oscillations in the flow.  The resulting vibration modes of disk,
corona, and disk/star and disk/magnetosphere boundary layers can lead
to QPOs or noise.  In a standard disk model several physical time
scales are of the order of the local orbital time scale $1/\nu_\phi$
(Shakura \& Sunyaev 1973, 1976, Thorne \& Price 1975, Pringle 1981;
\S13.2.3\editnote{check}).  Many analytic as well as numerical
explorations have been made of disk oscillation modes.  Often the
free-particle orbital and epicyclic frequencies can still be
recognized (e.g., Wagoner 1999), but combination frequencies with
(magneto-)hydrodynamic frequencies (e.g., of sound waves) occur as
well (e.g., Psaltis 2000).  In \S\S\myref{s:ns}--\myref{s:hmfns} some
examples of models in this class proposed for specific phenomena have
already been mentioned.  Here we emphasize models relevant to kHz QPOs
and black-hole HF QPOs.

In the strong-field gravity region global oscillation modes with
frequencies similar to the free-particle frequencies can occur which
are 'trapped' in (can not propagate outside) particular disk annuli
(Kato et al.\ 1998, Nowak and Lehr 1998, Wagoner 1999, Kato 2001, Lamb
2003, for useful summaries).  This trapping is a predicted
strong-field gravity effect; it does not occur in Newtonian gravity.
Of these 'diskoseismic' modes, the $g$, $c$ and $p$ modes occur at
approximately constant frequencies, which like the orbital frequency
at the ISCO scale as $1/M$.  The $w$ modes depend on the inner disk
radius and hence can vary in frequency. Diskoseismic modes were not
found in simulations including MHD turbulent effects (Hawley \& Krolik
2001).

If there is a sharp transition in the disk, according to Psaltis \&
Norman (2000) \editnote{what about alpar} the transition radius can
act as a filter with a response that has strong resonances near the
epicyclic (and orbital) frequencies of that radius; predicted
frequencies are similar to those of the relativistic precession model
(\S\myref{s:rpm}), with hydrodynamic corrections and additional
frequencies that might allow a better match to the LF QPO, kHz QPO and
sideband data (but see also Markovi\'c \& Lamb 2000).  Oscillations
associated with a viscous transition layer between Keplerian disk and
neutron star, some involving the star's magnetic field but {\it not}
requiring strong field gravity, and hence not involving the epicyclic
frequencies, have been proposed to apply to high-frequency and other
QPO and noise phenomena in neutron stars, black holes and white
dwarfs, with at least 6 different observed frequencies and their
correlations being predicted (Titarchuk et al.\ 1998, 1999, Osherovich
\& Titarchuk 1999, Titarchuk \& Osherovich 1999, Titarchuk 2002;
Titarchuk \& Wood 2002, but see also Miller 2003 for a theoretical
discussion of these ideas, and van Straaten et al.\ 2000, 2003 and
Jonker et al.\ 2000a, 2002a for discrepancies with the observations).
Oscillations associated with instabilities in a disk-magnetosphere
boundary have been explored as a model for high-frequency QPOs (Li \&
Narayan 2004, Kato et al.\ 2001).  Hydrodynamic disk modes, some
involving the effects of radiation and magnetic fields have further
been explored by various authors (e.g., Hanami 1988, Fukue \& Okada
1990, Spruit \& Taam 1990, Okuda \& Mineshige 1991, Alpar et al.\
1992, Luo \& Liang 1994, Chen \& Taam 1994, 1995, Wallinder 1995,
Abramowicz et al.\ 1995, Ipser 1996, Milsom \& Taam 1996, 1997,
Markovi\'c \& Lamb 1998, Lai 1999, Kato et al.\ 2001, Gnedin \&
Kiikov 2001, Varniere et al.\ 2002, Shirakawa \& Lai 2002a, Das et
al.\ 2003, Mukhopadhyay et al.\ 2003, Watarai \& Mineshige
2003b). Damping, or the superposition of many local frequencies (e.g.,
Nowak 1994) can turn intrinsically periodic disk oscillations into QPO
or broad band noise.

Fluctuations in the density or temperature of a Comptonizing medium
can produce a modulation in spectra formed by Comptonization (e.g.,
Boyle et al.\ 1986, Stollman et al.\ 1987, Alpar et al.\ 1992, Lee \&
Miller 1998), as can modulation of the rate at which photons are
injected into the medium (e.g., Kazanas \& Hua 1999, Laurent \&
Titarchuk 2001, Titarchuk \& Shrader 2002).
Various types of intrinsic flow instabilities can modulate the
accretion rate onto the central object (e.g., hydrodynamic; Taam \&
Lin 1984, general-relativistic; Paczy\'nski 1987, Kato 1990, and
radiation-feedback instabilities; Fortner et al.\ 1989, Miller \& Lamb
1992, Miller \& Park 1995) and in this way produce relatively
large-amplitude modulations in the X-ray flux.  All disk oscillations
to some extent directly affect the disk luminosity and the
photon-energy dependence of the disk emissivity (references above and,
e.g., Nowak \& Wagoner 1993), yet many of the models that have been
proposed essentially are variability-frequency models, predicting the
frequency, or the power spectrum, of the fluctuations only in some
physical flow parameter rather than in any observable quantity.
Clearly, in order to allow further tests of disk oscillation models
and to discriminate between them, such predictions of observables of
the oscillations are essential. \hide{Ghosh 1998 and Moderski and
Czerny 1999 comment is about a stability criterion, not a disk model
-- Paczy\'nski 1987 Nature refers to Kluz & Wagoner 1985 for ns
smaller than isco and predicts no freq.}
\subsection{Energy dependencies: amplitude, phase, cross-coherence}
\mylabel{s:Edep}
To the extent that the intrinsically periodic models discussed above
and in \S\myref{s:generic} include explicit X-ray flux modulation
mechanisms they are in principle severely constrained by the observed
photon-energy dependencies of amplitude and phase (\S\myref{s:timing})
of the modulation.  To explain why the variability often has a harder
spectrum than the average flux (i.e., larger fractional rms,
\S\myref{s:timing}, towards higher photon energies, see
e.g., \S\S\myref{s:nskhz} and \myref{s:nslfc}, but see also
\S\myref{s:bhlfc}) in models where the frequencies arise in the disk,
which usually emits a {\it softer} than average flux, two main types
of mechanism have been discussed: (i) modulation of the rate at which
the disk is providing soft seed photons to the Comptonizing corona
(references below), and (ii) modulation of the accretion rate into an
inner region, such as a neutron-star boundary layer or black-hole
inner corona whose X-ray spectrum is harder that that of the disk
(e.g., Churazov et al.\ 2001, Gilfanov et al.\ 2003).  Of course,
another possible explanation for this is that the frequencies do not
arise in the disk in the first place.  QPOs and noise in black holes
in the IMS tend to be more strongly energy dependent than in other
states or in neutron stars; this may be related to the large
differences between the X-ray spectral components that are
simultaneously prominent in black holes in this state.

Propagation of an X-ray signal through a hot cloud of Comptonizing
electrons produces hard lags: on average, photons emerging later have
undergone more scatterings and hence gained more energy; the time lag
between bands $E_1$ and $E_2$ scales roughly as $\log(E_2/E_1)$ (Payne
1980, Miller 1995, Nowak \& Vaughan 1996).  This mechanism has been
applied many times (see references in \S\S\myref{s:timing},
\myref{s:modulation} and \myref{s:diskoscillations} and, e.g., Wijers
et al.\ 1987, Miyamoto et al.\ 1988, Schulz \& Wijers 1993, Nowak
1994, Nowak et al.\ 1999b, Hua \& Titarchuk 1996, Hua et al.\ 1997,
1999, B\"ottcher \& Liang 1998, , Cullen 2000, Laurent \& Titarchuk
2001, Titarchuk \& Shrader 2002, Nobili et al.\ 2000), with detailed
calculations of the effects of the cloud oscillating (e.g., Fortner et
al.\ 1989, Stollman et al.\ 1987, Miller \& Lamb 1992, Miller 1995,
Lee \& Miller 1998, Lee et al.\ 2001), and estimates of the effects of
other Comptonizing geometries (e.g., Miller et al.\ 1998a, Shibazaki
et al.\ 1988, Reig et al.\ 2003c, Nobili 2003).  The applicability of
these models to black-hole BLN lags has been addressed in
\S\myref{s:bhlfc}; lags naturally arising from shot-noise models
involving propagating clumps or waves (\S\myref{s:shots}) may be more
appropriate to explain these.  Reprocessing by the disk of X-rays
emitted above the disk plane ('reflection'; Poutanen 2002, see also
Kaaret 2000), or by the companion star (Vikhlinin 1999) is another way
to affect the energy dependencies in the variability.  K\"ording \&
Falcke (2004) provide a detailed analysis of time lags produced by a
pivoting power-law spectrum. \hide{alternating rms, lags and coherence
odd and even: bottcher liang 0003139, was never published}

The observed energy dependencies in variability amplitude and phase
are very constraining, and none of the models cited are able to
explain everything observed.  The unknown geometry of the emitting and
scattering regions contributes to the considerable uncertainty with
respect to the correct model.  Clearly, this is an area where the
observations are needed to guide the modeling, yet observational
capabilities are not yet at the required level.  If, for example,
luminous blobs would be orbiting the compact object in the region of
the disk where the relativistically broadened Fe line is generated,
with sufficient sensitivity this would produce a very obvious signal
due to Doppler-shifted features moving up and down through the line
profile, but present instrumentation is not able to reliably detect
signals of this kind.  With the larger instruments currently under
consideration, such work may become possible.
\subsection{Intrinsically aperiodic models}\mylabel{s:shots}
The intrinsically aperiodic (\S\myref{s:generic}) models most often
used are in the class of shot-noise models (see also
\S\myref{s:timing}).  Shots are sensible in various physical settings,
e.g., magnetic flares (usually, storage of magnetic energy followed by
release in a reconnection event, e.g., Wheeler 1977, Galeev et
al.\ 1979, Pudritz \& Fahlman 1982, Aly \& Kuijpers 1990, Haardt et
al.\ 1994, Mineshige et al.\ 1995, Poutanen \& Fabian 1999, di Matteo
et al.\ 1999, Merloni et al.\ 2000, Kato et al.\ 2001, Rodriguez et
al.\ 2002c, Varniere et al.\ 2002, Krishan et al.\ 2003), infalling
blobs or waves producing a flare ('propagation models', Nowak et
al\,1999b: e.g., Miyamoto et al.\ 1988, Miyamoto \& Kitamoto 1989, Kato
1989, Manmoto et al.\ 1996, B\"ottcher \& Liang 1999, Misra 2000,
B\"ottcher 2001) and finite-lifetime orbiting blobs (references see
\S\myref{s:timing} and \S\myref{s:rpm}).  In particular the
propagation models may be able to produce the frequency-dependent lags
and relatively energy-independent high-frequency amplitudes (Lin et
al.\ 2000a, Maccarone et al.\ 2000) appropriate to black-hole BLN
(\S\myref{s:bhlfc}).

The observation that variability amplitude is linearly related to flux
over a wide range of time scales, {\it including} those of the
variability itself in SAX\,J1808.4-3658 and Cyg X-1 (Uttley \& McHardy
2001, Uttley 2004, Gleissner et al.\ 2004, see also
Ba{\l}uci\'nska-Church et al.\ 1997) is incompatible with
straightforward shot-noise models and instead suggests that slower
variations are produced first, and are then operated on by faster
processes to produce the higher frequency variability.  In this
context models seem appropriate where slow accretion rate variations
originate far out in the disk (where correspondingly long dynamical
time scales apply) and become observable after disturbances have
propagated inward to the emitting regions, undergoing further
modulation producing the faster variations on the way (Vikhlinin et
al.\ 1994, Lyubarskii 1997, see also {\. Z}ycki 2002, 2003, King et al.\
2004).

Among chaos models (see also \S\myref{s:timing}) for which some
physical justification was put forward are the 'dripping handrail'
model of Scargle et al.\ (1993), and models involving self-organized
criticality of a disk as described in a cellular automaton model
(Mineshige et al.\ 1994a,b, Abramowicz \& Bao 1994, Takeuchi et
al.\ 1995, Xiong et al.\ 2000).  Autonomous flow instabilities are
certainly suggested by direct and to some extent 'first principle'
magnetohydrodynamic simulation of disks, which tend to produce broad
noise spectra (Kawaguchi et al.\ 2000, Hawley \& Krolik 2001,
2002, Armitage \& Reynolds 2003).
\subsection{Non-flow models: boundary layer, stellar surface}\mylabel{s:nonflow}
In addition to models associated with various aspects of the accretion
flow, neutron-star oscillations or processes occurring on the
neutron-star surface or in a boundary layer could produce variability.
Clearly, such models exclude any possibility that the same phenomenon
also occurs in a black hole.  Processes like these are thought to
cause the burst oscillations (\S3.4), but they could also underlie
some of the observed variability in the persistent emission.  Radial
(e.g., Shibazaki \& Ebisuzaki 1989) and non-radial oscillations of
various kinds (e.g., McDermott \& Taam 1987, McDermott et al.\ 1988,
Epstein 1988, Bildsten \& Cutler 1995, Bildsten et al.\ 1996,
Strohmayer \& Lee 1996, Bildsten \& Cumming 1998), can cause a
variation in the surface emission properties; emission would be
further modulated as the spin periodically modulates aspect and
visibility of the 'spots' thus formed.  Temporary surface features due
to nuclear-burning processes, which were specifically proposed as a
model for neutron-star VLFN (\S\myref{s:nsotherrxrv}; Bildsten 1993,
1995, see \S3) or magnetohydrodynamic effects (Hameury et al.\ 1985)
have also been considered (see also Inogamov \& Sunyaev 1999, Popham
\& Sunyaev 2001) and a \about0.01\,Hz quasi-periodicity in nuclear
burning rate has been suggested by Revnivtsev et al.\ (2001a;
\S\myref{s:nsotherrxrv}).  Klein et al.\ (1996a,b) proposed a model
where photon bubbles rise up by buoyancy through the accreted material
and produce a flash of radiation when they burst at the top as a model
for kHz QPOs and for QPOs in high-B neutron stars (\S\myref{s:hmfns}).
\section{Final remarks}
Since the previous issue of this book (Lewin et al.\ 1995) \rxrv\
studies have come a long way towards strong-field gravity physics.
For the first time we are seeing, and are able to study in some
detail, variability at the dynamical time scale of the strong-field
region in accreting low magnetic-field neutron stars as well as
stellar-mass black holes.  Strong-field gravity is an integral part
of many of the models proposed.  The aim of future timing work will be
to turn the {\it diagnostics} of strong-field gravity and dense matter
we now have into true {\it tests} of GR and {\it determinations} of
the EOS.  A theoretical framework for interpreting the observables of
motion in the strong-field region in detail and for testing
strong-field gravity theories is currently emerging (e.g., Weinberg et
al.\ 2001, DeDeo \& Psaltis 2004).  What is needed on the observational
side is a considerable increase in timing sensitivity coupled with
good spectral capabilities.  As timing sensitivity in the relevant
regime is proportional to collecting area (not the square root of it;
\S\myref{s:timing}), this is an attainable goal.  With future
large-area timing instrumentation (\about10\,m$^2$, e.g., XEUS, see
Barret 2004, or a dedicated timing array) the predicted patterns of
weak sidebands and harmonics (``fingerprints'') of the variability
phenomena would be mapped out as a spectrum of interrelated
frequencies.  This would make models that successfully predicted them
unassailable, as of course they should be in order to be accepted as
true tests of general relativity in the strong-field regime.  Current
observations are still clearly limited by the drop in QPO amplitude
towards the extreme frequencies (Fig.\,\myref{f:mendezamplitudes}).
With more sensitivity the frequency range over which QPOs are detected
would be considerably widened, likely making it possible to follow kHz
QPOs up to the ISCO and check for the predicted frequency saturation
there.  A considerable synergy in terms of testing gravitation
theories may occur with the results of gravitational wave instruments

Combining timing and spectroscopy is another way to clinch the models.
For example, in an orbital motion model for a QPO phenomenon in the Fe
line region, to zeroeth order the frequency provides the orbital
period, and the line profile the orbital velocity, so that we can
solve for orbital radius $r$ and central mass $M$.  So, combining
spectral and timing measurements will provide strong tests of the
models and allow to start using them to learn more about the curved
spacetime near compact objects.  By measuring the line-profile changes
on short time scales, or equivalently the amplitude and phase
differences between QPOs in several spectral bands within the line
profile exciting tests are possible.  The line widths are $\sim$keV,
so moderate-resolution millisecond spectroscopy is sufficient to do
this. Clearly, entirely different signals are expected from QPOs
caused by luminous blobs orbiting in the strong-field region and from,
e.g., spiral-flow modulated-accretion QPOs: hence, such measurements
will decide the emission geometry and constrain the modulation model.

Finally, depending on the precise phenomenon, large area detectors
will make it possible to detect the QPOs either within one cycle
$\nu^{-1}$, or one coherence time $(\pi\Delta\nu)^{-1}$, allowing to
study them in the time domain. Wave-form studies will allow to
quantitatively constrain compact object mass, radius and angular
momentum, orbital velocity and gravitational ray bending by modeling
approaches such as described by Weinberg et al.\ (2001) for
neutron-star surface hot spots.  The opportunities provided by
large-area detectors for doing strong-gravity and dense-matter physics
by timing X-ray binaries are clearly excellent.
\paragraph{\bf Acknowledgements:} It is a pleasure to acknowledge the help of
the many colleagues who either made data available before publication,
provided, or produced new, figures, read versions of the manuscript
or provided insightful discussion at various stages during the writing
of this chapter: Marek Abramowicz, Didier Barret, Tomaso Belloni,
Deepto Chakrabarty, Eric Ford, Jeroen Homan, Peter Jonker, Fred Lamb,
Erik Kuulkers, Marc Klein-Wolt, Wlodek Klu\'zniak, Tom Maccarone,
Craig Markwardt, Mariano M\'endez, Simone Migliari, Dimitrios Psaltis,
Thomas Reerink, Pablo Reig, Roald Schnerr, Luigi Stella, Steve van
Straaten, Phil Uttley, Rudy Wijnands, Wenfei Yu.  I am particularly
indebted to Cole Miller and Diego Altamirano.  All errors are mine.
This work was supported in part by the Netherlands Organization for
Scientific Research (NWO) and the Netherlands Research School for
Astronomy (NOVA).
\def\aj{{AJ}}                   
\def\araa{{\it ARA\&A}}      
\def\apj{{\it ApJ}}          
\def\apjl{{\it ApJ}}         
\def\apjs{{\it ApJS}}        
\def\apss{{Ap\&SS}}             
\def\aap{{\it A\&A}}         
\def\aapr{{\it A\&A\,Rev.}}   
\def\aaps{{\it A\&AS}}         
\def\azh{{AZh}}                 
\def\baas{{BAAS}}               
\def\jrasc{{JRASC}}             
\def\memras{{MmRAS}}            
\def\mnras{{\it MNRAS}}             
\def\pra{{Phys.\,Rev.\,A}}        
\def\prb{{Phys.\,Rev.\,B}}        
\def\prc{{\it Phys.\,Rev.\,C}}        
\def\prd{{\it Phys.\,Rev.\,D}}        
\def\pre{{Phys.\,Rev.\,E}}        
\def\prl{{\it Phys.\,Rev.\,Lett.}}    
\def\pasp{{PASP}}               
\def\pasj{{\it PASJ}}               
\def\qjras{{QJRAS}}             
\def\skytel{{S\&T}}             
\def\solphys{{Sol.\,Phys.}}      
\def\sovast{{\it Soviet\,Ast.}}      
\def\ssr{{\it Space\,Sci.\,Rev.}}     
\def\zap{{ZAp}}                 
\def\nat{{\it Nature}}              
\def\iaucirc{{\it IAU\,Circ. No.}}       
\def\aplett{{Astrophys.\,Lett.}} 
\def\apspr{{Astrophys.\,Space\,Phys.\,Res.}}
\def\bain{{Bull.\,Astron.\,Inst.\,Netherlands}} 
\def\fcp{{Fund.\,Cosmic\,Phys.}}  
\def\gca{{Geochim.\,Cosmochim.\,Acta}}   
\def\grl{{Geophys.\,Res.\,Lett.}} 
\def\jcp{{J.\,Chem.\,Phys.}}      
\def\jgr{{J.\,Geophys.\,Res.}}    
\def\jqsrt{{J.\,Quant.\,Spec.\,Radiat.\,Transf.}}
\def\memsai{{Mem.\,Soc.\,Astron.\,Italiana}}
\def\nphysa{{Nucl.\,Phys.\,A}}   
\def\nphysb{{\it Nucl.\,Phys.\,B}}   
\def\physrep{{Phys.\,Rep.}}   
\def\physscr{{Phys.\,Scr}}   
\def\planss{{Planet.\,Space\,Sci.}}   
\def\procspie{{Proc.\,SPIE}}   
\parindent=0pt


\begin{thereferences}{}
\scriptsize
%


\bibitem{} Abramowicz, M.~\& Bao, G.\ 
1994, 
\pasj, 46, 523 

\bibitem{} Abramowicz, M.~A.~\& Klu{\' z}niak, W.\ 
2001, 
\aap, 374, L19 

\bibitem{} Abramowicz, M.~A.~\& Klu{\' z}niak, W.\ 
2004, 
in Rossi and beyond, AIP Conf. Proc. 714, 21; 
astro-ph/0312396 

\bibitem{} Abramowicz, M., Jaroszynski, M., \& Sikora, M.\ 
1978, 
\aap, 63, 221 

\bibitem{} Abramowicz, M.~A., Lanza, A., Spiegel, E.~A., \& Szuszkiewicz, E.\ 
1992, 
\nat, 356, 41 

\bibitem{} Abramowicz, M.~A., Chen, X., \& Taam, R.~E.\ 
1995, 
\apj, 452, 379 

\bibitem{} Abramowicz, M.A., Almergren, G.J.E., Klu\'zniak, W., \& Thampan, A.V.\ 
2003a,
gr-gc/0312070 

\bibitem{} Abramowicz, M.~A., Karas, V., Klu\'zniak, W., Lee, W.~H., \& 
Rebusco, P.\ 
2003b, 
\pasj, 55, 467 

\bibitem{} Abramowicz, M.~A., Bulik, T., Bursa, M., \& Klu{\' z}niak, W.\ 
2003c, 
\aap, 404, L21 

\bibitem{} Abramowicz, M.A., Klu\'zniak, W., Stuchlik, Z., \& Torok, G.\ 
2004,
astro-ph/0401464 

\bibitem{} Agrawal, V.~K.~\& Bhattacharyya, S.\ 
2003, 
\aap, 398, 223 

\bibitem{} Agrawal, V.~K.~\& Sreekumar, P.\ 
2003, 
\mnras, 346, 933 

\bibitem{} Akmal, A., Pandharipande, V.R., \& Ravenhall, D.G.\ 
1998,
\prc, 58, 1804 

\bibitem{} Alpar, M.~A.\ 
1986, 
\mnras, 223, 469 

\bibitem{} Alpar, M.~A.~\& Shaham, J.\ 
1985,
\nat, 316, 239 

\bibitem{} Alpar, M.~A.~\& Y\i lmaz, A.\ 
1997, 
New Astronomy, 2, 225 

\bibitem{} Alpar, M.~A., Hasinger, G., Shaham, J., \& Yancopoulos, S.\ 
1992, 
\aap, 257, 627 

\bibitem{} Aly, J.~J.~\& Kuijpers, J.\ 
1990, 
\aap, 227, 473 

\bibitem{} Andersson, N., Kokkotas ,K.D., \& Stergioulas, N.\ 
1999,
\apj, 516, 307 

\bibitem{} Andersson, N., Jones, D.~I., Kokkotas, K.~D., \& Stergioulas, N.\ 
2000, 
\apjl, 534, L75 

\bibitem{} Angelini, L., Stella, L., \& Parmar, A.~N.\ 
1989, 
\apj, 346, 906 

\bibitem{} Armitage, P.~J.~\& Natarajan, P.\ 
1999, 
\apj, 525, 909 

\bibitem{} Armitage, P.~J.~\& Reynolds, C.~S.\ 
2003, 
\mnras, 341, 1041 

\bibitem{} Asai, K., Dotani, T., Nagase, F., Mitsuda, K., Kitamoto, S., et al.\ 
1993, 
\pasj, 45, 801 

\bibitem{} Asai, K., Dotani, T., Mitsuda, K., Nagase, F., Kamado, Y., et al.\ 
1994, 
\pasj, 46, 479 

\bibitem{} Asaoka, I.~\& Hoshi, R.\ 
1989, 
\pasj, 41, 1049 

\bibitem{} Augusteijn, T., Karatasos, K., Papadakis, M., Paterakis, G. et  al.\ 
1992, 
\aap, 265, 177 

\bibitem{} Ba\l uci\'nska-Church, M., Takahashi, T., Ueda, Y., Church, M.~J., 
et al.\ 
1997, 
\apjl, 480, L115 

\bibitem{} Bao, G.~\& \O stgaard, E.\ 
1994, 
\apjl, 422, L51 

\bibitem{} Bao, G.~\& \O stgaard, E.\ 
1995, 
\apj, 443, 54 

\bibitem{} Bardeen, J.~M.~\&  Petterson, J.~A.\ 
1975, 
\apjl, 195, L65 

\bibitem{} Bardeen, J.~M., Press, W.~H., \& Teukolsky, S.~A.\ 
1972, 
\apj, 178, 347 

\bibitem{} Barnard, R., Church, M.~J., \&  Ba{\l}uci{\' n}ska-Church, M.\ 
2003a, 
\aap, 405, 237 

\bibitem{} Barnard, R., Kolb, U., \& Osborne, J.~P.\ 
2003b, 
\aap, 411, 553 

\bibitem{} Barr, P., White, N.~E., \& Page, C.~G.\ 
1985, 
\mnras, 216, 65P 

\bibitem{} Barret, D., 
2004,
in Rossi and beyond, AIP Conf. Proc. 714, 405; 
astro-ph/0401099 

\bibitem{} Barret, D.~\& Olive, J.\ 
2002, 
\apj, 576, 391 

\bibitem{} Barret, D.~\& Vedrenne, G.\ 
1994, 
\apjs, 92, 505 

\bibitem{} Barret, D., McClintock, J.~E., \& Grindlay, J.~E.\ 
1996, 
\apj, 473, 963 

\bibitem{} Barret, D., Olive, J.~F., Boirin, L., Done, C., Skinner, G.~K., \& 
Grindlay, J.~E.\ 
2000, 
\apj, 533, 329 

\bibitem{} Barret, D., Olive, J.~F., \& Oosterbroek, T.\ 
2003, 
\aap, 400, 643 

\bibitem{} Belloni, T.\ 
1998, 
New Astronomy Review, 42, 585 

\bibitem{} Belloni, T.\ 
2004, 
in The restless high-energy universe, Nuclear Phys. B, 132, 337;
astro-ph/0309028 

\bibitem{} Belloni, T.~\& Hasinger, G.\ 
1990a, 
\aap, 227, L33 

\bibitem{} Belloni, T.~\& Hasinger, G.\ 
1990b, 
\aap, 230, 103 

\bibitem{} Belloni, T., M\'endez, M., van der Klis, M., Hasinger, G., Lewin, 
W.~H.~G., et al.\ 
1996, 
\apjl, 472, L107 

\bibitem{} Belloni, T., van der Klis, M., Lewin, W.~H.~G., van Paradijs, J., 
Dotani, T., et al.\ 
1997, 
\aap, 322, 857 

\bibitem{} Belloni, T., M{\' e}ndez, M., van der Klis, M., Lewin, W.~H.~G., 
\& Dieters, S.\ 
1999a, 
\apjl, 519, L159 

\bibitem{} Belloni, T., Dieters, S., van den Ancker, M. E., Fender, R. P., et al.\ 
1999b, 
\apj, 527, 345 

\bibitem{} Belloni, T., Klein-Wolt, M., M{\' e}ndez, M., van der Klis, M., \& 
van Paradijs, J.\ 
2000, 
\aap, 355, 271 

\bibitem{} Belloni, T., M{\' e}ndez, M., \& S{\' a}nchez-Fern{\' a}ndez, C.\ 
2001, 
\aap, 372, 551 

\bibitem{} Belloni, T., Psaltis, D., \& van der Klis, M.\  
2002a, 
\apj, 572, 392 

\bibitem{} Belloni, T., Colombo, A.~P., Homan, J., Campana, S., \& 
van der Klis, M.\ 
2002b, 
\aap, 390, 199 

\bibitem{} Belloni, T., Homan, J., Nespoli, E., Casella, P., et al.\ 
2004,
\aap, in prep. 

\bibitem{} Benlloch, S., Wilms, J., Edelson, R., Yaqoob, T., \& Staubert, R.\ 
2001, 
\apjl, 562, L121 

\bibitem{} Berger, M.~\& van der Klis, M.\ 
1994, 
\aap, 292, 175 

\bibitem{} Berger, M.~\& van der Klis, M.\ 
1998, 
\aap, 340, 143 

\bibitem{} Berger, M., van der Klis, M., van Paradijs, J., Lewin, W.H.G., Lamb, F., 
et al.\ 
1996,
\apjl, 469, L13 

\bibitem{} Biehle, G.~T.~\& Blandford, R.~D.\ 
1993, 
\apj, 411, 302 

\bibitem{} Bildsten, L.\ 
1993, 
\apjl, 418, L21 

\bibitem{} Bildsten, L.\ 
1995, 
\apj, 438, 852 

\bibitem{} Bildsten, L.\  
1998, 
\apjl, 501, L89 

\bibitem{} Bildsten, L.~\& Cumming, A.\ 
1998, 
\apj, 506, 842 

\bibitem{} Bildsten, L.~\& Cutler, C.\ 
1995, 
\apj, 449, 800 

\bibitem{} Bildsten, L.~\& Ushomirsky, G.\ 
2000, 
\apjl, 529, L33 

\bibitem{} Bildsten, L., Ushomirsky, G., \& Cutler, C.\ 
1996, 
\apj, 460, 827 

\bibitem{} Blom, J.~J., in't Zand, J.~J.~M., Heise, J., Arefev, V., 
Borozdin, K., et al.\ 
1993, 
\aap, 277, 77 

\bibitem{} Bloser, P.~F., Grindlay, J.~E., Kaaret, P., Zhang, W., Smale, 
A.~P., \& Barret, D.\ 
2000a,
\apj, 542, 1000 

\bibitem{} Bloser, P.~F., Grindlay, J.~E., Barret, D., \& Boirin, L.\ 
2000b, 
\apj, 542, 989 

\bibitem{} Boirin, L., Barret, D., Olive, J.~F., Bloser, P.~F., \& Grindlay, 
J.~E.\ 
2000, 
\aap, 361, 121 

\bibitem{} Borozdin, K.~N.~\& Trudolyubov, S.~P.\ 
2000, 
\apjl, 533, L131 

\bibitem{} B\"ottcher, M.\ 
2001, 
\apj, 553, 960 

\bibitem{} B\"ottcher, M.~\& Liang, E.~P.\ 
1998, 
\apj, 506, 281 

\bibitem{} B\"ottcher, M.~\& Liang, E.~P.\ 
1999, 
\apjl, 511, L37 

\bibitem{} Boyd, P.~T., Smale, A.~P., Homan, J., Jonker, P.~G., van der Klis, 
M., et al.\ 
2000, 
\apjl, 542, L127 

\bibitem{} Boyd, P.~T., Smale, A.~P., \& Dolan, J.~F.\ 
2001, 
\apj, 555, 822 

\bibitem{} Boyle, C.~B., Fabian, A.~C., \& Guilbert, P.~W.\ 
1986, 
\nat, 319, 648 

\bibitem{} Bracewell, R.N.\ 
1986,
The Fourier Transform and its Applications, 2nd. ed., McGraw-Hill

\bibitem{} Bradshaw, C.~F., Geldzahler, B.~J., \& Fomalont, E.~B.\ 
2003, 
\apj, 592, 486 

\bibitem{} Bradt, H.~V., Rothschild, R.~E., \& Swank, J.~H.\ 
1993, 
\aaps, 97, 355 

\bibitem{} Brainerd, J.~\& Lamb, F.~K.\ 
1987, 
\apjl, 317, L33 

\bibitem{} Branduardi, G., Kylafis, N.~D., Lamb, D.~Q., \& Mason, K.~O.\ 
1980, 
\apjl, 235, L153 

\bibitem{} Brinkman, A.~C., Parsignault, D.~R., Schreier, E., Gursky, H., 
Kellogg, E.~M., et al.\ 
1974, 
\apj, 188, 603 

\bibitem{} Brocksopp, C., Jonker, P.~G., Fender, R.~P., Groot, P.~J., 
et al.\ 
2001, 
\mnras, 323, 517 

\bibitem{} Brocksopp, C., Fender, R. P., McCollough, M., Pooley, G. G., et al.\ 
2002, 
\mnras, 331, 765 

\bibitem{} Brocksopp, C., Bandyopadhyay, R.~M., \& Fender, R.~P.\ 
2004, 
New Astronomy, 9, 249 

\bibitem{} Brown, E.~F.~\& Ushomirsky, G.\ 
2000, 
\apj, 536, 915 

\bibitem{} Brugmans, F., 
1983,
Earth Surface Processes and Landforms 8, 527 

\bibitem{} Bulik, T., Gondek-Rosi{\' n}ska, D., \& Klu{\' z}niak, W.~{\L}.\ 
1999, 
\aap, 344, L71 

\bibitem{} Bulik, T., Klu{\' z}niak, W. \& Zhang, W.\ 
2000, 
\aap, 361, 153 

\bibitem{} Bussard, R.~W., Weisskopf, M.~C., Elsner, R.~F., \& Shibazaki, N.\ 
1988, 
\apj, 327, 284 

\bibitem{} Campana, S.\ 
2000, 
\apjl, 534, L79 

\bibitem{} Campana, S., Stella, L., Belloni, T., Israel, G.~L., Santangelo, 
A., et al.\ 
2002, 
\aap, 384, 163 

\bibitem{} Canizares, C.~R., Clark, G. W., Li, F. K., Murthy, G. T.,
Bardas, D., et al.\ 
1975, 
\apj, 197, 457 

\bibitem{} Cannizzo, J.~K.\ 
1997, 
\apj, 482, 178 

\bibitem{} Carpenter, G.~F., Eyles, C.~J., Skinner, G.~K., Willmore, A.~P., 
et al.\ 
1976, 
\mnras, 176, 397 


\bibitem{} Chagelishvili, G.~D., Lominadze, J.~G., \& Rogava, A.~D.\ 
1989, 
\apj, 347, 1100 

\bibitem{} Chakrabarti, S.~K.~\& Manickam, S.~G.\ 
2000, 
\apjl, 531, L41 

\bibitem{} Chakrabarty, D., Morgan, E.~H., Muno, M.~P., Galloway, D.~K., 
et al.\ 
2003, 
\nat, 424, 42 

\bibitem{} Chaput, C., Bloom, E., Cominsky, L., Godfrey, G., Hertz,
P., Scargle, J., et al.\ 
2000, 
\apj, 541, 1026 

\bibitem{} Chen, X.~\& Taam, R.~E.\ 
1994, 
\apj, 431, 732 

\bibitem{} Chen, X.~\& Taam, R.~E.\ 
1995, 
\apj, 441, 354 

\bibitem{} Chen, X., Swank, J.~H., \& Taam, R.~E.\ 
1997, 
\apjl, 477, L41 

\bibitem{} Chiappetti, L., et al.\ 
1990, 
\apj, 361, 596 

\bibitem{} Churazov, E., Gilfanov, M., \& Revnivtsev, M.\ 
2001, 
\mnras, 321, 759 

\bibitem{} Corbet, R.~H.~D., Smale, A. P., Charles, P. A., Lewin,
W. H. G., et al.\ 
1989,
\mnras, 239, 533 

\bibitem{} Cowley, A.~P., et al.\ 
1991, 
\apj, 381, 526 

\bibitem{} Crary, D.~J., Kouveliotou, C., van Paradijs, J., van der Hooft, F., et al.\ 
1996, 
\apjl, 462, L71 

\bibitem{} Cropper, M., Soria, R., Mushotzky, R.~F., Wu, K.,Markwardt, C.~B., et al.\ 
2004, 
\mnras, 349, 39 

\bibitem{} Cui, W.\ 
1999, 
\apjl, 524, L59 

\bibitem{} Cui, W.\ 
2000, 
\apjl, 534, L31 

\bibitem{} Cui, W., Heindl, W.~A., Rothschild, R.~E., Zhang, S.~N., Jahoda, 
K., \& Focke, W.\ 
1997a, 
\apjl, 474, L57 

\bibitem{} Cui, W., Zhang, S.~N., Focke, W., \& Swank, J.~H.\ 
1997b, 
\apj, 484, 383 

\bibitem{} Cui, W., Heindl, W.~A., Swank, J.~H., Smith, D.~M., Morgan, E.~H., 
et al.\ 
1997c, 
\apjl, 487, L73 

\bibitem{} Cui, W., Barret, D., Zhang, S.N., Chen, W., Boirin, L., Swank, J.\ 
1998a,
\apjl, 502, L49 

\bibitem{} Cui, W., Zhang, S.~N., \& Chen, W.\ 
1998b, 
\apjl, 492, L53 

\bibitem{} Cui, W., Zhang, S.~N., Chen, W., \& Morgan, E.~H.\ 
1999, 
\apjl, 512, L43 

\bibitem{} Cui, W., Zhang, S.~N., \& Chen, W.\ 
2000a, 
\apjl, 531, L45 

\bibitem{} Cui, W., Shrader, C.~R., Haswell, C.~A., \& Hynes, R.~I.\ 
2000b, 
\apjl, 535, L123 

\bibitem{} Cullen, J.\ 
2000, 
Publications of the Astronomical Society of Australia, 17, 48 

\bibitem{} Cunningham, C.~T.~\& Bardeen, J.~M.\ 
1972, 
\apjl, 173, L137 

\bibitem{} Czerny, B., Niko{\l}ajuk, M., Piasecki, M., \& Kuraszkiewicz, J.\ 
2001, 
\mnras, 325, 865 

\bibitem{} dal Fiume, D., Robba, N.~R., Frontera, F., \& Stella, L.\ 
1990, 
Nuovo Cimento C, 13, 463 

\bibitem{} Damen, E., Wijers, R.~A.~M.~J., van Paradijs, J., Penninx,
W., et al.\ 
1990, 
\aap, 233, 121 

\bibitem{} Das, T.~K., Rao, A.~R., \& Vadawale, S.~V.\ 
2003, 
\mnras, 343, 443 

\bibitem{} Datta, B., Thampan, A.V., \& Bombaci, I.\ 
1998, 
\aap, 334, 943 

\bibitem{} Datta, B., Thampan, A.~V., \& Bombaci, I.\ 
2000, 
\aap, 355, L19 

\bibitem{} Deeter, J.~E.\ 
1984, 
\apj, 281, 482 

\bibitem{} DeDeo, S.~\& Psaltis, D.\ 
2004, 
astro-ph/0405067 

\bibitem{} den Hartog, P.~R., in't Zand, J. J. M., Kuulkers, E., Cornelisse, R., Heise, J., et al.\
2003, 
\aap, 400, 633 

\bibitem{} Denis, M., Olive, J.-F., Mandrou, P., Roques, J. P.,
Ballet, J., Goldwurm, A., et al.\ 
1994, 
\apjs, 92, 459 

\bibitem{} Dieters, S.~W.~\& van der Klis, M.\ 
2000, 
\mnras, 311, 201 

\bibitem{} Dieters, S.~W., Vaughan, B.~A., Kuulkers, E., Lamb, F.~K., \& 
van der Klis, M.\ 
2000a, 
\aap, 353, 203 

\bibitem{} Dieters, S.~W., Belloni, T., Kuulkers, E., Woods, P., Cui,
W., Zhang, S. N, et al.\ 
2000b, 
\apj, 538, 307 

\bibitem{} di Matteo, T.~\& Psaltis, D.\ 
1999, 
\apjl, 526, L101 

\bibitem{} di Matteo, T., Celotti, A., \& Fabian, A.~C.\ 
1999, 
\mnras, 304, 809 

\bibitem{} Ding, G.~Q., Qu, J.~L., \& Li, T.~P.\ 
2003, 
\apjl, 596, L219 

\bibitem{} Di Salvo, T., \& Stella, L.\ 
2002,
astro-ph/0207219 

\bibitem{} Di Salvo, T., Stella, L., Robba, N. R., van der Klis, M.,
Burderi, L., et al.\ 
2000, 
\apjl, 544, L119 

\bibitem{} Di Salvo, T., M{\'e}ndez, M., van der Klis, M., Ford, E., 
\& Robba, N.~R.\ 
2001a, 
\apj, 546, 1107 

\bibitem{} Di Salvo, T., Robba, N.~R., Iaria, R., Stella, L., Burderi, L., \& 
Israel, G.~L.\ 
2001b,
\apj, 554, 49 

\bibitem{} Di Salvo, T., Farinelli, R., Burderi, L., Frontera, F.,
Kuulkers, E., et al.\ 
2002, 
\aap, 386, 535 

\bibitem{} Di Salvo, T., M\'endez, M., \& van der Klis, M.\ 
2003, 
\aap, 406, 177 

\bibitem{} Doi, K.\ 
1978, 
\nat, 275, 197 

\bibitem{} Done, C.~\& Gierli{\' n}ski, M.\ 
2003, 
\mnras, 342, 1041 

\bibitem{} Done, C., {\. Z}ycki, P.~T., \& Smith, D.~A.\ 
2002, 
\mnras, 331, 453 

\bibitem{} Dotani, T., Mitsuda, K., Makishima, K., \& Jones, M.~H.\ 
1989, 
\pasj, 41, 577 

\bibitem{} Dotani, T., Mitsuda, K., Inoue, H., Tanaka, Y., Kawai, N.,
Tawara, Y., et al.\ 
1990, 
\apj, 350, 395 

\bibitem{} Dubus, G., Kern, B., Esin, A.~A., Rutledge, R.~E., \& 
Martin, C.\ 
2003, 
astro-ph/0310615 

\bibitem{} Ebisawa, K., Mitsuda, K., \& Inoue, H.\ 
1989, 
\pasj, 41, 519 

\bibitem{} Ebisawa, K., Ogawa, M., Aoki, T., Dotani, T., Takizawa, M.,
Tanaka, Y., et al.\ 
1994, 
\pasj, 46, 375 

\bibitem{} Edelson, R.~\& Nandra, K.\ 
1999, 
\apj, 514, 682 

\bibitem{} Elsner, R.~F., Weisskopf, M.~C., Darbro, W., Ramsey, B.~D., 
Williams, A.~C., et al.\ 
1986, 
\apj, 308, 655 

\bibitem{} Elsner, R.~F., Shibazaki, N., \& Weisskopf, M.~C.\ 
1987, 
\apj, 320, 527 

\bibitem{} Elsner, R.~F., Shibazaki, N., \& Weisskopf, M.~C.\ 
1988, 
\apj, 327, 742 

\bibitem{} Epstein, R.~I.\ 
1988, 
\apj, 333, 880 

\bibitem{} Esin, A.~A., McClintock, J.~E., \& Narayan, R.\ 
1997, 
\apj, 489, 865 

\bibitem{} Esin, A.~A., Kuulkers, E., McClintock, J.~E., \& Narayan, R.\ 
2000, 
\apj, 532, 1069 

\bibitem{} Esin, A.~A., McClintock, J.~E., Drake, J.~J., Garcia, M.~R., 
Haswell, C.~A., et al.\ 
2001, 
\apj, 555, 483 

\bibitem{} Fabian, A.~C., Iwasawa, K., Reynolds, C.~S., \& Young, A.~J.\ 
2000, 
\pasp, 112, 1145 

\bibitem{} Farinelli, R., Frontera, F., Masetti, N., Amati, L.,
Guidorzi, C., et al.\ 
2003, 
\aap, 402, 1021 

\bibitem{} Feng, Y.~X., Li, T.~P., \& Chen, L.\ 
1999, 
\apj, 514, 373 

\bibitem{} Feroci, M., Matt, G., Pooley, G., Costa, E., Tavani, M., \& 
Belloni, T.\ 
1999, 
\aap, 351, 985 

\bibitem{} Finger, M.~H., Wilson, R.~B., \& Harmon, B.~A.\ 
1996, 
\apj, 459, 288 

\bibitem{} Focke, W.~B.\ 
1996, 
\apjl, 470, L127 

\bibitem{} Ford, E.C., \& van der Klis, M.\ 
1998,
\apjl, 506, L39 

\bibitem{} Ford, E., Kaaret, P., Tavani, M., Barret, D., Bloser, P.,
Grindlay, J., et al.\ 
1997a, 
\apjl, 475, L123 

\bibitem{} Ford, E.C., Kaaret, P., Chen, K., Tavani, M., Barret,
D., Bloser, P., 
et al.\ 
1997b,
\apjl, 486, L47 

\bibitem{} Ford, E.C., van der Klis, M., \& Kaaret, P.\ 
1998a
\apjl, 498 L41 

\bibitem{} Ford, E.C., van der Klis, M., van Paradijs, J., M\'endez,
M., Wijands, R., et al.\   
1998b,
\apjl, 508, L155 

\bibitem{} Ford, E.~C., van der Klis, M., M{\' e}ndez, M., van Paradijs, J., 
\& Kaaret, P.\ 
1999, 
\apjl, 512, L31 

\bibitem{} Ford, E.~C., van der Klis, M., M{\' e}ndez, M., Wijnands, R., 
Homan, J., et al.\ 
2000, 
\apj, 537, 368 

\bibitem{} Fortner, B., Lamb, F.~K., \& Miller, G.~S.\ 
1989, 
\nat, 342, 775 

\bibitem{} Fox, D.~W., Lewin, W.H.G., Rutledge, R.E., Morgan, E.H.,
Guerriero, R., et al.\ 
2001, 
\mnras, 321, 776 

\bibitem{} Fragile, P.~C., Mathews, G.~J., \& Wilson, J.~R.\ 
2001, 
\apj, 553, 955 

\bibitem{} Franco, L.~M.\ 
2001, 
\apj, 554, 340 

\bibitem{} Frontera, F., dal Fiume, D., Robba, N.~R., Manzo, G., Re, S., \& 
Costa, E.\ 
1987, 
\apjl, 320, L127 

\bibitem{} Frontera, F., Palazzi, E., Zdziarski, A. A., Haardt, F.,
Perola, G. C., et al.\ 
2001, 
\apj, 546, 1027 

\bibitem{} Frontera, F., Amati, L., Zdziarski, A.~A., Belloni, T., Del Sordo, 
S., et al.\ 
2003, 
\apj, 592, 1110 

\bibitem{} Fukue, J.~\& Okada, R.\ 
1990, 
\pasj, 42, 533 

\bibitem{} Galeev, A.~A., Rosner, R., \& Vaiana, G.~S.\ 
1979, 
\apj, 229, 318 

\bibitem{} Galloway, D.~K., Chakrabarty, D., Muno, M.~P., \& Savov,
P.\ 
2001, 
\apjl, 549, L85 

\bibitem{} Galloway, D.~K., Chakrabarty, D., Morgan, E.~H., \& Remillard, 
R.~A.\ 
2002, 
\apjl, 576, L137 

\bibitem{} Gierli{\' n}ski, M.~\& Done, C.\ 
2002a, 
\mnras, 337, 1373 

\bibitem{} Gierli{\' n}ski, M.~\& Done, C.\ 
2002b, 
\mnras, 331, L47 

\bibitem{} Gierli{\' n}ski, M.~\& Done, C.\ 
2003, 
\mnras, 342, 1083 

\bibitem{} Gierli{\' n}ski, M.~\& Zdziarski, A.~A.\ 
2003, 
\mnras, 343, L84 

\bibitem{} Gierli{\' n}ski, M., Zdziarski, A.~A., Poutanen, J., Coppi, P.~S., 
et al.\ 
1999, 
\mnras, 309, 496 

\bibitem{} Gierli{\' n}ski, M., Done, C., \& Barret, D.\ 
2002, 
\mnras, 331, 141 

\bibitem{} Giles, A.~B.\ 
1981, 
\mnras, 195, 721 

\bibitem{} Gilfanov, M., Churazov, E., \& Revnivtsev, M.\ 
1999, 
\aap, 352, 182 

\bibitem{} Gilfanov, M., Churazov, E., \& Revnivtsev, M.\ 
2000, 
\mnras, 316, 923 

\bibitem{} Gilfanov, M., Revnivtsev, M., \& Molkov, S.\ 
2003, 
\aap, 410, 217 

\bibitem{} Gilfanov, M., Churazov, E., \& Revnivtsev, M.\ 
2004, 
in Rossi and beyond, AIP Conf. Proc. 714, 97; 
astro-ph/0312445 

\bibitem{} Gleissner, T., Wilms, J., Pottschmidt, K., Uttley, P., Nowak, 
M.~A., et al.\ 
2004, 
\aap, 414, 1091 

\bibitem{} Glendenning, N.~K.~\& Weber, F.\ 
2001, 
\apjl, 559, L119 

\bibitem{} Gnedin, Y.~N.~\& Kiikov, S.~O.\ 
2001, 
Astronomy Letters, 27, 507 

\bibitem{} Gondek-Rosi{\' n}ska, D., Stergioulas, N., Bulik, T., 
Klu{\' z}niak, W., \& Gourgoulhon, E.\ 
2001, 
\aap, 380, 190 

\bibitem{} Grebenev, S.~A., Syunyaev, R.~A., Pavlinskii, M.~N., et al.\ 
1991, 
Soviet Astronomy Letters, 17, 413 

\bibitem{} Grindlay, J.~E., Band, D., Seward, F., Leahy, D., Weisskopf, 
M.~C., et al.\ 
1984, 
\apj, 277, 286 

\bibitem{} Grove, J.~E., Strickman, M.~S., Matz, S.~M., Hua, X.-M., 
Kazanas, D.,et al.\ 
1998, 
\apjl, 502, L45 

\bibitem{} Gruzinov, A.\  
1999,
astro-ph/9910335

\bibitem{} Haardt, F., Maraschi, L., \& Ghisellini, G.\ 
1994, 
\apjl, 432, L95 

\bibitem{} Haardt, F., Galli, M. R., Treves, A., Chiappetti, L., Dal
Fiume, D., et al.\ 
2001, 
\apjs, 133, 187 

\bibitem{} Halpern, J.~P., Leighly, K.~M., \& Marshall, H.~L.\ 
2003, 
\apj, 585, 665 

\bibitem{} Hameury, J.-M., King, A.~R., \& Lasota, J.-P.\ 
1985, 
\nat, 317, 597 

\bibitem{} Hanami, H.\ 
1988, 
\mnras, 233, 423 

\bibitem{} Hanawa, T., Hirotani, K., \& Kawai, N.\ 
1989, 
\apj, 336, 920 

\bibitem{} Hartle, J.~B.~\& Thorne, K.~S.\ 
1968, 
\apj, 153, 807 

\bibitem{} Hasinger, G.\ 
1987, 
in The origin and evolution of neutron stars, IAU Symp. 125, 333 

\bibitem{} Hasinger, G.\ 
1988, 
in Physics of Neutron Stars and Black Holes, Tokyo, Tanaka (ed.), 
p. 97 

\bibitem{} Hasinger, G., \& van der Klis, M.\ 
1989,
\aap, 225, 79 

\bibitem{} Hasinger, G.,  Langmeier, A., Sztajno, M., Truemper, J., \& 
Lewin, W.~H.~G.\ 
1986, 
\nat, 319, 469 

\bibitem{} Hasinger, G., Priedhorsky, W.~C., \& Middleditch, J.\ 
1989, 
\apj, 337, 843 

\bibitem{} Hasinger, G., van der Klis, M., Ebisawa, K., Dotani, T., \& 
Mitsuda, K.\ 
1990, 
\aap, 235, 131 

\bibitem{} Hawley, J.~F.~\& Krolik, J.~H.\ 
2001, 
\apj, 548, 348 

\bibitem{} Hawley, J.~F.~\& Krolik, J.~H.\ 
2002, 
\apj, 566, 164 

\bibitem{} Heiselberg, H., \& Hjorth-Jensen, M.\ 
1999,
\apjl, 525, L45 

\bibitem{} Hertz, P., Vaughan, B., Wood, K.~S., Norris, J.~P., Mitsuda, K., 
et al.\ 
1992, 
\apj, 396, 201 

\bibitem{} Hirano, A., Kitamoto, S., Yamada, T.~T., Mineshige, S., \& 
Fukue, J.\ 
1995, 
\apj, 446, 350 

\bibitem{} Hjellming, R.~M., Han, X.~H., Cordova, F.~A., \& Hasinger, G.\ 
1990a, 
\aap, 235, 147 

\bibitem{} Hjellming, R.~M., Stewart, R. T., White, G. L., Strom, R.,
Lewin, W. H. G., et al.\ 
1990b, 
\apj, 365, 681 

\bibitem{} Homan, J.~\& van der Klis, M.\ 
2000, 
\apj, 539, 847 

\bibitem{} Homan, J., van der Klis, M., Wijnands, R., Vaughan, B., \& 
Kuulkers, E.\ 
1998, 
\apjl, 499, L41 

\bibitem{} Homan, J., M\'endez, M., Wijnands, R., van der Klis, M., \& van Paradijs, J.\ 
1999a,
\apjl, 513, L119 

\bibitem{} Homan, J., Jonker, P.~G., Wijnands, R., van der Klis, M., \& 
van Paradijs, J.\ 
1999b, 
\apjl, 516, L91 

\bibitem{} Homan, J., Wijnands, R., van der Klis, M., Belloni, T., 
van Paradijs, J.,et al.\ 
2001, 
\apjs, 132, 377 

\bibitem{} Homan, J., van der Klis, M., Jonker, P.~G., Wijnands, R., 
Kuulkers, E., et al.\ 
2002, 
\apj, 568, 878 

\bibitem{} Homan, J., Miller, J.~M., Wijnands, R., Steeghs, D., Belloni, T., 
et al.\ 
2003a, 
ATel, 162, 1 

\bibitem{} Homan, J., Klein-Wolt, M., Rossi, S., Miller, J.~M., Wijnands, R., 
et al.\ 
2003b, 
\apj, 586, 1262 

\bibitem{} Homan, J., Wijnands, R., Rupen, M.~P., Fender, R., Hjellming, R.~M., et al.\  
2004, 
\aap, 418, 255 

\bibitem{} Hoshino, M.~\& Takeshima, T.\ 
1993, 
\apjl, 411, L79 

\bibitem{} Hua, X.~\& Titarchuk, L.\ 
1996, 
\apj, 469, 280 

\bibitem{} Hua, X., Kazanas, D., \& Titarchuk, L.\ 
1997, 
\apjl, 482, L57 

\bibitem{} Hua, X., Kazanas, D., \& Cui, W.\ 
1999, 
\apj, 512, 793 

\bibitem{} Hynes, R.~I., O'Brien, K., Horne, K., Chen, W., \& Haswell, C.~A.\ 
1998, 
\mnras, 299, L37 

\bibitem{} Hynes, R.~I., Charles, P.~A., Casares, J., Haswell, C.~A., 
et al.\ 
2003a, 
\mnras, 340, 447 

\bibitem{} Hynes, R.~I., Haswell, C. A., Cui, W., Shrader, C. R.,
O'Brien, K., et al.\ 
2003b, 
\mnras, 345, 292 

\bibitem{} Iaria, R., Di Salvo, T., Robba, N.~R., Burderi, L., Stella, L., 
Frontera, F., et al.\ 
2004, 
\apj, 600, 358 

\bibitem{} Ikegami, T.\ 
1986,
PhD thesis, University of Tokyo \editnote{=== check year} 

\bibitem{} Ilovaisky, S.~A., Chevalier, C., White, N.~E., Mason, K.~O., 
Sanford, P.~W.,et al.\ 
1980, 
\mnras, 191, 81 

\bibitem{} Imamura, J.~N., Kristian, J., Middleditch, J., \& Steiman-Cameron, 
T.~Y.\ 
1990, 
\apj, 365, 312 

\bibitem{} Inogamov, N.~A.~\& Sunyaev, R.~A.\ 
1999, 
Astronomy Letters, 25, 269 

\bibitem{} in 't Zand, J.~J.~M., Verbunt, F., Strohmayer, T. E.,
Bazzano, A., et al.\ 
1999a, 
\aap, 345, 100 

\bibitem{} in 't Zand, J.~J.~M., Heise, J., Kuulkers, E., Bazzano, A., 
Cocchi, M., et al.\ 
1999b, 
\aap, 347, 891 

\bibitem{} in 't Zand, J.~J.~M., Bazzano, A., Cocchi, M., Cornelisse,
R., Heise, J., et al.\ 
2000, 
\aap, 355, 145 

\bibitem{} in 't Zand, J.~J.~M., Cornelisse, R., Kuulkers, E., Heise,
J., Kuiper, L., et al.\ 
2001, 
\aap, 372, 916 

\bibitem{} in 't Zand, J.~J.~M., Markwardt, C. B., Bazzano, A., Cocchi, M., et al.\ 
2002, 
\aap, 390, 597 

\bibitem{} Ipser, J.~R.\ 
1996, 
\apj, 458, 508 

\bibitem{} Israel, G.~L., Mereghetti, S., \& Stella, L.\ 
1994, 
\apjl, 433, L25 

\bibitem{} Jernigan, J.~G., Klein, R.~I., \& Arons, J.\ 
2000, 
\apj, 530, 875 

\bibitem{} Ji, J.~F., Zhang, S.~N., Qu, J.~L., \& Li, T.\ 
2003, 
\apjl, 584, L23 

\bibitem{} Jones, C., Giacconi, R., Forman, W., \& Tananbaum, H.\ 
1974, 
\apjl, 191, L71 

\bibitem{} Jonker, P.G., Wijnands, R., van der Klis, M., Psaltis, D., Kuulkers, E., 
et al.\   
1998,
\apj, 499, L191 

\bibitem{} Jonker, P.~G., van der Klis, M., \& Wijnands, R.\ 
1999, 
\apjl, 511, L41 

\bibitem{} Jonker, P.G., van der Klis, M., Wijnands, R., Homan, J., van Paradijs, J., 
et al.\  
2000a, 
\apj, 537, 374 

\bibitem{} Jonker, P.~G., M{\' e}ndez, M., \& van der Klis, M.\ 
2000b, 
\apjl, 540, L29 

\bibitem{} Jonker, P.~G., van der Klis, M., Homan, J., Wijnands, R., 
van Paradijs, J., et al.\ 
2000c, 
\apj, 531, 453 

\bibitem{} Jonker, P.~G., van der Klis, M., Homan, J., M\'endez, M., van Paradijs, J., et al.\ 
2001, 
\apj, 553, 335 

\bibitem{} Jonker, P.~G., van der Klis, M., Homan, J., M{\' e}ndez, M., 
et al.\ 
2002a, 
\mnras, 333, 665 

\bibitem{} Jonker, P.~G., M{\' e}ndez, M., \& van der Klis, M.\ 
2002b, 
\mnras, 336, L1 

\bibitem{} Jonker, P.~G., M{\' e}ndez, M., Nelemans, G., Wijnands, R., \& 
van der Klis, M.\ 
2003, 
\mnras, 341, 823 

\bibitem{} Kaaret, P.\
2000,
in Stellar endpoints, AIP-Conf. Proc. 599, 406;
astro-ph/0008424 

\bibitem{} Kaaret, P., Ford, E., \& Chen, K.\ 
1997,
\apjl, 480, L27 

\bibitem{} Kaaret, P., Yu, W., Ford, E.C., \& Zhang, S.N.\ 
1998,
\apjl, 497, L93 

\bibitem{} Kaaret, P., Piraino, S., Ford, E.C., \& Santangelo, A.\ 
1999a,
\apjl, 514, L31 

\bibitem{} Kaaret, P., Piraino, S., Bloser, P.F., Ford, E.C., Grindlay, J.E., et al.\ 
1999b,
\apjl, 520, L37 

\bibitem{} Kaaret, P., Zand, J.~J.~M.~i., Heise, J., \& Tomsick, J.~A.\ 
2002, 
\apj, 575, 1018 

\bibitem{} Kaaret, P., Zand, J.~J.~M.~i., Heise, J., \& Tomsick, J.~A.\ 
2003, \
\apj, 598, 481 

\bibitem{} Kahn, S.~M.~\& Blissett, R.~J.\ 
1980, 
\apj, 238, 417 

\bibitem{} Kalemci, E., Tomsick, J.~A., Rothschild, R.~E., Pottschmidt, K., 
\& Kaaret, P.\ 
2001, 
\apj, 563, 239 

\bibitem{} Kalemci, E., Tomsick, J.~A., Rothschild, R.~E., Pottschmidt, K., 
Corbel, S., et al.\ 
2003, 
\apj, 586, 419 

\bibitem{} Kalemci, E., Tomsick, J.~A., Rothschild, R.~E., Pottschmidt, K., \& 
Kaaret, P.\ 
2004, 
\apj, 603, 231 

\bibitem{} Kallman, T., Boroson, B., \& Vrtilek, S.~D.\ 
1998, 
\apj, 502, 441 

\bibitem{} Kalogera, V.~\&  Psaltis, D.\ 
2000, 
\prd, 61, 024009 

\bibitem{} Kamado, Y., Kitamoto, S., \& Miyamoto, S.\ 
1997, 
\pasj, 49, 589 

\bibitem{} Kanbach, G., Straubmeier, C., Spruit, H.~C., \& Belloni, T.\ 
2001, 
\nat, 414, 180 

\bibitem{} Karas, V.\ 
1999a, 
\apj, 526, 953 

\bibitem{} Karas, V.\ 
1999b, 
\pasj, 51, 317 

\bibitem{} Kato, S.\ 
1989, 
\pasj, 41, 745 

\bibitem{} Kato, S.\ 
1990, 
\pasj, 42, 99 

\bibitem{} Kato, S.\ 
2001, 
\pasj, 53, L37 

\bibitem{} Kato, S., Fukue, J., \& Mineshige, S.\ 
1998, 
in Black-hole accretion disks, Kyoto University Press 

\bibitem{} Kato, Y., Hayashi, M.~R., Miyaji, S., \& Matsumoto, R.\ 
2001, 
Advances in Space Research, 28, 505 

\bibitem{} Kawaguchi, T., Mineshige, S., Machida, M., Matsumoto, R., \& 
Shibata, K.\ 
2000, 
\pasj, 52, L1 

\bibitem{} Kawai, N., Matsuoka, M., Inoue, H., Ogawara, Y., Tanaka, Y., 
Kunieda, H., et al.\ 
1990, 
\pasj, 42, 115 

\bibitem{} Kazanas, D.~\& Hua, X.\ 
1999, 
\apj, 519, 750 

\bibitem{} Kazanas, D., Hua, X., \& Titarchuk, L.\  
1997, 
\apj, 480, 735 

\bibitem{} King, A.~R., Pringle, J.~E., West, R.~G., \& Livio, M.\ 
2004, 
\mnras, 348, 111 

\bibitem{} Kitamoto, S., Tsunemi, H., \& Roussel-Dupre, D.\ 
1992, 
\apj, 391, 220 

\bibitem{} Klein, R.~I., Arons, J., Jernigan, G., \& Hsu, J.~J.-L.\ 
1996a, 
\apjl, 457, L85 

\bibitem{} Klein, R.~I., Jernigan, J.~G., Arons, J., Morgan, E.~H., \& Zhang, W.\ 
1996b, 
\apjl, 469, L119 

\bibitem{} Klein-Wolt, M., Homan, J. \& van der Klis, M.\ 
2004a,
PhD thesis, Univ. of Amsterdam, p. 113 

\bibitem{} Klein-Wolt, M., van Straaten, S. \& van der Klis, M.\ 
2004b,
PhD thesis, Univ. of Amsterdam, p. 165 

\bibitem{} Klu\'zniak, W.\ 
1998,
\apjl, 509, L37 

\bibitem{} Klu\'zniak, W.\& Abramowicz, M.~A.\ 
2001, 
astro-ph/0105057 

\bibitem{} Klu\'zniak, W.\& Abramowicz, M.~A.\ 
2002, 
astro-ph/0203314 

\bibitem{} Klu\'zniak, W.\& Abramowicz, M.~A.\ 
2003, 
astro-ph/0304345 

\bibitem{} Klu\'zniak, W.~\& Wagoner, R.~V.\ 
1985, 
\apj, 297, 548 

\bibitem{} Klu\'zniak, W., Michelson, P., \& Wagoner, R.~V.\ 
1990, 
\apj, 358, 538 

\bibitem{} Klu\'zniak, W., Abramowicz, M.~A., Kato, S., Lee, W.~H., \& 
Stergioulas, N.\ 
2004, 
\apjl, 603, L89 

\bibitem{} Kommers, J.~M., Fox, D.~W., Lewin, W.~H.~G., Rutledge, R.~E., 
et al.\ 
1997, 
\apjl, 482, L53 

\bibitem{} Kommers, J.~M., Chakrabarty, D., \& Lewin, W.~H.~G.\ 
1998, 
\apjl, 497, L33 

\bibitem{} Kong, A.~K.~H., Charles, P.~A., Kuulkers, E., \& Kitamoto, S.\ 
2002, 
\mnras, 329, 588 

\bibitem{} K\"ording, E.~\& Falcke, H.\ 
2004, 
\aap, 414, 795 

\bibitem{} Kotov, O., Churazov, E., \& Gilfanov, M.\ 
2001, 
\mnras, 327, 799 

\bibitem{} Krishan, V., Ramadurai, S., \& Wiita, P.~J.\ 
2003, 
\aap, 398, 819 

\bibitem{} Krolik, J.~H.~\& Hawley, J.~F.\ 
2002, 
\apj, 573, 754 

\bibitem{} Kubota, A.~\& Done, C.\ 
2004, 
\mnras, 353, 980  

\bibitem{} Kubota, A.~\& Makishima, K.\ 
2004, 
\apj, 601, 428 

\bibitem{} Kuulkers, E.\ 
1995,
PhD thesis, Univ. of Amsterdam 

\bibitem{} Kuulkers, E.\ 
1998, 
New Astronomy Review, 
42, 1 

\bibitem{} Kuulkers, E.~\& van der Klis, M.\ 
1995, 
\aap, 303, 801 

\bibitem{} Kuulkers, E., \& van der Klis, M.\ 
1998,
\aap, 332, 845 
 
\bibitem{} Kuulkers, E., van der Klis, M., Oosterbroek, T., Asai, K., 
Dotani, T., et al.\ 
1994, 
\aap, 289, 795 

\bibitem{} Kuulkers, E., van der Klis, M., \& van Paradijs, J.\ 
1995, 
\apj, 450, 748 

\bibitem{} Kuulkers, E., van der Klis, M., \& Vaughan, B.~A.\ 
1996, 
\aap, 311, 197 

\bibitem{} Kuulkers, E., van der Klis, M., Oosterbroek, T., et al.\ 
1997a, 
\mnras, 287, 495 

\bibitem{} Kuulkers, E., van der Klis, M., \& Parmar, A.~N.\ 
1997b, 
\apjl, 474, L47 

\bibitem{} Kuulkers, E., Wijnands, R., \& van der Klis, M.\ 
1999, 
\mnras, 308, 485  

\bibitem{} Kuulkers, E., Homan, J., van der Klis, M., Lewin, W.~H.~G., \& 
M{\' e}ndez, M.\ 
2002, 
\aap, 382, 947 

\bibitem{} Kuznetsov, S.~I.\ 
2001, 
Astronomy Letters, 27, 790 

\bibitem{} Kuznetsov, S.~I.\ 
2002a, 
Astronomy Letters, 28, 73 

\bibitem{} Kuznetsov, S.~I.\ 
2002b, 
Astronomy Letters, 28, 811 

\bibitem{} Kylafis, N.~D.~\& Klimis, G.~S.\ 
1987, 
\apj, 323, 678 

\bibitem{} Lai, D.\ 
1998, 
\apj, 502, 721 

\bibitem{} Lai, D.\ 
1999, 
\apj, 524, 1030 

\bibitem{} Lamb, F.~K.\ 
2003,
in Jan van Paradijs memorial meeting, ASP Conf. Series 308, 221 

\bibitem{} Lamb, F.~K.~\& Miller, M.~C.\ 
2001, 
\apj, 554, 1210 

\bibitem{} Lamb, F.~K.~\& Miller, M.~C.\ 
2003, 
astro-ph/0308179 

\bibitem{} Lamb, F.~K., Shibazaki, N., Alpar, M.~A., \& Shaham, J.\ 
1985, 
\nat, 317, 681 

\bibitem{} Langmeier, A., Sztajno, M., Hasinger, G., Truemper, J., \& Gottwald, M.\ 
1987, 
\apj, 323, 288 

\bibitem{} Langmeier, A., Hasinger, G., \& Truemper, J.\ 
1989, 
\apjl, 340, L21 

\bibitem{} Langmeier, A., Hasinger, G., \& Truemper, J.\ 
1990, 
\aap, 228, 89 

\bibitem{} Laurent, P.~\& Titarchuk, L.\ 
2001, 
\apjl, 562, L67 

\bibitem{} Lazzati, D.~\& Stella, L.\ 
1997, 
\apj, 476, 267 

\bibitem{} Lee, H.~C.~\& Miller, G.~S.\ 
1998, 
\mnras, 299, 479 

\bibitem{} Lee, H.~C., Misra, R., \& Taam, R.~E.\ 
2001, 
\apjl, 549, L229 

\bibitem{} Lee, W.~H., Abramowicz, M.~A., \& Klu{\' z}niak, W.\ 
2004, 
\apjl, 603, L93 

\bibitem{} Lense, J., \& Thirring, H.\ 
1918,
Phys. Z. 19, 156

\bibitem{} Levin, Y.\ 
1999,
\apj, 517, 328 

\bibitem{} Lewin, W.~H.~G., et al.\ 
1987, 
\mnras, 226, 383 

\bibitem{} Lewin, W.~H.~G., van Paradijs, J., \& van der Klis, M.\ 
1988, 
Space Science Reviews, 46, 273 

\bibitem{} Lewin, W.~H.~G., Lubin, L.~M., Tan, J., van der Klis, M., 
et al.\ 
1992, 
\mnras, 256, 545 

\bibitem{} Lewin, W.~H.~G., van Paradijs, J., \& Taam, R.E.\ 
1993, 
Space Science Reviews, 62, 223 

\bibitem{} Lewin, W.H.G., van Paradijs, J., \& van den Heuvel, E.P.J.\ 
1995, 
X-ray binaries, CUP

\bibitem{} Lewin, W.~H.~G., Rutledge, R.~E., Kommers, J.~M., van Paradijs, 
J., et al.\ 
1996, 
\apjl, 462, L39 

\bibitem{} Li, F.~K., McClintock, J.~E., Rappaport, S., Wright, E.~L., \& 
Joss, P.~C.\ 
1980, 
\apj, 240, 628 

\bibitem{} Li, X., Ray, S., Dey, J., Dey, M., \& Bombaci, I.\ 
1999, 
\apjl, 527, L51 

\bibitem{} Li, T.~P.~\& Muraki, Y.\ 
2002, 
\apj, 578, 374 

\bibitem{} Li, L.~\& Narayan, R.\ 
2004, 
\apj, 601, 414 

\bibitem{} Liang, E.~P.~T.~\& Thompson, K.~A.\ 
1980, 
\apj, 240, 271 

\bibitem{} Lin, D., Smith, I.~A., B\"ottcher, M., \& Liang, E.~P.\ 
2000a, 
\apj, 531, 963 

\bibitem{}Lin, D., Smith, I. A., Liang, E. P., Bridgman, T., Smith,
D. M., et al.\ 
2000b, 
\apj, 532, 548 

\bibitem{} Lin, D., Smith, I.~A., Liang, E.~P., \& B\"ottcher, M.\ 
2000c, 
\apjl, 543, L141 

\bibitem{} Ling, J.~C., Mahoney, W.~A., Wheaton, W.~A., Jacobson, A.~S., \& 
Kaluzienski, L.\ 
1983, 
\apj, 275, 307 

\bibitem{} Lochner, J.~C., Swank, J.~H., \& Szymkowiak, A.~E.\ 
1989, 
\apj, 337, 823 

\bibitem{} Lochner, J.~C., Swank, J.~H., \& Szymkowiak, A.~E.\ 
1991,
\apj, 376, 295 

\bibitem{} Lubin, L.~M., Lewin, W.~H.~G., Tan, J., Stella, L., \& 
van Paradijs, J.\ 
1991, 
\mnras, 249, 300 

\bibitem{} Lubin, L.~M., Lewin, W.~H.~G., Dotani, T., Oosterbroek, T., 
et al.\ 
1992a, 
\mnras, 256, 624 

\bibitem{} Lubin, L.~M., Lewin, W.~H.~G., Rutledge, R.~E., van Paradijs, J., 
et al.\ 
1992b, 
\mnras, 258, 759 

\bibitem{} Lubin, L.~M., Lewin, W.~H.~G., van Paradijs, J., \& van der Klis, 
M.\ 
1993, 
\mnras, 261, 149 

\bibitem{} Luo, C.~\& Liang, E.~P.\ 
1994, 
\mnras, 266, 386 

\bibitem{} Lutovinov, A.~A.~\& Revnivtsev, M.~G.\ 
2003, 
Astronomy Letters, 29, 719 

\bibitem{} Lyubarskii, Y.~E.\ 
1997, 
\mnras, 292, 679 

\bibitem{} Maccarone, T.~J.~\& Coppi, P.~S.\ 
2002a, 
\mnras, 335, 465 

\bibitem{} Maccarone, T.~J.~\& Coppi, P.~S.\ 
2002b, 
\mnras, 336, 817 

\bibitem{} Maccarone, T.~J.~\& Coppi, P.~S.\  
2003, 
\mnras, 338, 189 

\bibitem{} Maccarone, T.~J., Coppi, P.~S., \& Poutanen, J.\ 
2000,
\apjl, 537, L107 

\bibitem{} Maejima, Y., Makishima, K., Matsuoka, M., Ogawara, Y., Oda, M., 
et al.\ 
1984, 
\apj, 285, 712 

\bibitem{} Main, D.~S., Smith, D.~M., Heindl, W.~A., Swank, J., Leventhal, M., 
et al.\ 
1999, 
\apj, 525, 901 

\bibitem{} Makino, Y.\ 
1993, 
PhD thesis, University of Tokyo 

\bibitem{} Makishima, K., Mitsuda, K., Inoue, H., Koyama, K., Matsuoka, M., et al.\ 
1983, 
\apj, 267, 310 

\bibitem{} Makishima, K., Maejima, Y., Mitsuda, K., Bradt, H.~V., Remillard, 
R.~A., et al..\ 
1986, 
\apj, 308, 635 

\bibitem{} Makishima, K., Ishida, M., Ohashi, T., Dotani, T., Inoue,
H., Mitsuda, K., et al.\ 
1989, 
\pasj, 41, 531 

\bibitem{} Malzac, J., Belloni, T., Spruit, H.~C., \& Kanbach, G.\ 
2003, 
\aap, 407, 335 

\bibitem{} Manmoto, T., Takeuchi, M., Mineshige, S., Matsumoto, R., \& 
Negoro, H.\ 
1996, 
\apjl, 464, L135 

\bibitem{} Markovi\'c, D.,
2000,
astro-ph/0009450 

\bibitem{} Markovi\'c, D.~\& Lamb, F.~K.\ 
1998, 
\apj, 507, 316 

\bibitem{} Markovi\'c, D.~\& Lamb, F.~K.\ 
2000,
astro-ph/0009169 

\bibitem{} Markowitz, A., Edelson, R., Vaughan, S., Uttley, P., et al.\ 
2003, 
\apj, 593, 96 

\bibitem{} Markwardt, C.\ 
2001, 
Astrophysics and Space Science Supplement, 276, 209 

\bibitem{} Markwardt, C.~B.~\& Swank, J.~H.\ 
2003, 
\iaucirc, 8144 

\bibitem{} Markwardt, C.B., Strohmayer, T.E., \& Swank, J.H.\  
1999a, 
\apjl, 512, L125 

\bibitem{} Markwardt, C.~B., Swank, J.~H., \& Taam, R.~E.\ 
1999b, 
\apjl, 513, L37 

\bibitem{} Markwardt, C.~B., Swank, J.~H., Strohmayer, T.~E., 
in 't Zand, J.~J.~M., et al.\ 
2002, 
\apjl, 575, L21 

\bibitem{} Markwardt, C.~B., Smith, E., \& Swank, J.~H.\ 
2003, 
\iaucirc, 8080, 2 

\bibitem{} Marshall, F.E., \& Markwardt, C.B.\  
1999,
\iaucirc, 7103 

\bibitem{} Matsuba, E., Dotani, T., Mitsuda, K., Asai, K., Lewin,
W.H.G., et al.\ 
1995,
\pasj, 47, 575 

\bibitem{} Mauche, C.~W.\ 
2002, 
\apj, 580, 423 

\bibitem{} McClintock, J.~E., Haswell, C. A., Garcia, M. R., Drake,
J. J., Hynes, R. I., et al.\ 
2001, 
\apj, 555, 477 

\bibitem{} McDermott, P.~N.~\& Taam, R.~E.\ 
1987, 
\apj, 318, 278 

\bibitem{} McDermott, P.~N., van Horn, H.~M., \& Hansen, C.~J.\ 
1988, 
\apj, 325, 725 

\bibitem{} McGowan, K.~E., Charles, P.~A., O'Donoghue, D., \& Smale, A.~P.\ 
2003, 
\mnras, 345, 1039 

\bibitem{} McHardy, I.~M., Papadakis, I.~E., Uttley, P., Page, M.~J., \& 
Mason, K.~O.\ 
2004, 
\mnras, 348, 783 

\bibitem{} McNamara, B.~J., Harrison, T. E., Zavala, R. T., Galvan,
E., et al.\
2003, 
\aj, 125, 1437 

\bibitem{} Meekins, J.~F., Wood, K.~S., Hedler, R.~L., Byram, E.~T., 
Yentis, D.~J., et al.\ 
1984, 
\apj, 278, 288 

\bibitem{} M\'endez M.
1999,
in 19th Texas Symp., Paris; astro-ph/9903469

\bibitem{} M\'endez M.
2002a,
in Jan van Paradijs memorial meeting, ASP Conf. Proc. 308, 289;
astro-ph/0207279  

\bibitem{} M{\' e}ndez, M.\ 
2002b, 
in The Ninth Marcel Grossmann Meeting, 2319;
astro-ph/0207278 

\bibitem{} M\'endez, M.~\& van der Klis, M.\ 
1997, 
\apj, 479, 926 

\bibitem{} M\'endez, M., \& van der Klis, M.\ 
1999,
\apjl, 517, L51 

\bibitem{} M{\'e}ndez, M.~\& van der Klis, M.\ 
2000, 
\mnras, 318, 938 

\bibitem{} M\'endez, M., van der Klis, M., van Paradijs, J., Lewin, W.H.G., 
Lamb, F.K., et al.\ 
1997,
\apjl, 485, L37 

\bibitem{} M\'endez, M., van der Klis, M., van Paradijs, J., Lewin, W.H.G., 
et al.\ 
1998a, 
\apjl, 494, L65 

\bibitem{} M\'endez, M., van der Klis, M., Wijnands, R., Ford, E.C.,
van Paradijs, J., et al.\  
1998b,
\apjl, 505, L23 

\bibitem{} M\'endez, M., van der Klis, M., \& van Paradijs, J.\ 
1998c,
\apjl, 506, L117 

\bibitem{} M\'endez, M., Belloni, T., \& van der Klis, M.\ 
1998d, 
\apjl, 499, L187 

\bibitem{} M\'endez, M., van der Klis, M., Ford, E.C., Wijnands, R., 
\& van Paradijs, J.\   
1999, 
\apjl, 511, L49 

\bibitem{} M{\' e}ndez, M., van der Klis, M., \& Ford, E.~C.\ 
2001, 
\apj, 561, 1016 

\bibitem{} M{\' e}ndez, M., Cottam, J., \& Paerels, F.\ 
2002,
astro-ph/0207277 

\bibitem{} Menna, M.~T., Burderi, L., Stella, L., Robba, N., \& van der Klis, 
M.\ 
2003, 
\apj, 589, 503 

\bibitem{} Merloni, A., Vietri, M., Stella, L., \& Bini, D.\ 
1999, 
\mnras, 304, 155 

\bibitem{} Merloni, A., Di Matteo, T., \& Fabian, A.~C.\ 
2000, 
\mnras, 318, L15 

\bibitem{} Middleditch, J.~\& Priedhorsky, W.~C.\ 
1986, 
\apj, 306, 230 

\bibitem{} Migliari, S., Fender, R.~P., Rupen, M., Jonker, P.~G., 
Klein-Wolt, M., et al.\ 
2003a, 
\mnras, 342, L67 

\bibitem{} Migliari, S., van der Klis, M., \& Fender, R.~P.\ 
2003b, 
\mnras, 345, L35 

\bibitem{} Migliari, S., Fender, R.~P., Rupen, M., Wachter, S.,
Jonker, P.~G. et al.\ 
2004, 
\mnras, 351, 186 

\bibitem{} Miller, M.~C.\ 
1995, 
\apj, 441, 770 

\bibitem{} Miller, M.~C.\ 
1999, 
\apj, 520, 256 

\bibitem{} Miller, M.~C.\ 
2000, 
\apj, 537, 342 

\bibitem{} Miller, M.~C.\ 
2003, 
in Rossi and beyond, AIP Conf. Proc. 714, 365; 
astro-ph/0312449 

\bibitem{} Miller, G.~S.~\& Lamb, F.~K.\ 
1992, 
\apj, 388, 541 

\bibitem{} Miller, G.~S.~\& Park, M.\ 
1995, 
\apj, 440, 771 

\bibitem{} Miller, M.C., Lamb, F.K., \& Psaltis, D.\  
1998a,
\apj, 508, 791 

\bibitem{} Miller, M.C., Lamb, F.K., \& Cook, G.B.\  
1998b,
\apj, 509, 793 

\bibitem{} Miller, J.~M., Wijnands, R., Homan, J., Belloni, T.,
Pooley, D., et al.\ 
2001, 
\apj, 563, 928 

\bibitem{} Miller, J.~M., Fabian, A. C., Wijnands, R., Reynolds, C. S., Ehle, M., et al.\ 
2002, 
\apjl, 570, L69 

\bibitem{} Milsom, J.~A.~\& Taam, R.~E.\ 
1996, 
\mnras, 283, 919 

\bibitem{} Milsom, J.~A.~\& Taam, R.~E.\ 
1997, 
\mnras, 286, 358 

\bibitem{} Mineshige, S., Takeuchi, M., \& Nishimori, H.\ 
1994a, 
\apjl, 435, L125 

\bibitem{} Mineshige, S., Ouchi, N.~B., \& Nishimori, H.\ 
1994b, 
\pasj, 46, 97 

\bibitem{} Mineshige, S., Kusnose, M., \& Matsumoto, R.\ 
1995, 
\apjl, 445, L43 

\bibitem{} Misra, R.\ 
2000, 
\apjl, 529, L95 

\bibitem{} Mitsuda, K.~\& Dotani, T.\ 
1989, 
\pasj, 41, 557 

\bibitem{} Mitsuda, K., Inoue, H., Nakamura, N., \& Tanaka, Y.\ 
1989, 
\pasj, 41, 97 

\bibitem{} Mitsuda, K., Dotani, T., Yoshida, A., Vaughan, B., \& 
Norris, J.~P.\ 
1991, 
\pasj, 43, 113 

\bibitem{} Miyamoto, S.,\ 
1994,
ISAS RN 548 

\bibitem{} Miyamoto, S.~\& Kitamoto, S.\ 
1989, 
\nat, 342, 773 

\bibitem{} Miyamoto, S., Kitamoto, S., Mitsuda, K., \& Dotani, T.\ 
1988, 
\nat, 336, 450 

\bibitem{} Miyamoto, S., Kimura, K., Kitamoto, S., Dotani, T., \& Ebisawa, K.\ 
1991, 
\apj, 383, 784 

\bibitem{} Miyamoto, S., Kitamoto, S., Iga, S., Negoro, H., \& Terada, K.\ 
1992, 
\apjl, 391, L21 

\bibitem{} Miyamoto, S., Iga, S., Kitamoto, S., \& Kamado, Y.\ 
1993, 
\apjl, 403, L39 

\bibitem{} Miyamoto, S., Kitamoto, S., Iga, S., Hayashida, K., \& Terada, K.\ 
1994, 
\apj, 435, 398 

\bibitem{} Miyamoto, S., Kitamoto, S., Hayashida, K., \& Egoshi, W.\ 
1995,
\apjl, 442, L13 

\bibitem{} Moon, D.~\& Eikenberry, S.~S.\ 
2001a, 
\apjl, 549, L225  

\bibitem{} Moon, D.~\& Eikenberry, S.~S.\ 
2001b, 
\apjl, 552, L135 

\bibitem{} Morgan, E.~H., Remillard, R.~A., \& Greiner, J.\ 
1997, 
\apj, 482, 993 

\bibitem{} Morsink, S.~M.~\& Stella, L.\ 
1999, 
\apj, 513, 827 

\bibitem{} Motch, C., Ilovaisky, S.~A., \& Chevalier, C.\ 
1982, 
\aap, 109, L1 

\bibitem{} Motch, C., Ricketts, M.~J., Page, C.~G., Ilovaisky, S.~A., \& 
Chevalier, C.\ 
1983, 
\aap, 119, 171 

\bibitem{} Mukhopadhyay, B., Ray, S., Dey, J., \& Dey, M.\ 
2003, 
\apjl, 584, L83 

\bibitem{} Muno, M.~P., Morgan, E.~H., \& Remillard, R.~A.\ 
1999,
\apj, 527, 321 

\bibitem{} Muno, M.~P., Remillard, R.~A., Morgan, E.~H., Waltman, E.~B., 
et al.\ 
2001, 
\apj, 556, 515 

\bibitem{} Muno, M.~P., Remillard, R.~A., \& Chakrabarty, D.\ 
2002, 
\apjl, 568, L35 

\bibitem{} Muno, M.~P., Galloway, D.~K., \& Chakrabarty, D.\ 
2004, 
\apj, 608, 930  

\bibitem{} Murdin, P., Jauncey, D.~L., Lerche, I., Nicolson, G.~D., 
Kaluzienski, L.~J., et al.\ 
1980, 
\aap, 87, 292 

\bibitem{} Naik, S.~\& Rao, A.~R.\ 
2000, 
\aap, 362, 691 

\bibitem{} Naik, S., Agrawal, P.~C., Paul, B., et al.\ 
2000, 
Journal of Astrophysics and Astronomy, 21, 29 

\bibitem{} Nandi, A., Manickam, S.~G., Rao, A.~R., \& Chakrabarti, S.~K.\  
2001, 
\mnras, 324, 267 

\bibitem{} Narita, T., Grindlay, J.~E., Bloser, P.~F., \& Chou, Y.\ 
2003, 
\apj, 593, 1007 

\bibitem{} Negoro, H., Miyamoto, S., \& Kitamoto, S.\ 
1994, 
\apjl, 423, L127 

\bibitem{} Negoro, H., Kitamoto, S., Takeuchi, M., \& Mineshige, S.\ 
1995, 
\apjl, 452, L49 

\bibitem{} Negoro, H., Kitamoto, S., \& Mineshige, S.\ 
2001, 
\apj, 554, 528 

\bibitem{} Nespoli, E., Belloni, T., Homan, J., Miller, J.~M., Lewin, W.~H.~G.,
et al.\ 
2003, 
\aap, 412, 235 

\bibitem{} Nobili, L.\ 
2003, 
\apj, 582, 954 

\bibitem{} Nobili, L., Turolla, R., Zampieri, L., \& Belloni, T.\ 
2000, 
\apjl, 538, L137 

\bibitem{} Nolan, P.~L., Gruber, D. E., Matteson, J. L., Peterson,
L. E.,Rothschild, R. E.,  al.\ 
1981, 
\apj, 246, 494 

\bibitem{} Norris, J.~P.~\& Wood, K.~S.\ 
1987, 
\apj, 312, 732 

\bibitem{} Norris, J.~P., Hertz, P., Wood, K.~S., Vaughan, B.~A., 
Michelson, P.~F.,et al.\ 
1990, 
\apj, 361, 514 

\bibitem{} Nowak, M.~A.\ 
1994, 
\apj, 422, 688 

\bibitem{} Nowak, M.~A.\ 
1995, 
\pasp, 107, 1207 

\bibitem{} Nowak, M.~A.\ 
2000, 
\mnras, 318, 361 

\bibitem{} Nowak, M.~\& Lehr, D.\ 
1998, 
in Theory of Black Hole Accretion Disks, Abramowicz et al.\ (eds.),
Cambridge Univ. Press, 233 

\bibitem{} Nowak, M.~A.~\& Vaughan, B.~A.\ 
1996, 
\mnras, 280, 227 

\bibitem{} Nowak, M.~A.~\& Wagoner, R.~V.\ 
1993, 
\apj, 418, 187 

\bibitem{} Nowak, M.~A.~\& Wilms, J.\ 
1999, 
\apj, 522, 476 

\bibitem{} Nowak, M.~A., Wagoner, R.~V., Begelman, M.~C., \& Lehr, D.~E.\ 
1997, 
\apjl, 477, L91 

\bibitem{} Nowak, M.~A., Vaughan, B.~A., Wilms, J., Dove, J.~B., \& Begelman,
M.~C.\ 
1999a, 
\apj, 510, 874 

\bibitem{} Nowak, M.~A., Wilms, J., Vaughan, B.~A., Dove, J.~B., \& Begelman, 
M.~C.\ 
1999b,
\apj, 515, 726 

\bibitem{} Nowak, M.~A., Wilms, J., \& Dove, J.~B.\ 
1999c,
\apj, 517, 355 

\bibitem{} Nowak, M.~A., Wilms, J., Heindl, W.~A., Pottschmidt, K.,et al.\ 
2001, 
\mnras, 320, 316 

\bibitem{} Nowak, M.~A., Wilms, J., \& Dove, J.~B.\ 
2002, 
\mnras, 332, 856 

\bibitem{} O'Brien, K., Horne, K., Gomer, R.~H., Oke, J.~B., \& van
der Klis, M.\ 
2004,
\mnras, 350, 587 

\bibitem{} Oda, M., Gorenstein, P., Gursky, H., Kellogg, E., Schreier, E., 
et al.\ 
1971, 
\apjl, 166, L1 

\bibitem{} Ogawara, Y., Doi, K., Matsuoka, M., Miyamoto, S., \& Oda, M.\ 
1977, 
\nat, 270, 154 

\bibitem{} Okuda, T.~\& Mineshige, S.\ 
1991, 
\mnras, 249, 684 

\bibitem{} Olive, J.~F., Barret, D., Boirin, L., Grindlay, J.~E., Swank, 
J.~H., \& Smale, A.~P.\ 
1998, 
\aap, 333, 942 

\bibitem{} Olive, J., Barret, D., \& Gierli{\' n}ski, M.\ 
2003, 
\apj, 583, 416 

\bibitem{} O'Neill, P.~M., Kuulkers, E., Sood, R.~K., \& Dotani, T.\ 
2001, 
\aap, 370, 479 

\bibitem{} O'Neill, P.~M., Kuulkers, E., Sood, R.~K., \& van der Klis, M.\ 
2002, 
\mnras, 336, 217 

\bibitem{} Oosterbroek, T., Penninx, W., van der Klis, M., van Paradijs, J., \& 
et al.\ 
1991, 
\aap, 250, 389 

\bibitem{} Oosterbroek, T., Lewin, W.~H.~G., van Paradijs, J., 
van der Klis, M.,et al.\ 
1994, 
\aap, 281, 803 

\bibitem{} Oosterbroek, T., van der Klis, M., Kuulkers, E., van Paradijs, J., 
et al.\ 
1995, 
\aap, 297, 141 

\bibitem{} Oosterbroek, T., van der Klis, M., Vaughan, B., van
Paradijs, J., et al.\ 
1996, 
\aap, 309, 781 

\bibitem{} Oosterbroek, T., van der Klis, M., van Paradijs, J.,
Vaughan, B., et al.\ 
1997, 
\aap, 321, 776 

\bibitem{} Oosterbroek, T., Barret, D., Guainazzi, 
M., \& Ford, E.~C.\ 
2001, 
\aap, 366, 138 

\bibitem{} Orlandini, M.~\& Boldt, E.\ 
1993, 
\apj, 419, 776 

\bibitem{} Orlandini, M.~\& Morfill, G.~E.\ 
1992, 
\apj, 386, 703 

\bibitem{} Ortega-Rodr{\'{\i}}guez, M.~\& Wagoner, R.~V.\ 
2000, 
\apj, 537, 922 

\bibitem{} Osherovich, V.~\& Titarchuk, L.\ 
1999, 
\apjl, 522, L113 

\bibitem{} Ouyed, R.\ 
2002, 
\aap, 382, 939 

\bibitem{} Paczy\'nski, B.\ 
1987, 
\nat, 327, 303 

\bibitem{} Park, S.~Q., Miller, J. M., McClintock, J. E., Remillard,
R. A., et al.\ 
2003, 
astro-ph/0308363 

\bibitem{} Parmar, A.~N., Stella, L., \& White, N.~E.\ 
1986, 
\apj, 304, 664 

\bibitem{} Parmar, A.~N., Angelini, L., Roche, P., \& White, N.~E.\ 
1993, 
\aap, 279, 179 

\bibitem{} Parmar, A.~N., Oosterbroek, T., Boirin, L., \& Lumb, D.\ 
2002, 
\aap, 386, 910 

\bibitem{} Patterson, J.\ 
1979, 
\apj, 234, 978 

\bibitem{} Paul, B., Agrawal, P.~C., Rao, A.~R., Vahia, M.~N., Yadav, J.~S., 
et al.\ 
1997, 
\aap, 320, L37 

\bibitem{} Payne, D.~G.\ 
1980, 
\apj, 237, 951 

\bibitem{} Penninx, W., Lewin, W.~H.~G., Zijlstra, A.~A., Mitsuda, K.,
et al.\ 
1988, 
\nat, 336, 146 

\bibitem{} Penninx, W., Hasinger, G., Lewin, W.~H.~G., van Paradijs, J., et al.\ 
1989, 
\mnras, 238, 851 

\bibitem{} Penninx, W., Lewin, W.~H.~G., Mitsuda, K., van der Klis, M., 
et al.\ 
1990, 
\mnras, 243, 114 

\bibitem{} Penninx, W., Lewin, W.~H.~G., Tan, J., Mitsuda, K., van der Klis, 
M., et al.\ 
1991, 
\mnras, 249, 113 

\bibitem{} P\'erez, C.~A., Silbergleit, A.~S., Wagoner, R.~V., \& Lehr, D.~E.\ 
1997, 
\apj, 476, 589 

\bibitem{} Piraino, S., Santangelo, A., \& Kaaret, P.\ 
2000, 
\aap, 360, L35 

\bibitem{} Piraino, S., Santangelo, A., \& Kaaret, P.\ 
2002, 
\apj, 567, 1091 

\bibitem{} Ponman, T.~J., Cooke, B.~A., \& Stella, L.\ 
1988, 
\mnras, 231, 999 

\bibitem{} Popham, R.~\& Sunyaev, R.\ 
2001, 
\apj, 547, 355 %

\bibitem{} Pottschmidt, K., Koenig, M., Wilms, J., \& Staubert, R.\ 
1998, 
\aap, 334, 201 

\bibitem{} Pottschmidt, K., Wilms, J., Nowak, M. A., Pooley, G. G.,
Gleissner, T., et al.\ 
2003, 
\aap, 407, 1039 

\bibitem{} Poutanen, J.\ 
2002, 
\mnras, 332, 257 

\bibitem{} Poutanen, J.~\& Fabian, A.~C.\ 
1999, 
\mnras, 306, L31 

\bibitem{} Press, W.~H.~\& Schechter, P.\ 
1974, 
\apj, 193, 437 

\bibitem{} Priedhorsky, W., Garmire, G.~P., Rothschild, R., Boldt, E., 
Serlemitsos, P., et al.\ 
1979, 
\apj, 233, 350 

\bibitem{} Priedhorsky, W., Hasinger, G., Lewin, W.~H.~G., Middleditch, J., 
Parmar, A., et al.\ 
1986, 
\apjl, 306, L91 

\bibitem{} Pringle, J.~E.\ 
1981, 
\araa, 19, 137 

\bibitem{} Prins, S.~\& van der Klis, M.\ 
1997, 
\aap, 319, 498 

\bibitem{} Psaltis, D.\ 
2000, 
astro-ph/0010316 

\bibitem{} Psaltis, D.\ 
2001, 
Advances in Space Research, 28, 481 

\bibitem{} Psaltis, D., \& Norman, C.\ 
2000,
astro-ph/0001391 

\bibitem{} Psaltis, D., M\'endez, M., Wijnands, R., Homan, J., Jonker, P.G., et al.\ 
1998, 
\apjl, 501, L95 

\bibitem{} Psaltis, D., Belloni, T., \& van der Klis, M.\ 
1999a, 
\apjl, 520, 262 

\bibitem{} Psaltis, D., Wijnands, R., Homan, J., Jonker, P.G., van der Klis, M., et al.\ 
1999b, 
\apj  520, 763 

\bibitem{} Pudritz, R.~E.~\& Fahlman, G.~G.\ 
1982, 
\mnras, 198, 689 

\bibitem{} Qu, J.~L., Yu, W., \& Li, T.~P.\ 
2001, 
\apj, 555, 7 

\bibitem{} Rao, A.~R., Naik, S., Vadawale, S.~V., \& Chakrabarti, S.~K.\ 
2000a, 
\aap, 360, L25 

\bibitem{} Rao, A.~R., Yadav, J.~S., \& Paul, B.\ 
2000b, 
\apj, 544, 443 

\bibitem{} Rebusco, P.\ 
2004, 
\pasj, 56, 553 

\bibitem{} Rebusco, P.\ 
2004, 
astro-ph/0403341 

\bibitem{} Reerink, T., Schnerr, R., van der Klis, M., \& van Straaten, S.\ 
2004, 
\aap, submitted 

\bibitem{} Reig, P., M{\' e}ndez, M., van der Klis, M., \& Ford, E.~C.\ 
2000a, 
\apj, 530, 916 

\bibitem{} Reig, P., Belloni, T., van der Klis, M., M{\' e}ndez, M., Kylafis, 
N.~D., \& Ford, E.~C.\ 
2000b, 
\apj, 541, 883 

\bibitem{} Reig, P., Papadakis, I., \& Kylafis, N.~D.\  
2002, 
\aap, 383, 202 

\bibitem{} Reig, P., Papadakis, I., \& Kylafis, N.~D.\ 
2003a, 
\aap, 398, 1103 

\bibitem{} Reig, P., Belloni, T., \& van der Klis, M.\ 
2003b, 
\aap, 412, 229 

\bibitem{} Reig, P., Kylafis, N.~D., \& Giannios, D.\ 
2003c, 
\aap, 403, L15 

\bibitem{} Reig, P., van Straaten, S., \& van der Klis, M.\ 
2004, 
\apj, 602, 918 

\bibitem{} Reilly, K.~T., Bloom, E. D., Focke, W., Giebels, B.,
Godfrey, G., et al.\ 
2001, 
\apjl, 561, L183 

\bibitem{} Remillard, R.~A.~\& Morgan, E.~H.\ 
1998, 
Nuclear Physics B, (Proc. Suppl.), 69, 316 

\bibitem{} Remillard, R.~A.~\& Morgan, E.~H.\ 
1999, 
BAAS, 31, 1421 

\bibitem{} Remillard, R.~A., McClintock, J.~E., Sobczak, G.~J., Bailyn, C.~D., 
et al.\ 
1999a, 
\apjl, 517, L127 

\bibitem{} Remillard, R., Morgan, E., Levine, A., Muno, M., McClintock, J., 
et al.\ 
1999b, 
BAAS, 31, 731 

\bibitem{} Remillard, R.~A., Morgan, E.~H., McClintock, J.~E., Bailyn, 
C.~D., \& Orosz, J.~A.\ 
1999c, 
\apj, 522, 397 

\bibitem{} Remillard, R.~A., Muno, M.~P., McClintock, J.~E., \& Orosz, J.~A.\ 
2002a, 
\apj, 580, 1030 

\bibitem{} Remillard, R., Muno, M., McClintock, J.~E., \& Orosz, J.\  
2002b, 
in 4th Microquasars Workshop, Carg\`ese, 2002, Durouchoux et al.\
(eds.), p. 49; astro-ph/0208402 

\bibitem{} Remillard, R.~A., Sobczak, G.~J., Muno, M.~P., \& McClintock, 
J.~E.\ 
2002c, 
\apj, 564, 962 

\bibitem{} Remillard, R.~A., Muno, M.~P., McClintock, J.~E., \& Orosz, J.~A.\ 
2003, 
abstract AAS/HEAD 7, 30.03  

\bibitem{} Revnivtsev, M., Gilfanov, M., Churazov, E., Sunyaev, R.,
Borozdin, K., et al.\ 
1998a, 
\aap, 331, 557 

\bibitem{} Revnivtsev, M., Gilfanov, M., \& Churazov, E.\ 
1998b, 
\aap, 339, 483 

\bibitem{} Revnivtsev, M., Borozdin, K., \& Emelyanov, A.\ 
1999a, 
\aap, 344, L25 

\bibitem{} Revnivtsev, M., Gilfanov, M., \& Churazov, E.\ 
1999b, 
\aap, 347, L23 

\bibitem{} Revnivtsev, M.~G., Borozdin, K.~N., 
Priedhorsky, W.~C., \& Vikhlinin, A.\ 
2000a, 
\apj, 530, 955 

\bibitem{} Revnivtsev, M., Sunyaev, R., \& Borozdin, K.\ 
2000b, 
\aap, 361, L37 

\bibitem{} Revnivtsev, M.~G., Trudolyubov, S.~P., \& Borozdin, K.~N.\ 
2000c, 
\mnras, 
312, 151 see also 315, 655 

\bibitem{} Revnivtsev, M., Churazov, E., Gilfanov, 
M., \& Sunyaev, R.\ 
2001a, 
\aap, 372, 138 

\bibitem{} Revnivtsev, M., Gilfanov, M., \& Churazov, E.\ 
2001b, 
\aap, 380, 520 

\bibitem{} Revnivtsev, M., Gilfanov, M., Churazov, E., \& Sunyaev, R.\ 
2002, 
\aap, 391, 1013 

\bibitem{} Reynolds, C.~S.~\& Nowak, M.~A.\ 
2003, 
\physrep, 377, 389 

\bibitem{} Rezzolla, L., Lamb, F.~K., \& Shapiro, S.~L.\ 
2000, 
\apjl, 531, L139 

\bibitem{} Rezzolla, L., Yoshida, S., Maccarone, T.~J., \& Zanotti, O.\ 
2003, 
\mnras, 344, L37 

\bibitem{} Ricci, D., Israel, G.~L., \& Stella, L.\ 
1995, 
\aap, 299, 731 

\bibitem{} Robinson, E.~L.~\& Warner, B.\ 
1972, 
\mnras, 157, 85 

\bibitem{} Rodriguez, J., Durouchoux, P., Mirabel, I.~F., Ueda, Y., Tagger, 
M., et al.\ 
2002a, 
\aap, 386, 271 

\bibitem{} Rodriguez, J., Corbel, S., Kalemci, E., \& Tomsick, J.A.\ 
2002b,
astro-ph/0205341 

\bibitem{} Rodriguez, J., Varni{\` e}re, P., Tagger, M., \& Durouchoux, P.\ 
2002c, 
\aap, 387, 487 

\bibitem{} Rodriguez, J., Corbel, S., \& Tomsick, J.~A.\ 
2003, 
\apj, 595, 1032 

\bibitem{} Rossi, S., Homan, J., Miller, J.~M., \& Belloni, T.\ 
2004, 
in The restless high-energy universe, Nuclear Phys. B, 132, 416;
astro-ph/0309129 

\bibitem{} Rothschild, R.~E., Boldt, E.~A., Holt, S.~S., \& Serlemitsos, 
P.~J.\ 
1974, 
\apjl, 189, L13 

\bibitem{} Rothschild, R.~E., Boldt, E.~A., Holt, S.~S., \& Serlemitsos, P.~J.\ 
1977, 
\apj, 213, 818 

\bibitem{} Rutledge, R.~E., Lubin, L.~M., Lewin, W.~H.~G., Vaughan, B., 
et al.\ 
1995, 
\mnras, 277, 523 

\bibitem{} Rutledge, R.~E., Lewin, W. H. G., van der Klis, M., van
Paradijs, J., et al.\ 
1999, 
\apjs, 124, 265 

\bibitem{} Samimi, J., Share, G. H., Wood, K., Yentis, D., Meekins,
J., et al.\ 
1979, 
\nat, 278, 434 

\bibitem{} Santolamazza, P., Fiore, F., Burderi, L., \& Di Salvo, T.\ 
2003, 
astro-ph/0311382 

\bibitem{} Scargle, J.~D., Steiman-Cameron, T., Young, K., Donoho, D.~L., 
et al.\ 
1993, 
\apjl, 411, L91 

\bibitem{} Schaab, C., \& Weigel, M.K.\ 
1999, 
\mnras, 308, 718 

\bibitem{} Schmidtke, P.~C., Ponder, A.~L., \& Cowley, A.~P.\ 
1999, 
\aj, 117, 1292 

\bibitem{} Schnerr, R.~S., Reerink, T., van der Klis, M., Homan, J.,
M{\' e}ndez, M., et al.\ 
2003, 
\aap, 406, 221 

\bibitem{} Schnerr, R.~S., Reerink, T., van der Klis, M., Homan, J., 
M{\' e}ndez, M., et al.\ 
2003, 
astro-ph/0305161 

\bibitem{} Schnittman, J.~D.~\& Bertschinger, E.\ 
2004, 
\apj, 606, 1098 

\bibitem{} Schulz, N.~S.~\& Wijers, R.~A.~M.~J.\ 
1993, 
\aap, 273, 123 

\bibitem{} Schulz, N.~S., Hasinger, G., \& Truemper, J.\ 
1989, 
\aap, 225, 48 

\bibitem{} Sellmeijer, H. \& van der Klis, M.\ 
1999,
unpublished 

\bibitem{} Shakura, N.~I.~\& Sunyaev, R.~A.\ 
1973, 
\aap, 24, 337 

\bibitem{} Shakura, N.~I.~\& Sunyaev, R.~A.\ 
1976, 
\mnras, 175, 613 

\bibitem{} Shibata, M.~\& Sasaki, M.\ 
1998, 
\prd, 58, 104011 

\bibitem{} Shibazaki, N.~\& Ebisuzaki, T.\ 
1989, 
\pasj, 41, 641 

\bibitem{} Shibazaki, N.~\& Lamb, F.~K.\ 
1987, 
\apj, 318, 767 

\bibitem{} Shibazaki, N., Elsner, R.~F., \& Weisskopf, M.~C.\ 
1987, 
\apj, 322, 831 

\bibitem{} Shibazaki, N., Elsner, R.~F., Bussard, R.~W., Ebisuzaki, T., \& 
Weisskopf, M.~C.\ 
1988, 
\apj, 331, 247 

\bibitem{} Shirakawa, A.~\& Lai, D.\ 
2002a, 
\apj, 564, 361 

\bibitem{} Shirakawa, A.~\& Lai, D.\ 
2002b, 
\apj, 565, 1134 

\bibitem{} Shirey, R.E., Bradt, H.V., Levine, A.M., \& Morgan, E.H.\ 
1996, 
\apj, 469, L21 

\bibitem{} Shirey, R.~E., Bradt, H.~V., Levine, A.~M., \& Morgan, E.~H.\ 
1998, 
\apj, 506, 374 

\bibitem{} Shirey, R.~E., Bradt, H.~V., \& Levine, A.~M.\ 
1999, 
\apj, 517, 472 

\bibitem{} Shvartsman, V.~F.\ 
1971, 
Soviet Astronomy, 15, 377 

\bibitem{} Sibgatullin, N.~R.\ 
2002, 
Astronomy Letters, 28, 83 

\bibitem{} Silbergleit, A.~S., Wagoner, R.~V., \& Ortega-Rodr{\'{\i}}guez, 
M.\ 
2001, 
\apj, 548, 335 

\bibitem{} Singh, K.~P.~\& Apparao, K.~M.~V.\ 
1994, 
\apj, 431, 826 

\bibitem{} Smale, A.~P.\ 
1998, 
\apjl, 498, L141 

\bibitem{} Smale, A.~P.~\&  Kuulkers, E.\ 
2000, 
\apj, 528, 702 

\bibitem{} Smale, A.P., Zhang, W., \& White, N.E.\ 
1997, 
\apjl, 483, L119 

\bibitem{} Smale, A.~P., Homan, J., \& Kuulkers, E.\ 
2003, 
\apj, 590, 1035 

\bibitem{} Smith, I.~A.~\& Liang, E.~P.\ 
1999, 
\apj, 519, 771 

\bibitem{} Smith, D.~M., Heindl, W.~A., Swank, J., Leventhal, M., Mirabel, 
I.~F., et al.\ 
1997, 
\apjl, 489, L51 

\bibitem{} Smith, D.~M., Heindl, W.~A., Markwardt, C.~B., \& Swank, J.~H.\ 
2001, 
\apjl, 554, L41 

\bibitem{} Smith, D.~M., Heindl, W.~A., \& Swank, J.~H.\ 
2002, 
\apj, 569, 362 

\bibitem{} Sobczak, G.~J., McClintock, J.~E., Remillard, R.~A., Bailyn, 
C.~D., \& Orosz, J.~A.\ 
1999, 
\apj, 520, 776 

\bibitem{} Sobczak, G.~J., McClintock, J.~E., Remillard, R.~A., Cui, W., 
Levine, A.~M., et al.\ 
2000a, 
\apj, 531, 537 

\bibitem{} Sobczak, G.~J., McClintock, J.~E., Remillard, R.~A., Cui, W., 
Levine, A.~M., et al.\ 
2000b, 
\apj, 544, 993 

\bibitem{} Spruit, H.~C.~\& Kanbach, G.\ 
2002, 
\aap, 391, 225 

\bibitem{} Spruit, H.~C.~\& Taam, R.~E.\ 
1990, 
\aap, 229, 475 

\bibitem{} Srinivasan, G.\ 
2002, 
\aapr, 11, 67 

\bibitem{} Steiman-Cameron, T.~Y., Scargle, J.~D., Imamura, J.~N., \& 
Middleditch, J.\  
1997, 
\apj, 487, 396 

\bibitem{} Stella, L.\ 
1988, 
Memorie della Societa Astronomica Italiana, 59, 185 

\bibitem{} Stella, L.~\& Vietri, M.\ 
1998, 
\apjl, 492, L59 

\bibitem{} Stella, L.~\& Vietri, M.\ 
1999, 
Physical Review Letters, 82, 17 

\bibitem{} Stella, L., White, N.~E., Davelaar, J., Parmar, A.~N., Blissett, 
R.~J., et al.\ 
1985, 
\apjl, 288, L45 

\bibitem{} Stella, L., Parmar, A.~N., \& White, N.~E.\ 
1987a, 
\apj, 321, 418 

\bibitem{} Stella, L., White, N.E., \& Priedhorsky, W.,
1987b,
\apjl, 315, L49 

\bibitem{} Stella, L., Haberl, F., Lewin, W.~H.~G., Parmar, A.~N., 
van der Klis, M., et al.\ 
1988a, 
\apjl, 327, L13 

\bibitem{} Stella, L., Haberl, F., Lewin, W.~H.~G., Parmar, A.~N., 
van Paradijs, J., et al.\ 
1988b, 
\apj, 324, 379 

\bibitem{} Stella, L., Vietri, M., \& Morsink, S.~M.\ 
1999, 
\apjl, 524, L63 

\bibitem{} Stergioulas, N., Klu{\' z}niak, W., \& Bulik, T.\ 
1999, 
\aap, 352, L116 

\bibitem{} Stoeger, W.~R.\ 
1980, 
\mnras, 190, 715 

\bibitem{} Stollman, G.~M., van Paradijs, J., Hasinger, G., Lewin, W.~H.~G., 
et al.\ 
1987, 
\mnras, 227, 7P 

\bibitem{} Strohmayer, T.~E.\ 
2001a, 
\apjl, 552, L49 

\bibitem{} Strohmayer, T.~E.\  
2001b, 
\apjl, 554, L169 

\bibitem{} Strohmayer, T.~E.~\& Lee, U.\ 
1996, 
\apj, 467, 773 

\bibitem{} Strohmayer, T.E., Zhang, W., Swank, J.H., Smale, A., Titarchuk,
L., \& Day, C.\   
1996, 
\apjl, 469, L9 

\bibitem{} Strohmayer, T.E., Jahoda, K., Giles, A.B., \& Lee, U.\  
1997, 
\apj, 486, 355 

\bibitem{} Strohmayer, T.~E., Markwardt, C.~B., Swank, J.~H., \& 
in't Zand, J.\ 
2003, 
\apjl, 596, L67 

\bibitem{} Strohmayer, T.~E.~\& Mushotzky, R.~F.\ 
2003, 
\apjl, 586, L61 

\bibitem{} Sunyaev, R.~A.\ 
1973, 
Soviet Astronomy, 16, 941 

\bibitem{} Sunyaev, R.~\&  Revnivtsev, M.\ 
2000, 
\aap, 358, 617 

\bibitem{} Sutherland, P.~G., Weisskopf, M.~C., \& Kahn, S.~M.\ 
1978, 
\apj, 219, 1029 

\bibitem{} Swank, J., Chen, X., Markwardt, C., \& Taam, R.\ 
1998,
in Some like it hot, AIP Conf. Proc. 431, 327 

\bibitem{} Taam, R.~E.~\& Lin, D.~N.~C.\ 
1984, 
\apj, 287, 761 

\bibitem{} Takeshima, T.\ 
1992, 
PhD thesis, Univ. of Tokyo 

\bibitem{} Takeshima, T., Dotani, T., Mitsuda, K., \& Nagase, F.\ 
1991, 
\pasj, 43, L43 

\bibitem{} Takeuchi, M., Mineshige, S., \& Negoro, H.\ 
1995, 
\pasj, 47, 617 

\bibitem{} Takizawa, M., Dotani, T., Mitsuda, K., Matsuba, E., Ogawa,
M., Aoki, T., et al.\ 
1997, 
\apj, 489, 272 

\bibitem{} Tan, J., Lewin, W.~H.~G., Lubin, L.~M., van Paradijs, J., 
Penninx, W., et al.\ 
1991, 
\mnras, 251, 1 

\bibitem{} Tan, J., Lewin, W.~H.~G., Hjellming, R.~M., Penninx, W., 
van Paradijs, J., et al.\ 
1992, 
\apj, 385, 314 

\bibitem{} Tanaka, Y. \& Tenma team,  
1983, 
IAU Circ. 3891 

\bibitem{} Tanaka, Y.\ 
1989,
in 23d ESLAB Symp., White et al. (eds.), p.~3

\bibitem{} Tananbaum, H., Gursky, H., Kellogg, E., Giacconi, R., \& Jones, C.\ 
1972, 
\apjl, 177, L5 

\bibitem{} Tawara, Y., Hayakawa, S., Hunieda, H., Makino, F., \& Nagase, F.\ 
1982, 
\nat, 299, 38 

\bibitem{} Taylor, J.~H., Wolszczan, A., Damour, T., \& Weisberg, J.~M.\ 
1992, 
\nat, 355, 132 

\bibitem{} Tennant, A.~F.\ 
1987, 
\mnras, 226, 971 

\bibitem{} Tennant, A.~F.\ 
1988, 
\mnras, 230, 403 

\bibitem{} Tennant, A.~F., Fabian, A.~C., \& Shafer, R.~A.\ 
1986, 
\mnras, 219, 871 

\bibitem{} Terada, K., Kitamoto, S., Negoro, H., \& Iga, S.\ 
2002, 
\pasj, 54, 609 

\bibitem{} Terrell, N.~J.~J.\ 
1972, 
\apjl, 174, L35 

\bibitem{} Thampan, A.V., Bhattacharya, D., \& Datta, B.\ 
1999, 
\mnras, 302, L69

\bibitem{} Thorne, K.~S.~\& Price, R.~H.\ 
1975, 
\apjl, 195, L101 

\bibitem{} Timmer, J., Schwarz, U., Voss, H.~U., Wardinski, I., Belloni, T., 
et al.\ 
2000, 
\pre, 61, 1342 

\bibitem{} Titarchuk, L.\ 
2002, 
\apjl, 578, L71 

\bibitem{} Titarchuk, L.\ 
2003, 
\apj, 591, 354 

\bibitem{} Titarchuk, L.~\& Osherovich, V.\ 
1999, 
\apjl, 518, L95 

\bibitem{} Titarchuk, L.~\& Osherovich, V.\ 
2000, 
\apjl, 542, L111 

\bibitem{} Titarchuk, L.~\& Shrader, C.~R.\ 
2002, 
\apj, 567, 1057 

\bibitem{} Titarchuk, L.~\& Wood, K.\ 
2002, 
\apjl, 577, L23 

\bibitem{} Titarchuk, L., Lapidus, I., \& Muslimov, A.\ 
1998, 
\apj, 499, 315 

\bibitem{} Titarchuk, L., Osherovich, V., \& Kuznetsov, S.\ 
1999, 
\apjl, 525, L129 

\bibitem{} Tomsick, J.~A.~\& Kaaret, P.\ 
2000, 
\apj, 537, 448 

\bibitem{} Tomsick, J.~A.~\& Kaaret, P.\ 
2001, 
\apj, 548, 401 

\bibitem{} Tomsick, J.A., Halpern, J.P., Kemp, J., \& Kaaret, P.\ 
1999,
\apj, 521, 341 

\bibitem{} Tomsick, J.~A., Kalemci, E., \& Kaaret, P.\ 
2004, 
\apj, 601, 439 

\bibitem{} Treves, A., Belloni, T., Corbet, R. H. D., Ebisawa, K.,
Falomo, R., et al.\ 
1990, 
\apj, 364, 266 

\bibitem{} Trudolyubov, S.~P.\ 
2001, 
\apj, 558, 276 

\bibitem{} Trudolyubov, S.~P., Churazov, E.~M., \& Gilfanov, M.~R.\ 
1999a, 
Astronomy Letters, 25, 718 

\bibitem{} Trudolyubov, S., Churazov, E., \& Gilfanov, M.\ 
1999b, 
\aap, 351, L15 

\bibitem{} Trudolyubov, S.~P., Borozdin, K.~N., \& Priedhorsky, W.~C.\ 
2001, 
\mnras, 322, 309 

\bibitem{} Ubertini, P., Bazzano, A., Cocchi, M., Natalucci, L., Heise, J., 
et al.\ 
1999, 
\apjl, 514, L27 

\bibitem{} Uemura, M., Kato, T., Ishioka, R., Tanabe, K., Kiyota, S.,
Monard, B., et al.\ 
2002, 
\pasj, 54, L79 

\bibitem{} Uemura, M., Kato, T., Ishioka, R., Tanabe, K., Torii, K.,
Santallo, R., et al.\ 
2004, 
\pasj, 56, 61 

\bibitem{} Unno, W., Yoneyama, T., Urata, K., Masaki, I., Kondo, M., \& 
Inoue, H.\ 
1990, 
\pasj, 42, 269 

\bibitem{} Ushomirsky, G., Cutler, C., \& Bildsten, L.\ 
2000, 
\mnras, 319, 902 

\bibitem{} Uttley, P.\ 
2004, 
\mnras, 347, L61 

\bibitem{} Uttley, P.~\& McHardy, I.~M.\ 
2001, 
\mnras, 323, L26 

\bibitem{} Uttley, P., McHardy, I.~M., \& Papadakis, I.~E.\ 
2002, 
\mnras, 332, 231 

\bibitem{} van der Hooft, F., et al.\ 
1996, 
\apjl, 458, L75 

\bibitem{} van der Hooft, F., et al.\ 
1999a, 
\apj, 513, 477 

\bibitem{} van der Hooft, F., et al.\ 
1999b, 
\apj, 519, 332 

\bibitem{} van der Klis, M.\ 
1986,
in The physics of accretion onto compact objects, Lect. Notes Phys. 266, 157

\bibitem{} van der Klis, M.\ 
1989a, 
\araa, 27, 517 

\bibitem{} van der Klis, M.\ 
1989b,
in Timing Neutron Stars, NATO ASI C262, p. 27

\bibitem{} van der Klis, M.\ 
1994a, 
\apjs, 92, 511 

\bibitem{} van der Klis, M.\ 
1994b, 
\aap, 283, 469 

\bibitem{} van der Klis, M.\ 
1995a,
in X-ray binaries, Lewin et al.\ (eds.), Cambridge Univ. Press, p. 252

\bibitem{} van der Klis, M.\ 
1995b, 
in The lives of the neutron stars, NATO ASI C450, p. 301 

\bibitem{} van der Klis, M.\ 
1997, 
in Astronomical Time Series, ASSL 218, 121 

\bibitem{} van der Klis, M.\ 
1998, 
in The Many Faces of Neutron Stars, NATO ASI C Proc.~515, p. 337 

\bibitem{} van der Klis, M.\ 
1999, 
in Stellar endpoints, AIP Conf. Proc. 599, 406 

\bibitem{} van der Klis, M.\ 
2000, 
\araa, 38, 717 

\bibitem{} van der Klis, M.\ 
2001, 
\apj, 561, 943 

\bibitem{} van der Klis, M.\ 
2002, 
in XEUS - studying the evolution of the hot universe, MPE rep. 281, 354

\bibitem{} van der Klis, M.~\& Jansen, F.~A.\ 
1985, 
\nat, 313, 768 

\bibitem{} van der Klis, M., Jansen, F., van Paradijs, J., Lewin, W.~H.~G., 
et al.\ 
1985, 
\nat, 316, 225 

\bibitem{} van der Klis, M., Jansen, F., van Paradijs, J., Lewin, W.~H.~G., 
Sztajno, M., et al.\ 
1987a, 
\apjl, 313, L19 

\bibitem{} van der Klis, M., Stella, L., White, N., Jansen, F., \& 
Parmar, A.~N.\ 
1987b, 
\apj, 316, 411 

\bibitem{} van der Klis, M., Hasinger, G., Stella, L., Langmeier, A., 
et al.\ 
1987c, 
\apjl, 319, L13 

\bibitem{} van der Klis, M., Hasinger, G., Damen, E., Penninx, W., et al.\  
1990,
\apj, 360, L19 

\bibitem{} van der Klis, M., Kitamoto, S., Tsunemi, H., \& Miyamoto, S.\ 
1991, 
\mnras, 248, 751 

\bibitem{} van der Klis, M., Swank, J.H., Zhang, W., Jahoda, K., Morgan, E.H., et al.\ 
1996, 
\apjl, 469, L1 

\bibitem{} van der Klis, M., Wijnands, R.A.D., Horne, K., \& Chen, W.\ 
1997, 
\apjl, 481, L97 

\bibitem{} van der Klis, M., Chakrabarty, D., Lee, J.~C., Morgan, E.~H., 
et al.\ 
2000, 
\iaucirc, 7358, 3 

\bibitem{} van Paradijs, J., Hasinger, G., Lewin, W.~H.~G., van der Klis, M.,
et al.\ 
1988a, 
\mnras, 231, 379 

\bibitem{} van Paradijs, J., Penninx, W., Lewin, W.~H.~G., Sztajno, M., \& Truemper, J.\ 
1988b, 
\aap, 192, 147 

\bibitem{} van Straaten, S., Ford, E.~C., van der Klis, M., M{\' e}ndez, M., 
\& Kaaret, P.\ 
2000, 
\apj, 540, 1049 

\bibitem{} van Straaten, S., van der Klis, M., Kuulkers, E., \& M{\' e}ndez, M.\ 
2001, 
\apj, 551, 907 

\bibitem{} van Straaten, S., van der Klis, M., Di Salvo, T., \& Belloni, T.\ 
2002, 
\apj, 568, 912 

\bibitem{} van Straaten, S., van der Klis, M., \& M{\' e}ndez, M.\ 
2003, 
\apj, 596, 1155 

\bibitem{} van Straaten, S., van der Klis, M., \& Wijnands, R.\ 
2004, 
\apj\ in press 

\bibitem{} Varni{\` e}re, P., Rodriguez, J., \& Tagger, M.\ 
2002, 
\aap, 387, 497 

\bibitem{} Vasiliev, L., Trudolyubov, S., \& Revnivtsev, M.\ 
2000, 
\aap, 362, L53 

\bibitem{} Vaughan, B.~A.~\& Nowak, M.~A.\ 
1997, 
\apjl, 474, L43 

\bibitem{} Vaughan, B., van der Klis, M., Lewin, W.~H.~G., 
Wijers, R.~A.~M.~J., et al.\ 
1994, 
\apj, 421, 738 

\bibitem{} Vaughan, B.A., van der Klis, M., M\'endez, M., van Paradijs, J., 
et al.\ 
1997, 
\apjl, 483, L115 

\bibitem{} Vaughan, B.A., van der Klis, M., M\'endez, M., van Paradijs, J., 
et al.\  
1998, 
\apjl, 509, L145 

\bibitem{} Vaughan, B.~A., van der Klis, M., Lewin, W.~H.~G., 
van Paradijs, J., et al.\ 
1999, 
\aap, 343, 197 

\bibitem{} Vietri, M.~\& Stella, L.\ 
1998, 
\apj, 503, 350 

\bibitem{} Vignarca, F., Migliari, S., Belloni, T., Psaltis, D., \& 
van der Klis, M.\ 
2003, 
\aap, 397, 729 

\bibitem{} Vikhlinin, A.\ 
1999, 
\apjl, 521, L45 

\bibitem{} Vikhlinin, A., Churazov, E., \& Gilfanov, M.\ 
1994, 
\aap, 287, 73 

\bibitem{} Vikhlinin, A., Churazov, E., Gilfanov, M., Sunyaev, R., et al.\ 
1995, 
\apj, 441, 779 

\bibitem{} Vrtilek, S.~D., Raymond, J.~C., Garcia, M.~R., Verbunt, F., 
Hasinger, G., et al.\ 
1990, 
\aap, 235, 162 

\bibitem{} Vrtilek, S.~D., Penninx, W., Raymond, J.~C., Verbunt, F., 
Hertz, P.,et al.\ 
1991, 
\apj, 376, 278 

\bibitem{} Vrtilek, S.~D., Raymond, J.~C., Boroson, B., McCray, R., 
Smale, A., et al.\ 
2003, 
\pasp, 115, 1124 

\bibitem{} Wagoner, R.~W.\ 
1999, 
\physrep, 311, 259 

\bibitem{} Wagoner, R.~V.\ 
2002, 
\apjl, 578, L63 

\bibitem{} Wallinder, F.~H.\ 
1995, 
\mnras, 273, 1133 

\bibitem{} Wang, Y.-M.~\& Schlickeiser, R.\ 
1987, 
\apj, 313, 200 

\bibitem{} Wang, D., Ma, R., Lei, W., \& Yao, G.\ 
2003, 
\mnras, 344, 473 

\bibitem{} Warner, B.~\& Woudt, P.~A.\ 
2002, 
\mnras, 335, 84 

\bibitem{} Watarai, K.~\& Mineshige, S.\ 
2003a, 
\pasj, 55, 959 

\bibitem{} Watarai, K.~\& Mineshige, S.\ 
2003b, 
\apj, 596, 421 

\bibitem{} Weinberg, N., Miller, M.~C., \& Lamb, D.~Q.\ 
2001, 
\apj, 546, 1098 

\bibitem{} Weisskopf, M.~C.~\& Sutherland, P.~G.\ 
1978, 
\apj, 221, 228 

\bibitem{} Weisskopf, M.~C., Kahn, S.~M., \& Sutherland, P.~G.\ 
1975, 
\apjl, 199, L147 

\bibitem{} Wen, L., Cui, W., \& Bradt, H.~V.\  
2001, 
\apjl, 546, L105 

\bibitem{} Wheeler, J.~C.\ 
1977, 
\apj, 214, 560 

\bibitem{} White, N.~E.~\& Marshall, F.~E.\ 
1984, 
\apj, 281, 354 

\bibitem{} White, N.E., \& Zhang, W.\ 
1997, 
\apjl, 490, L87 

\bibitem{} White, N.~E., Peacock, A., \& Taylor, B.~G.\ 
1985, 
\apj, 296, 475 

\bibitem{} Wijers, R.~A.~M.~J., van Paradijs, J., \& Lewin, W.~H.~G.\ 
1987, 
\mnras, 228, 17P 

\bibitem{} Wijnands, R.\ 
2004,
in Rossi and beyond, AIP Conf. Proc. 714, 97; 
astro-ph/0403409 

\bibitem{} Wijnands, R.~\& Homan, J.\ 
2003, 
ATel, 165, 1 

\bibitem{} Wijnands, R.~\& Miller, J.~M.\ 
2002, 
\apj, 564, 974 

\bibitem{} Wijnands, R.A.D., \& van der Klis, M.\  
1997, 
\apjl, 482, L65 

\bibitem{} Wijnands, R.~\& van der Klis, M.\ 
1998a, 
\apjl, 507, L63 

\bibitem{} Wijnands, R. \& van der Klis, M.\ 
1998b, 
\nat, 394, 344 

\bibitem{} Wijnands, R. \& van der Klis, M.\  
1999a, 
\apjl, 514, 939 

\bibitem{} Wijnands, R.~\& van der Klis, M.\ 
1999b, 
\aap, 345, L35 

\bibitem{} Wijnands, R.~\& van der Klis, M.\ 
1999c, 
\apj, 522, 965 

\bibitem{} Wijnands, R.~\& van der Klis, M.\ 
2000, 
\apjl, 528, L93 

\bibitem{} Wijnands, R.~\& van der Klis, M.\ 
2001, 
\mnras, 321, 537 

\bibitem{} Wijnands, R.~A.~D., van der Klis, M., Psaltis, D., Lamb, F.~K., 
Kuulkers, E., et al.\ 
1996, 
\apjl, 469, L5 

\bibitem{} Wijnands, R.A.D., van der Klis, M., van Paradijs, J., Lewin, W.H.G., 
et al.\  
1997a, 
\apjl, 479, L141 

\bibitem{} Wijnands, R., Homan, J., van der Klis, M., M\'endez, M.,
Kuulkers, E., et al.\ 
1997b, 
\apjl, 490, L157 

\bibitem{} Wijnands, R.~A.~D., van der Klis, M., Kuulkers, E., Asai, K., \& 
Hasinger, G.\ 
1997c, 
\aap, 323, 399 

\bibitem{} Wijnands, R., Homan, J., van der Klis, M., Kuulkers, E.,  
van Paradijs, J., et al.\  
1998a,
\apjl, 493, L87  

\bibitem{} Wijnands, R.A.D., van der Klis, M., M\'endez, M., van Paradijs, J., 
et al.\ 
1998b, 
\apjl, 495, L39 

\bibitem{} Wijnands, R., M\'endez, M., van der Klis, M., Psaltis, D., Kuulkers, E., 
et al.\ 
1998c, 
\apjl, 504, 35  

\bibitem{} Wijnands, R., van der Klis, M., \& Rijkhorst, E.\ 
1999a, 
\apjl, 512, L39 

\bibitem{} Wijnands, R., Homan, J., \& van der Klis, M.\ 
1999b, 
\apjl, 526, L33 

\bibitem{} Wijnands, R., M{\' e}ndez, M., Miller, J.~M., \& Homan, J.\ 
2001a,
\mnras, 328, 451 

\bibitem{} Wijnands, R., Strohmayer, T., \& Franco, L.~M.\ 
2001b, 
\apjl, 549, L71 

\bibitem{} Wijnands, R., Miller, J.~M., \& van der Klis, M.\ 
2002a, 
\mnras, 331, 60 

\bibitem{} Wijnands, R., Muno, M.~P., Miller, J.~M., Franco, L.~M., 
Strohmayer, T., et al.\ 
2002b, 
\apj, 566, 1060 

\bibitem{} Wijnands, R., van der Klis, M., Homan, J., Chakrabarty, D., 
et al.\ 
2003, 
\nat, 424, 44 

\bibitem{} Wilms, J., Nowak, M.~A., Pottschmidt, K., Heindl, W.~A., et al.\ 
2001, 
\mnras, 320, 327 

\bibitem{} Wood, K.~S., Ray, P. S., Bandyopadhyay, R. M., Wolff,
M. T., et al.\ 
2000, 
\apjl, 544, L45 

\bibitem{} Woods, P.~M., Kouveliotou, C., Finger, M.~H., Gogus, E., Swank, J., 
et al.\ 
2002, 
\iaucirc, 7856, 1 

\bibitem{} Xiong, Y., Wiita, P.~J., \& Bao, G.\ 
2000, 
\pasj, 52, 1097 

\bibitem{} Yamaoka, K., Ueda, Y., Inoue, H., Nagase, F., Ebisawa, K.,
et al.\ 
2001, 
\pasj, 53, 179 

\bibitem{} Yoshida, S.~\& Lee, U.\ 
2001, 
\apj, 546, 1121 

\bibitem{} Yoshida, K., Mitsuda, K., Ebisawa, K., Ueda, Y., Fujimoto, R., et al.\ 
1993, 
\pasj, 45, 605 

\bibitem{} Yu, W.~\& van der Klis, M.\ 
2002, 
\apjl, 567, L67 

\bibitem{} Yu, W., Zhang, S.N., Harmon, B.A., Paciesas, W.S., Robinson, C.R., et al.\ 
1997, 
\apj, 490, L153 

\bibitem{} Yu, W., Li, T.P., Zhang, W., \& Zhang, S.N.\ 
1999, 
\apjl, 512, L35 

\bibitem{} Yu, W., van der Klis, M., \& Jonker, P.~G.\ 
2001, 
\apjl, 559, L29 

\bibitem{} Yu, W., Klein-Wolt, M., Fender, R., \& van der Klis, M.\ 
2003, 
\apjl, 589, L33 

\bibitem{} Zdunik, J.~L., Haensel, P., Gondek-Rosi{\' n}ska, D., \& 
Gourgoulhon, E.\ 
2000, 
\aap, 356, 612 

\bibitem{} Zdziarski, A.~A., Poutanen, J., Paciesas, W.~S., \& Wen, L.\ 
2002, 
\apj, 578, 357 

\bibitem{} Zhang, W., Lapidus, I., White, N.E., \& Titarchuk, L.\  
1996a, 
\apjl, 469, L17 

\bibitem{} Zhang, W., Lapidus, I., White, N.E., \& Titarchuk, L.\  
1996b, 
\apjl, 473, L135 

\bibitem{} Zhang, W., Morgan, E.~H., Jahoda, K., Swank, J.~H.,
Strohmayer, T.~E., et al.\ 
1996c, 
\apjl, 469, L29 

\bibitem{} Zhang, S.~N., Cui, W., Harmon, B.~A., Paciesas, W.~S., 
Remillard, R.~E.,et al.\ 
1997a, 
\apjl, 477, L95 

\bibitem{} Zhang, W., Strohmayer, T.E., \& Swank, J.H.\  
1997b, 
\apjl, 482, L167 

\bibitem{} Zhang, S.~N., Ebisawa, K., Sunyaev, R., Ueda, Y., Harmon,
B. A., et al.\ 
1997c, 
\apj, 479, 381 

\bibitem{} Zhang, W., Strohmayer, T.E., \& Swank, J.H.\ 
1998a, 
\apj, 500, L167 

\bibitem{} Zhang, W., Smale, A.P., Strohmayer, T.E., \& Swank, J.H.\ 
1998b, 
\apj, 500, L171 

\bibitem{} Zhang, W., Jahoda, K., Kelley, R.L., Strohmayer, T.E., Swank, J.H., 
\& Zhang, S.N.\   
1998c, 
\apj, 495, L9 

\bibitem{} {\. Z}ycki, P.~T.\ 
2002, 
\mnras, 333, 800 

\bibitem{} {\. Z}ycki, P.~T.\ 
2003, 
\mnras, 340, 639 

\bibitem{} {\. Z}ycki, P.~T., Done, C., \& Smith, D.~A.\ 
1999a, 
\mnras, 305, 231 

\bibitem{} {\. Z}ycki, P.~T., Done, C., \& Smith, D.~A.\ 
1999b, 
\mnras, 309, 561 

\end{thereferences}
\end{document}